\documentclass[a4paper,fleqn,usenatbib,useAMS]{mnras}

\usepackage{graphicx}   
\usepackage{amsmath}    
\usepackage{amssymb}    
\usepackage{multicol}        
\usepackage{bm}         
\usepackage{pdflscape}  

\usepackage[T1]{fontenc}
\usepackage{ae,aecompl}

\title[NGC 7538]{The stellar contents and star formation in the NGC 7538 region}

\author[Sharma et al.]{
Saurabh Sharma,$^{1}$\thanks{E-mail: saurabh@aries.res.in}
A. K. Pandey,$^{1}$ D. K. Ojha,$^{2}$ Himali Bhatt,$^{3}$ K. Ogura,$^{4}$
\newauthor
N. Kobayashi,$^{5}$ R. Yadav$^{6}$ and J. C. Pandey$^{1}$
\\\\
$^{1}$Aryabhatta Research Institute of Observational Sciences (ARIES), Manora Peak, Nainital, 263 002, India, saurabh@aries.res.in\\
$^{2}$Tata Institute of Fundamental Research, Homi Bhabha Road, Colaba, Mumbai - 400 005, India\\
$^{3}$INSPIRE Faculty, Department of Science \& Technology, New Delhi-110 016, India\\
$^{4}$Kokugakuin University, Higashi, Shibuya-ku, Tokyo 150-8440, Japan\\
$^{5}$Institute of Astronomy, University of Tokyo, 2-21-1 Osawa, Mitaka, Tokyo 181-0015, Japan\\
$^{6}$National Astronomical Research Institute of Thailand, Chiang Mai, Thailand 
}

\date{Accepted XXX. Received YYY; in original form ZZZ}

\pubyear{2016}

\begin{document}
\label{firstpage}
\pagerange{\pageref{firstpage}--\pageref{lastpage}}
\maketitle

\begin{abstract}
 Deep optical photometric data on the NGC 7538 region were collected and combined with archival data sets 
from $Chandra$, 2MASS and {\it Spitzer} surveys in order to generate a new catalog of young stellar 
objects (YSOs) including those not showing IR excess emission.
 This new catalog is complete down to 0.8 M$_\odot$.
The  nature of the YSOs associated with the NGC 7538 region and their spatial distribution are used to study 
the star formation process and the resultant mass function (MF) in the region.
Out of the 419 YSOs,  $\sim$91\%  have ages between 0.1 to 2.5 Myr
and $\sim$86\%  have masses between 0.5 to 3.5 M$_\odot$,
as derived by spectral energy distribution fitting analysis.
Around 24\%, 62\% and 2\% of these YSOs are classified to be the Class I, Class II and Class III sources, respectively.
The X-ray activity in the Class I, Class II and Class III objects is not significantly different from each other.
This result implies that the enhanced X-ray surface flux due to the increase in the
rotation rate may be compensated by the decrease in the stellar surface area during the pre-main sequence evolution.
Our analysis shows that the O3V type high mass star `IRS 6' might have 
triggered the formation of young low mass stars up to a radial distance of 3 pc.
The MF shows a turn-off at around 1.5 M$_\odot$ and the value of its slope `$\Gamma$'
in the mass range $1.5 <$M/M$_\odot < 6$ comes out to be $-1.76\pm0.24$, 
which is  steeper than the Salpeter value.
\end{abstract}



\section{Introduction}

Observational studies of bubbles associated with H\,{\sevensize II} 
regions suggest that their expansion probably triggers 14\% to 30\% of 
the star formation in our Galaxy \citep[e.g.,][]{2010A&A...523A...6D,2012MNRAS.421..408T,2012ApJ...755...71K}.
These observational results have revealed the importance of 
OB stars on star formation activity on the Galactic scale. 
Massive stars have a profound effect on their natal environment
creating wind-blown shells, cavities and  H\,{\sevensize II} regions. The immense
amount of energy released through their stellar winds and radiation 
disperses and destroys the remaining molecular gas and likely
inhibits further star formation. However, it has also been argued
that in some circumstances, the energy feedback by these massive stars
can promote and induce subsequent star formation in the
surrounding molecular gas before it disperses \citep{2012ApJ...744..130K}.
Identification and characterization of the young stellar objects (YSOs) in 
star-forming regions (SFRs) hosting massive stars are essential steps to examine the
physical processes that govern the star formation of the next generation in such regions.
One of the notable feedback effects of massive stars is
the triggering of star formation of new generations,
either by sweeping the neighboring molecular gas into a
dense shell which subsequently fragments into pre-stellar cores 
\citep[e.g.,][]{1977ApJ...214..725E,1994MNRAS.268..291W,1998ASPC..148..150E}
or by compressing pre-existing dense clumps 
\citep[e.g.,][]{1982ApJ...260..183S,1989ApJ...346..735B,1994A&A...289..559L}.
The former process is called `collect and collapse' and the latter 
`radiation driven implosion (RDI).'
The aligned elongated distribution of young stellar objects (YSOs) 
 with respect to the high mass star/stars
around the interface of an H\,{\sevensize II}  region and a molecular cloud is 
considered as an observational signature of the RDI process 
\citep{2002AJ....123.2597O,2005ApJ...624..808L},
whereas the neutral compressed layer  observed as a ring of molecular
line emission surrounding the  H\,{\sevensize II} region
is an observational signature of the collect and collapse process \citep[for details, cf.][]{2005A&A...433..565D}.
\citet{2015MNRAS.450.1199D} with the help of the  hydrodynamic simulations of star formation, 
including the feedback from the O-type stars, found that these observational markers 
 does not improve significantly the chances of correctly identifying a given star as triggered,
therefore  they urge caution in interpreting observations of star formation near feedback-driven structures in terms of triggering. Of course, systems where many putative triggering indicators can be satisfied simultaneously are more likely to be genuine sites of triggering.

\begin{figure*}
\centering
\includegraphics[height=12cm,width=12cm]{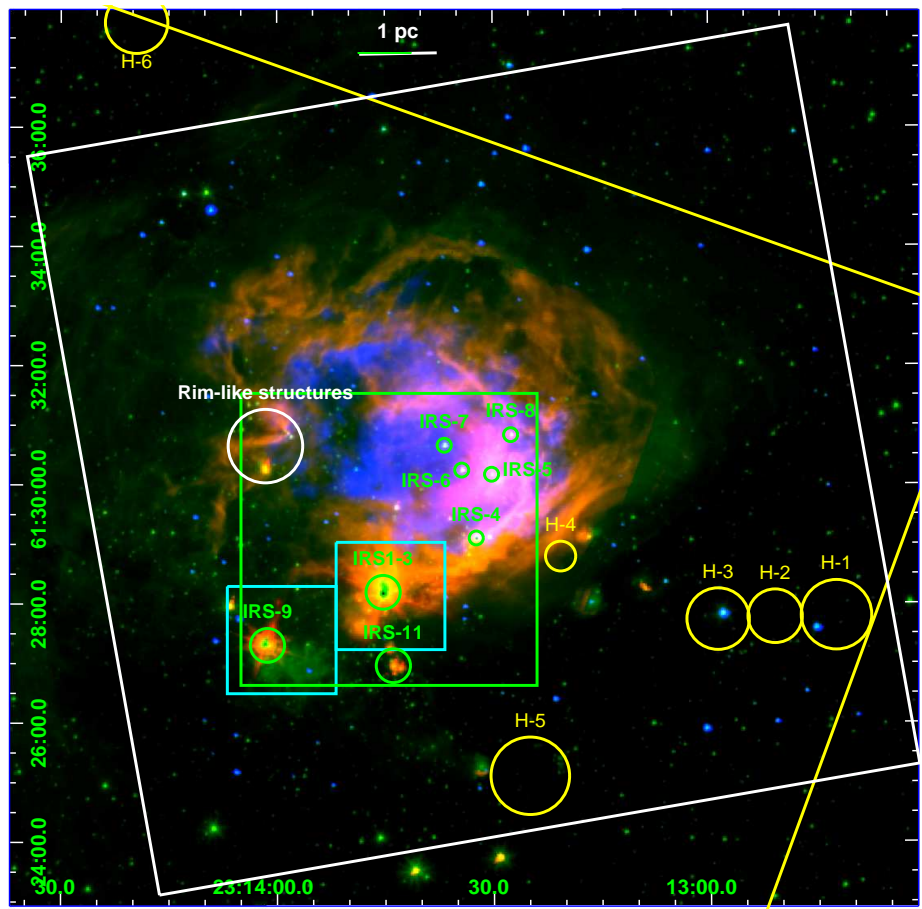}
\caption{\label{color} Color-composite image obtained by  using the H$\alpha$ (blue), 3.6 $\mu$m (green),
and 8.0 $\mu$m (red) images for an area of $\sim15\times 15$ arcmin$^2$ around NGC 7538. 
White box represents the region covered in the optical survey.
 Region below the slightly horizontal yellow line and westward of slightly vertical yellow line represent the area
covered in the $Chandra$ X-ray survey.
The green and cyan boxes represent the areas covered by \citet{2004ApJ...616.1042O} 
and  \citet{2014MNRAS.443.3218M} in their near-infrared (NIR) surveys, respectively.
Positions of various IR sources (IRS 1-11), rim-like structures and 
$Herschel$-identified massive SFRs (H1 - H6) are also shown. The x-axis and y-axis are RA and Dec, respectively. in J2000 epoch.} 
\end{figure*}

NGC 7538 (cf. Fig. \ref{color})  located at a distance of 2.65 kpc (cf. Appendix  A)
is an H\,{\sevensize II}  region which belongs to the Cas 
OB2 complex in the Perseus spiral arm \citep{2008ApJ...685L..51F} and 
it contains massive stars in
different evolutionary stages; main-sequence (MS) stars with spectral types between O3 and O9
\citep[IRS 6 and IRS 5,][]{2010A&A...517A...2P} which ionize the
H\,{\sevensize II} region NGC 7538, 
infra-red (IR) sources IRS 1 \citep[associated  with a disc and an outflow;][]{2009A&A...501..999P,2009ApJ...699L..31S}, 
IRS 2 and IRS 3 \citep{1974ApJ...187..473W} located south of NGC 7538 and associated  with UCH\,{\sevensize II}
regions and with the IR cluster NGC 7538S \citep{1993ApJ...407..657C,2003A&A...397..177B,2010ApJ...715..919S}, and
YSOs like IRS 9 and IRS 11 \citep{1979MNRAS.188..463W}. 
Thus, as a SFR associated with an H\,{\sevensize II}  region, it is ideally suited to study the impact of massive stars 
on the formation of high- and low-mass stars in its surroundings.
In this paper, we will study this region in continuation of our efforts to understand the star formation scenario in such SFRs
\citep{2007MNRAS.380.1141S,2008MNRAS.383.1241P,2008MNRAS.384.1675J, 2009MNRAS.396..964C,2011MNRAS.415.1202C,2012PASJ...64..107S,2012ApJ...759...48M, 2013ApJ...764..172P, 2013MNRAS.432.3445J}.

The NGC 7538 region has been studied by \citet{2004ApJ...616.1042O} by using NIR observations 
centered at IRS 1-3 (cf. Fig. \ref{color}), along with radio continuum observations at 1280 MHz. 
They have identified YSOs and classified them according to their evolutionary stages using NIR two-colour diagrams (TCDs)
and generated the $K_s$-band luminosity function (KLF) to 
discuss the age sequence and  mass spectrum of the YSOs in the region. 
They have discerned several substructures having different evolutionary stages. 
\citet{2004AJ....128.2942B} by using their NIR survey
found that most of the red sources in this region
are concentrated at the southern rim bounding the optical H\,{\sevensize II}  region 
and in the area around the IR sources IRS 1-3. 
\citet{2010A&A...517A...2P} have put an upper limit to the age of the central cluster  as 2.2 Myr
based on the lifetime of the most massive O3V star (60 M$_\odot$, IRS 6).
\citet{2004ApJ...600..269S}, using their  high spatial resolution submillimeter maps, found that
the three major centers of star-forming activity (IRS 1-3, IRS 11, and IRS 9) in the NGC 7538 
region are connected to each other  through filamentary dust ridges. 
Recently, \citet{2014MNRAS.443.3218M} have also observed in NIR 
the two comparatively smaller regions of NGC 7538 
centered on IRS 1-3 and IRS 9 (cf. Fig. \ref{color})
to study the luminosity function (LF) and initial mass function (IMF) of these regions. 
They have complemented their deep NIR observations with X-ray, radio and molecular line observations 
to study the  stellar population,  ionized gas characteristics and dense molecular gas morphology in the region. 
Recently, \citet{2014MNRAS.439.3719C} 
have presented a homogeneous IR and molecular data to study the spatial distribution of YSOs and the 
properties of their clustering and correlation with the surrounding molecular cloud structures. 
They have compared these properties of this massive SFR with those of low-mass SFRs.
However, they could not reach any concrete conclusion about the triggering mechanism 
due to the lack of the investigation of time causality between the expansion of the H\,{\sevensize II} region and the age
of the newly formed stars.

Most of the previous  studies related to the star formation in this region are concentrated mainly on 
the IRS 1-11 region directly associated with NGC 7538. 
In the present work, we will revisit this region, but not only study a wider area  but also
make use of deep optical data of our own along with the  multiwavelength 
archival data sets from various surveys ($Chandra$, {\it Spitzer}, 2MASS).
Whereas the previous studies have neither derived the physical parameters (age/masses) of the 
individual YSOs 
nor checked for their association with the NGC 7538 region,
we have done them 
and present a catalog of YSOs containing these information. For part of them spectral energy distribution (SED) fitting analyses have been applied.
The spatial distribution of these YSOs, along with those of the mid-infrared (MIR) and radio emission,
and the mass function (MF), will be used to constrain the star formation history in the region.

The rest of this paper is organized as follows: 
In Section 2, we describe the optical and archival data sets. 
In Section 3, we identify YSOs, catalog them according to their evolutionary stages,
and derive their physical parameters.
The X-ray spectral analysis of the YSOs is also explained in this section.
All of these analyses are discussed in Section 4  and
the conclusions are summarized in Section 5.

\section{Observation and data reduction}

\subsection{Optical data}

The CCD $UBV{(RI)}_c$ and $H\alpha$+continuum photometric data of the NGC 7538 region, 
centered at $\alpha_{J2000}$: 23$^h$13$^m$38$^s$, $\delta_{J2000}$: +61$^\circ$31$^\prime$23$^\prime$$^\prime$; 
$l=111.548^\circ$ and $b=0.832^\circ$,
were acquired on 06, 07 November 2004 and 25 October 2005, respectively, by using 
the $2048\times 2048$ pixel$^2$ CCD camera mounted on the f/13 Cassegrain focus of the 
104-cm Sampurnanand telescope of Aryabhatta Research Institute of Observational Sciences (ARIES), Nainital, India. 
In this set up, each pixel of the CCD corresponds to $0.37$ arcsec and the entire chip 
covers a field of $\sim 13\times13$ arcmin$^2$ on the sky. 
To improve the signal to noise ratio (SNR), the observations were carried out in 
the binning mode of $2\times2$ pixel. 
The read-out noise and gain of the CCD are 5.3 $e^-$ and 10 $e^-$/ADU respectively. 
The average FWHMs of the star images were $\sim2$ arcsec. 
The broad-band $UBV{(RI)}_c$ observations of the NGC 7538 region were standardized by observing stars in the 
SA95 field  \citep{1992AJ....104..340L} centered on SA 112  
($\alpha_{J2000}$: 03$^h$53$^m$40$^s$, $\delta_{J2000}$: -00$^\circ$00$^\prime$54$^\prime$$^\prime$)
on 06 November 2004. 
A reference field of $\sim 13\times13$ arcmin$^2$ located about
10 arcmin away towards south-west of the NGC 7538 region was also observed to estimate 
the contamination due to foreground/background field stars on 30 November 2005 in $VI_c$.
This reference field were standardized by using the common stars with the NGC 7538 region.
The log of the observations is given in Table \ref{log}.
In Fig. \ref{color}, we have shown the observed region (white box) on the colour-composite image 
of NGC 7538  obtained by combining the 3.6 $\mu$m (blue colour), H$\alpha$ (green colour) and 8.0 $\mu$m (red colour) images.

\begin{table}
\caption{\label{log}  Log of optical observations with the 104-cm Sampurnanand telescope, Nainital.}
\begin{tabular}{@{}rr@{}}
\hline
Date of observations/Filter& Exp. (sec)$\times$ No. of frames\\
\hline
&NGC 7538\\
06 November 2004\\
$U$   &  $300\times4$\\
$B$   &  $180\times4,60\times3$\\
$V$   &  $60\times4$\\
$R_c$ &  $30\times4$\\
$I_c$ &  $30\times4$\\
\\
07 November 2004\\
$U$   &  $900\times3$\\
$B$   &  $600\times3$\\
$V$   &  $600\times3$\\
$R_c$ &  $300\times4$\\
$I_c$ &  $300\times4$\\
\\
25 October 2005\\
$H\alpha$ &$900\times2,300\times3$\\
$Continuum$&$900\times1,300\times1$\\
\\
&Reference field\\
30 November 2005\\
$V$   &  $600\times3$\\
$I_c$   &  $300\times3$\\
\hline
\end{tabular}
\end{table}

\begin{figure}
\centering
\includegraphics[height=6cm,width=7cm]{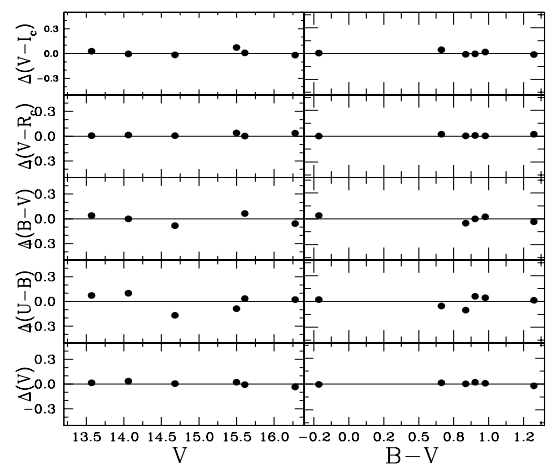}
\caption{\label{resd} Differences between standard and transformed magnitudes 
and colours of standard
stars plotted against the standard $V$ magnitude and $(B-V)$ colour.}
\end{figure}

Initial processing of the data frames was done by 
using the IRAF\footnote{IRAF is distributed by National Optical Astronomy
Observatories, USA} and ESO-MIDAS\footnote{ ESO-MIDAS is developed and
maintained by the  European Southern Observatory.} data reduction packages. Photometry  of
the cleaned frames was carried out by using DAOPHOT-II software \citep{1987PASP...99..191S}.
The point spread function (PSF) was obtained for each frame by using several uncontaminated
stars. Magnitudes obtained from different frames
were averaged. When brighter stars were saturated on deep exposure frames, their
magnitudes have been taken from short exposure frames.
We used the DAOGROW program for construction of an aperture growth curve required for
determining the difference between the aperture and PSF magnitudes.

\begin{figure}
\centering
\includegraphics[height=7cm,width=7cm]{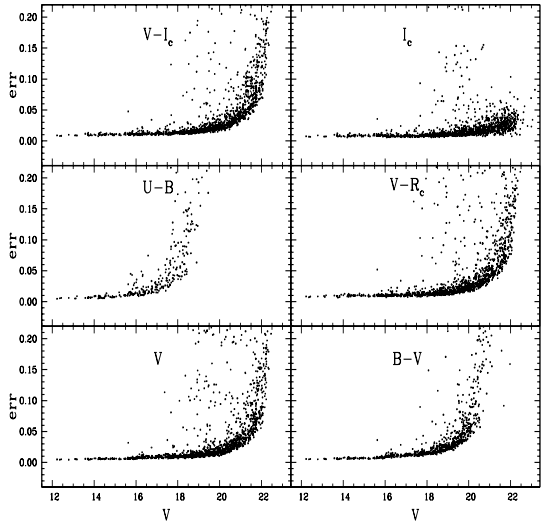}
\caption{\label{err}DAOPHOT errors in magnitudes and colours as a function
of $V$ magnitude.  } 
\end{figure}         

Calibration of the instrumental magnitudes to the standard system was
done by using the procedures outlined by \citet{1992ASPC...25..297S}.
The calibration equations
derived by the least-squares linear regression are as follows:\\

{\it  \small
\noindent
$u= U + (7.178\pm0.029) -(0.035\pm0.039)(U-B) + (0.467\pm0.046)X$,

\noindent
$b= B + (4.771\pm0.027) -(0.037\pm0.028)(B-V) + (0.255\pm0.025)X$,

\noindent
$v= V + (4.274\pm0.007) -(0.025\pm0.005)(V-I_c) + (0.143\pm0.005)X$,

\noindent
$r_c= R_c + (4.169\pm0.005) -(0.028\pm0.009)(V-R_c) + (0.102\pm0.003)X$,

\noindent
$i_c= I_c + (4.611\pm0.011) +(0.022\pm0.008)(V-I_c) + (0.072\pm0.008)X$ \\

}

where $U,B,V,R_c$ and $I_c$ are the standard magnitudes and $u,b,v,r_c$ and $i_c$ are the
instrumental aperture magnitudes normalized for 1 second of exposure time and $X$
is the airmass. We have ignored the second-order colour correction terms as they are generally
small in comparison to other errors present in the photometric data reduction.
Fig. \ref{resd} shows the standardization residual, $\Delta$, between standard and transformed
$V$ magnitudes and $(U-B), (B-V), (V-R_c)$, and $(V-I_c)$ colours of the standard stars as a function of $V$ magnitude and $(B-V)$ colour. 
As can be seen from the figure, the residuals are not showing any trends with colour or magnitude.
The typical DAOPHOT errors in magnitude and colour as a function of $V$ magnitude,
are shown in Fig. \ref{err}.
It can be seen that, in the $V$ band, the errors become large ($\ge$0.1 mag) for stars
fainter than $V\simeq22$ mag, so the measurements beyond this magnitude are not reliable.
In total 969 sources, with detection at least in the $V$ and $I_c$ bands and errors less than 0.1 mag,
have been identified in this study.

\subsection{Grism slitless spectroscopy}

Slitless grism spectroscopy has also been carried out for the NGC 7538 region in search for H$\alpha$ emission 
stars. The observations were made with the Himalayan Faint Object Spectrograph Camera (HFOSC) on the 2-m 
{\it Himalayan Chandra Telescope}  (HCT) of the Indian Astronomical Observatory (IAO), Hanle, India on 
2004 September 21.
A combination of a `wide H$\alpha$' interference filter (6300-6740 \AA) and grism 5 
(5200-10300 \AA) of HFOSC was used without slit. The central $2 K \times 2 K$ pixel of the $2 K \times 4 K$ CCD 
was used for data acquisition. The pixel size is 15 $\mu$m with an image scale of 0.297 arcsec pixel$^{-1}$. 
The resolution of the spectra is 870. We secured three spectroscopic
frames of 5-min and 1-min exposure each with the grism in as well as three direct
frames of 1-min exposure each with the grism out. The seeing size at the time of the 
observations was $\simeq$ 2 arcsec.

\begin{table}
\caption{\label{slitless} $H\alpha$ emission stars identified by grism spectroscopy.}
\begin{tabular}{@{}ccccc@{}}
\hline
   ID  &  $\alpha_{J2000}$ & $\delta_{J2000}$ &  EW (H$\alpha$)\\ 
       &   ($^h$:$^m$:$^s$)  & ($^\circ$:$^\prime$:$^\prime$$^\prime$) &  (\AA)\\
\hline
  G1& 23:13:16.66&  +61:28:01.4& 9.9  \\
  G2& 23:13:32.75&  +61:27:49.7&  91.2\\
  G3& 23:13:56.36&  +61:25:06.9&  3.0 \\
  G4& 23:14:06.64&  +61:27:57.7&  35.0\\
  G5& 23:14:13.04&  +61:34:52.4&  8.4 \\
  G6& 23:14:13.38&  +61:31:46.6&  10.6 \\
\hline
\end{tabular}
\end{table}

\subsection{Archival data}

\subsubsection{$Chandra$ X-ray data}

The NGC 7538 region has also been observed by $Chandra$ X-ray Observatory on 25 March, 2005 (cf. Fig. \ref{color}) 
centered
at $\alpha_{J2000}$: 23$^h$13$^m$47.8$^s$, $\delta_{J2000}$: +61$^\circ$28$^\prime$15.5$^\prime$$^\prime$.
The satellite roll angle (i.e. the orientation of the CCD array relative to north-south direction) during the observations was 19.2$^\circ$.
Exposures for 30 ks were obtained in the {\it very faint data mode}  with a 3.2 s frame time by 
using the ACIS-I imaging array as the primary 
detector. ACIS-I consists of four front illuminated $1024\times1024$ CCDs with a pixel size of 0.492 arcsec  and a combined 
FOV of $\sim$17$\times$17 arcmin$^2$. 
The observed region of NGC 7538 by $Chandra$ is shown in  Fig. \ref{color}.
More detailed information on $Chandra$  and its instrumentation can be found 
in the $Chandra$ Proposer's Guide\footnote{See http://asc.harvard.edu/proposer/POG}. 
\citet{2005prpl.conf.8307T} have used this data and presented a preliminary report
indicating 180 X-ray sources in this region. 123 of them have 2MASS NIR counterparts. 
They have studied IR sources in this region and found that IRS 2, 4, 5, 6, 9  and 11
are bright in X-ray and have soft X-ray spectra similar to early type field stars.
IRS 7 and 8  were identified as foreground stars due to absence of X-ray
and, IRS 1 and IRS 2 were not detected probably due to high extinction.

Recently, \citet{2014MNRAS.443.3218M} have reported 27 sources with NIR counterparts  
in the IRS 1-3 and IRS 9 regions (cf. Fig. \ref{color}).

\begin{figure}
\centering\includegraphics[height=7cm,width=7cm,angle=0]{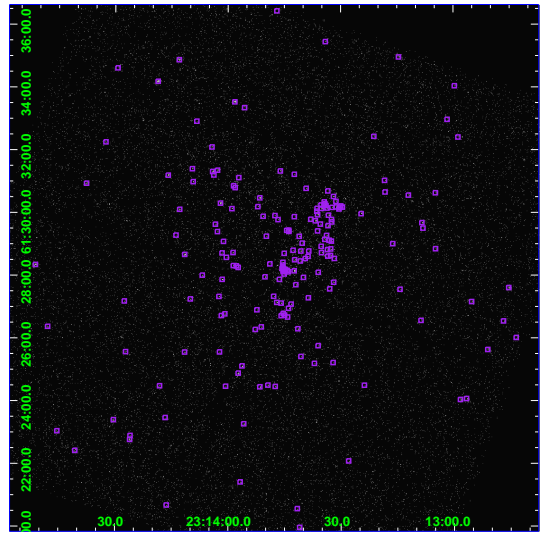}
\caption{\label{xray} X-ray image obtained from the $Chandra$ observations. 
The identified point sources are shown by purple squares. The x-axis and y-axis are RA and Dec. in J2000 epoch. }
\end{figure}

In this study, X-ray data will be used to study the X-ray properties of the YSOs.
We have used CIAO 4.8 data analysis software in combination with the $Chandra$ calibration database CALDB v.4.7.1
to reduce the X-ray data. 
The  light curves from on-chip  background regions were inspected for possible 
large background fluctuations that might have resulted from solar flares, 
however we did not find any such fluctuations. 
After filtering the data for the energy band 0.5 to 8.0 keV, 
the time integrated background was found to be 0.06 counts arcsec$^{-2}$.
Source detection was then performed with the Palermo Wavelet Detection, 
PWDetect\footnote{http://www.astropa.unipa.it/progetti\_ricerca/PWDetect/} code \citep{1997ApJ...483..350D}. 
It analyzes the data at different spatial scales, allowing the detection of both point-like and moderately extended sources, 
and efficiently resolving close pairs. The most important input parameter for this code is the detection threshold, 
which we estimated from the relationship between the background level and expected number of spurious detection 
due to Poisson noise \citep[for detail, see][]{1997ApJ...483..350D}. 
The background level was determined with the {\sc BACKGROUND} command of {\sc XIMAGE}. 
After filtering events in the energy band 0.5-8.0 keV,
the ACIS-I observations comprise a total of about 56 k counts.
This background level translates into a SNR threshold 
of 5.0 if we accept one spurious detection in the 
FOV, or into  SNR 4.6 if we accept 10 spurious detections.  
The first choice results in detection of 158 sources, whereas the second one detects 193 sources. 
We decided to adopt the second criterion; however, we manually rejected 3 sources because they were detected  either in the CCD 
gaps or twice.  Finally, 190 X-ray point sources have been adopted for further analyses.
The locations of these 190 X-ray point sources are shown in Fig. \ref {xray} overlaid on the $Chandra$ X-ray image.

\subsubsection{{\it Spitzer} and 2MASS IR data \label{obs-spit}}

This region was observed by $Spitzer$ spacebased telescope  on 14 August, 2006 (PI: Giovanni Fazio; program ID: 30784)
with Infrared Array Camera (IRAC). 
These images  have been mosaiced to create a $15\times15$ arcmin$^2$ FOV (cf. Fig. \ref{color})
in all IRAC bands by using the MOPEX software provided by $Spitzer$ Science Center (SSC),
containing the common area observed in the optical and X-ray wavelengths.
These mosaiced images are used to make a color composite image (cf. Fig. \ref{color}) 
of this region.

\citet{2004ApJ...616.1042O} have done NIR observations in a $5\times5$ arcmin$^2$ FOV
near IRS 1-3 (cf. Fig. \ref{color}) by using 2.2m University of Hawaii telescope. 
The NIR survey has 10$\sigma$ limiting magnitudes of 19.5, 18.4, and 17.3 in the $J$, $H$, and $K_s$ bands, respectively. 
Comparatively smaller regions ($1.8\times1.8$ arcmin$^2$  FOV) centered on IRS 1-3 and IRS 9 (cf. Fig \ref{color}) 
were observed by using the 8.2m Subaru telescope by \citet{2014MNRAS.443.3218M} in $J$, $H$, and $K$ bands. 
The 10$\sigma$ limiting magnitudes for these observations were $\sim$22, 21, and 20 in the $J$, $H$, and $K$ bands, respectively. 
Also, recently, \citet{2014MNRAS.439.3719C} have compiled $J$ and $K$ data (upto $K\sim17.5$ mag) in a widest FOV of $20\times20$ arcmin$^2$ 
in the NGC 7538 region using the 2.1m telescope at Kitt Peak National Observatory.  
These authors have published their YSOs catalog, 
which are  merged together to form a combined catalog of YSOs (cf. Section 3.1.2)
in the present study.

We have also used the 2MASS Point Source Catalog (PSC) \citep{2003yCat.2246....0C} for NIR (JHK$_s$)
photometry in the NGC 7538 region. This catalog is reported to be 99$\%$ complete down 
to the limiting magnitudes of 15.8, 15.1 and 14.3 in the $J$, $H$ and
$K_s$ band, respectively\footnote{http://tdc-www.harvard.edu/catalogs/tmpsc.html}.
We have selected only those sources which have NIR photometric accuracy $<$ 0.2 mag and detection
in at least $H$ and $K_s$ bands.

\section{Analysis and results}

\subsection{Identification of  YSOs \label{idf}}

YSOs are usually grouped into the evolutionary classes 0, I, II, and III, which represent
in-falling protostars, evolved protostars, classical T-Tauri stars (CTTSs) and weak-line 
T-Tauri stars (WTTSs), respectively \citep[cf.][]{1999ARAA..37..363F}. Class 0
YSOs are so deeply buried inside the molecular clouds that they are not visible at optical or NIR
wavelengths 
whereas Class I are characterized with the growth of an accretion disk surrounded by an envelope and are 
visible in IR and occasionally in optical if viewed in the pole-on direction 
\citep{2005A&A...429..543N,2007prpl.conf..117W,2011ARA&A..49...67W}.
The CTTSs feature disks from which the material is accreted and 
line emission, e.g., in H$\alpha$, can be seen due to this accreting material. These disks can also be probed through
their IR excess (over the normal stellar photospheres). WTTSs on the contrary have little
or no disk material left and hence have no strong H$\alpha$ emission and IR excess.
As the level of X-ray emission of pre-main sequence (PMS) stars are higher than that of MS stars 
\citep[e.g.,][]{1999ARA&A..37..363F,2002ApJ...574..258F,2003SSRv..108..577F,2004A&ARv..12...71G,2012A&A...539A..74C},
X-ray observations provide a very efficient means of
selecting WTTSs associated with the SFRs, which might be missed in the surveys 
based only on  H$\alpha$ emission and/or IR excess emission.
Below, we report identification of YSOs on the basis of their
H$\alpha$, IR and/or X-ray emission.

\begin{figure}
\centering\includegraphics[height=4cm,width=4cm,angle=0]{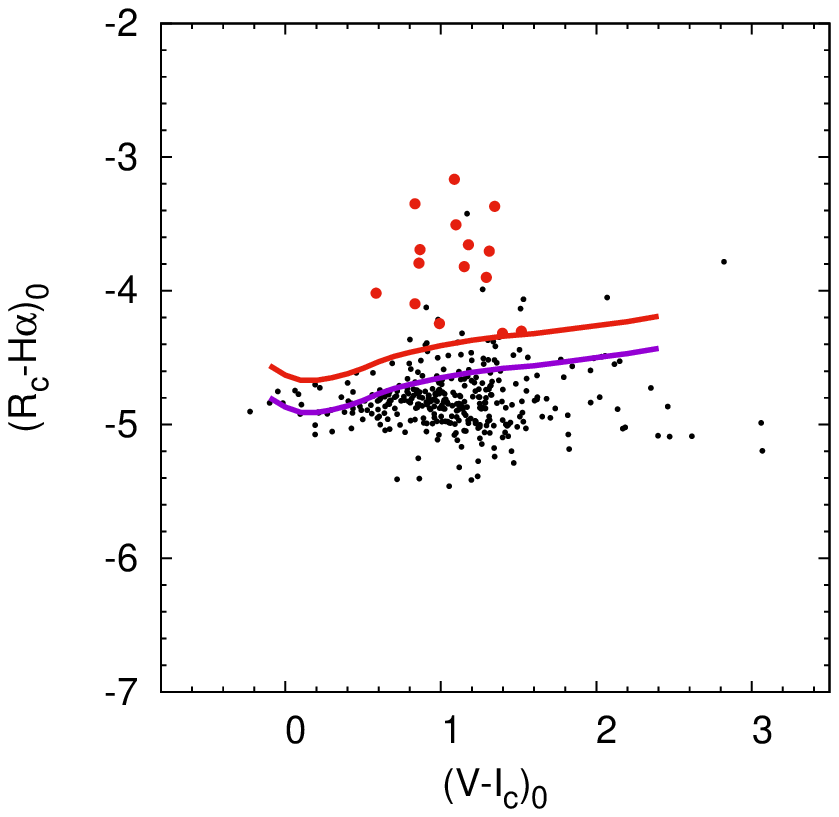}
\centering\includegraphics[height=4cm,width=4cm,angle=0]{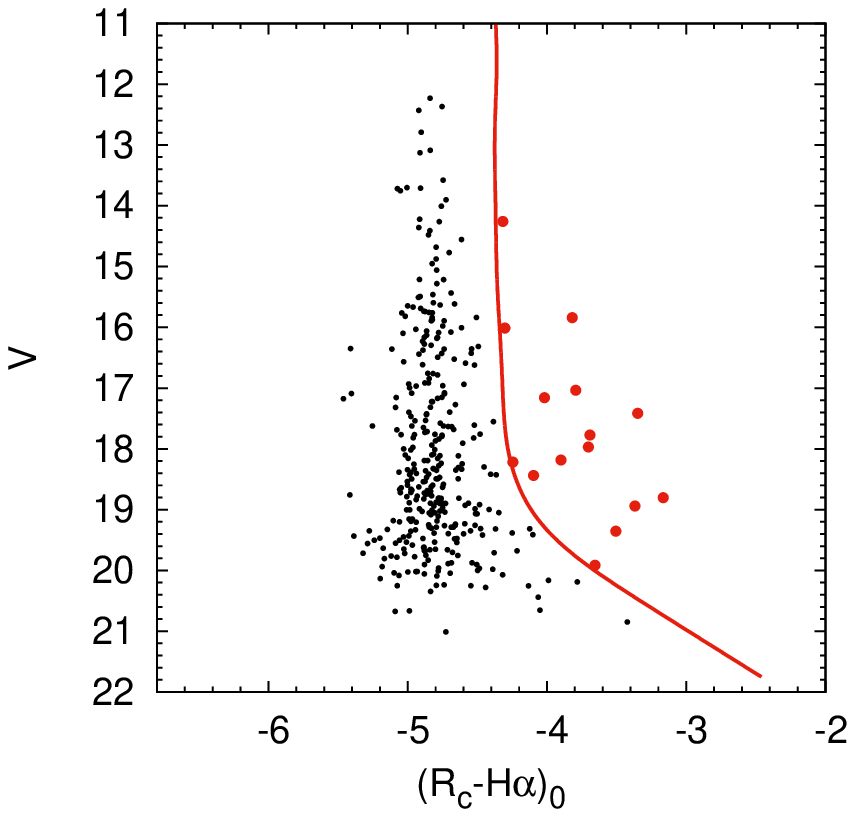}
\caption{\label{halp} {\it (Left-hand panel):} $(R_c- H\alpha)_0$ index is shown as a function of $(V-I_c)_0$
colour. The blue curve indicates the relation for MS stars as taken from
\citet{1997AJ....114.2644S} and reddened for $A_V$= 2.3 mag (cf. Appendix A). The red curve yields
the thresholds for H$\alpha$ emitter candidates (cf. Section 3.1.1). {\it (Right-hand panel):} $V$ versus $(R_c - H\alpha)_0$ CMD.
The solid curve demarcates the  H$\alpha$ emitter candidates from MS stars. 
The red dots represent selected H$\alpha$ emitting candidates.}
\end{figure}

\subsubsection{H$\alpha$ emission}

H$\alpha$ emission stars in a SFR are considered as PMS stars or candidates and
the level of its strength (expressed by equivalent width `EW$_{H\alpha}$')
could be a direct indicator of their evolutionary stages. The conventional distinction between
CTTS and WTTS is EW$_{H\alpha}$  = 10\AA~ and it is greater for the former \citep[cf.][]{1988cels.book.....H}.
The slitless grism frames were visually inspected to identify stars with H$\alpha$ enhancement.
Then the EW$_{H\alpha}$ of these stars has been estimated by using the APALL task of IRAF.
EWs in pixels is converted into angstroms by multiplying 3.8 to it, as has been calculated for grism 5 
having resolution R = 870\footnote{http://www.iiap.res.in/iao/hfosc.html} and 
assuming sampling of 2 pixels. 

We have identified six  H$\alpha$ emission stars spectroscopically and their position and EW$_{H\alpha}$ are given in Table \ref{slitless}. 
Based on the value of EWs, three each were classified as Class II and Class III sources, respectively. 

To identify H$\alpha$ emission stars we have also used narrow band H$\alpha$ photometry based on its excess emission.
In the study of NGC~6383, \citet{2010AA...511A..25R} found that H$\alpha$ equivalent width of 10\AA 
corresponds to $(R_c-H\alpha)$ index 
of 0.24 $\pm$ 0.04 mag  above the MS relation introduced by \citet{1997AJ....114.2644S}.
In the present study, we have used this condition to identify H$\alpha$ emission stars
and classified them as Class II sources.
Since our H$\alpha$ photometry is not calibrated, in order to compare this H$\alpha$ magnitude to that 
of \citet{1997AJ....114.2644S}, 
we estimated the zero point by comparing visually the observed $(R_c-H\alpha)$ and dereddened $(V - I_c)$ with the $(R_c-H\alpha)_0$
versus $(V - I_c)_0$ relation of emission-free MS stars as determined by \citet{1997AJ....114.2644S}
for the NGC 2264 region. We have to subtract 3.85  mag from the observed values to obtain $(R_c-H\alpha)$
index in the \citet{1997AJ....114.2644S} system. 
The $(V -I_c)$ colour is dereddened by the minimum extinction value derived in Appendix A (i.e., $E(B-V)_{min}$ = 0.75  mag and $R_V$ = 2.82).
In the left-hand panel of Fig.~\ref{halp}, we have plotted the $(R_c-H\alpha)_0$ versus $(V-I_c)_0$ distribution of
all the stars along with the MS relation (blue curve) given by \citet{1997AJ....114.2644S}.
All the sources above the red curve \citep[0.24 mag above the MS relation, cf.,][]{1997AJ....114.2644S} are assumed as H$\alpha$ excess emission sources of Class II evolutionary stage.
Since there is a large scatter in the distribution in Fig.~\ref{halp}; left-hand panel, there may be
some false identifications of H$\alpha$ emitters. To minimize false detections, we have introduced
another selection criterion using the $V$ versus $(R_c-H\alpha)_0$
CMD (cf. Fig.~\ref{halp}; right-hand panel). We define an envelope which
contains MS stars keeping in mind its broadening at faint magnitudes due to large photometric errors,
possible presence of binaries, field stars etc. \citep{1994ApJS...90...31P}.
The stars which are having values of
$(R_c-H\alpha)_0 - \sigma_{(R_c-H\alpha)_0}$  larger than that of the envelope of the MS can be safely assumed as
probable H$\alpha$ emitters \citep{2014A&A...567A.109K}. 
Our final criterion for photometrically selected H$\alpha$ emitters is to satisfy both of the above conditions. 
15 sources  (red dots in  Fig.~\ref{halp}) were identified
as H$\alpha$ emitting YSOs and were categorized as Class II YSOs.
None of them were identified through grism spectroscopy.

In total 21 stars were classified as YSOs (3 as Class III and 18 as Class II) 
on the basis of H$\alpha$ spectroscopy and photometry. 

\subsubsection{IR emission}

\begin{figure}
\centering\includegraphics[height=7cm,width=7cm,angle=0]{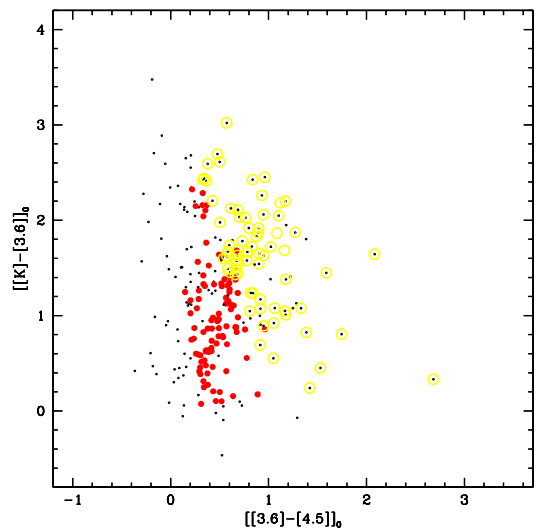}
\caption{\label{IRAC}  ${[[K] - [3.6]]}_0$ vs. ${[[3.6] - [4.5]]}_0$ TCD 
for those YSOs identified by \citet{2014MNRAS.439.3719C} that lack detection at
either 5.8 $\mu$m or 8.0 $\mu$m, but have NIR detection in the $K$ band (black dots). 
The YSOs classified as Class I and Class II, based on the colour criteria by \citet{2009ApJS..184...18G}, 
are shown by using yellow circles  and  red dots, respectively.
 }
\end{figure}

Recently, \citet{2014MNRAS.439.3719C} have identified 562 YSOs in a $15 \times 15$ arcmin$^2$
FOV of the NGC 7538 region on the basis of NIR/MIR excess emission.
They classified 234 YSOs which have detections in all four IRAC bands
on the basis of the  value of the slopes `$\alpha_{IRAC}$' of their SEDs. 
We further used  [[3.6] - [4.5]]$_0$ versus [[$K$] - [3.6]]$_0$  TCD \citep[for details see,][]{2009ApJS..184...18G} 
for the classification of remaining YSOs.
YSOs having $[J-H]$ color $ \ge 0.6$ mag and lying above the
CTTS locus or its extension are traced back to CTTS locus or its extension to get their intrinsic colors.
For YSOs that lack $J$ band photometry, baseline colors in the
$[H - K]$ versus [[3.6] - [4.5]] color-color space YSO locus, as measured by \citet{2005ApJ...632..397G} is used
to get their intrinsic colors.
43 Class I  and  108 Class II  sources were classified by using their location 
in the above mentioned TCD and are shown as yellow circles and red dots, respectively in  Fig. \ref{IRAC}.
For those YSOs which are not detected in the IRAC bands  but have NIR $(JHK)$ detections and have $(J-H)>0.6$ mag (13 sources), we have used 
the classical NIR TCDs \citep[cf.][]{2004ApJ...616.1042O,2012PASJ...64..107S} for their classifications.
Finally, out of 562 YSOs, 239 (234 IRAC four band sources plus 5 MIPS sources) were classified by  \citet[][]{2014MNRAS.439.3719C}
and 164 are classified in this study.
Remaining 159 YSOs could not be classified as 
either they have  no IRAC detections and $(J-H)< 0.6$ mag
or they do not fall at the location of Class I/II YSOs in IRAC [[3.6] - [4.5]]$_0$ versus [[$K$] - [3.6]]$_0$ TCD.

\begin{figure}
\centering\includegraphics[height=7cm,width=7cm,angle=0]{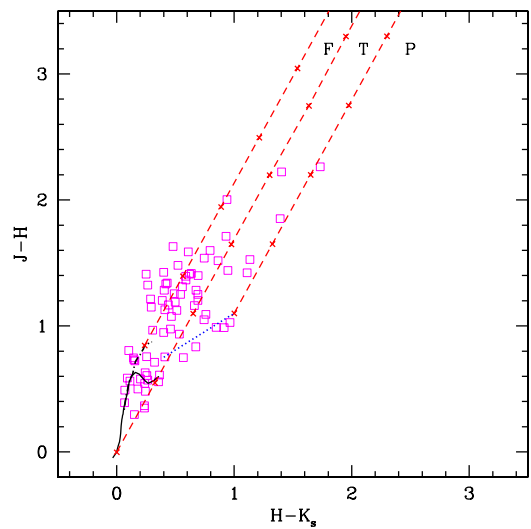}
\caption{\label{nirccd} NIR TCD for the sources having X-ray counterparts (purple squares) in the NGC 7538 region. 
The continuous and thick black dashed curves represent the dereddened MS and giant branches \citep{1988PASP..100.1134B}, 
respectively. The dotted blue line indicates the locus of dereddened CTTSs \citep{1997AJ....114..288M}. 
The parallel red dashed lines are the reddening lines drawn from the tip (spectral type M4) of the 
giant branch (left reddening line), from the base (spectral type A0) of the MS branch (middle reddening
line) and from the tip of the intrinsic CTTS line (right reddening line). 
The crosses on the reddening lines show an increment of $A_V$ = 5 mag. }
\end{figure}

\citet{2004ApJ...616.1042O} have  studied the NGC 7538 region and identified and classified 
286 YSOs (18 Class I and 268 Class II YSOs) based on their excess emission in IR
using their location in the NIR TCD 
(e.g. Fig. \ref{nirccd}).
The solid and thick black dashed curves in Fig. \ref{nirccd} represent the un-reddened MS and
giant branch \citep{1988PASP..100.1134B}, respectively. The dotted blue line indicates the locus
of un-reddened CTTSs \citep{1997AJ....114..288M}. 
The parallel dashed red lines are the reddening lines drawn from the tip
(spectral type M4) of the giant branch (``upper reddening line"), from the base
(spectral type A0) of the MS branch (``middle reddening line") and from the tip of the
intrinsic CTTS line (``lower reddening line"). The extinction ratios
$A_J/A_V = 0.265, A_H/A_V = 0.155$ and $A_K/A_V=0.090$ have been taken from \citet{1981ApJ...249..481C}. 
These extinction ratios are for the normal Galactic medium (i.e., $R_V$= 3.1),
since for $\lambda > \lambda_I$ the reddening law can 
be taken as a universal quantity \citep{1989ApJ...345..245C,1995ApJS..101..335H}. 
The sources can be classified according to the three regions in this diagram \citep[cf.][]{2004ApJ...608..797O}.
`F' sources are located between the upper and middle reddening lines and are considered
to be either field stars (MS stars, giants) or Class III sources and Class II sources with small
NIR excess. `T' sources are located between the middle and lower reddening lines. These sources
are considered to be mostly CTTSs (or Class II objects) with large NIR excess. There may be an
overlap of Herbig Ae/Be stars in the `T' region \citep{1992ApJ...397..613H}. `P' sources are
those located in the region red-ward of the lower reddening line and are most likely Class I
objects (protostellar-like objects; \citet{2004ApJ...608..797O}). 
Most recently, \citet{2014MNRAS.443.3218M} also identified and classified 168 YSOs (24 Class I and 144 Class II) similarly
in the IRS 1-3 region. 
We combined these two NIR data sets with a match radius of 1 arcsec and cataloged  408 YSOs (40 Class I and 368 Class II; 46 were in both catalogs).
After combining this catalog with that of \citet[][562 YSOs]{2014MNRAS.439.3719C} 
with the same  match radius of 1 arcsec, we compiled altogether 
890 YSOs (80 YSOs were in both catalogs) in the selected region of NGC 7538.  
169, 569 and 3 are Class I, Class II and Class III objects, respectively.
The details are given in  Table \ref{data1_yso}.

\begin{figure}
\centering\includegraphics[height=6cm,width=7cm,angle=0]{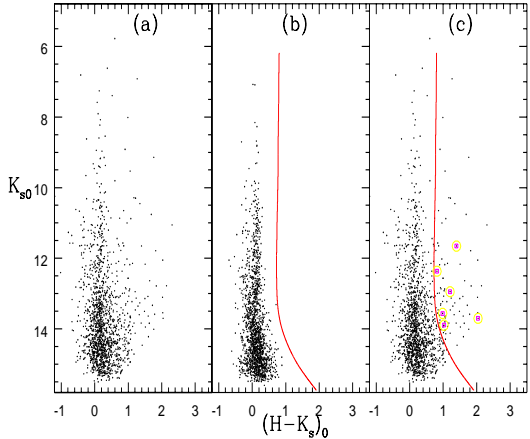}
\caption{\label{nirband} $K_s{_0}/(H-K_s){_0}$ CMD for (a) stars in the NGC 7538 region, (b) stars in the reference field
 and (c) identified X-ray emitting IR excess stars (purple squares inside yellow circles).
The red thick line demarcates the X-ray emitting IR excess stars from MS stars. }
\end{figure}

\subsubsection{X-ray emission\label{x-ray}}

The sample of the YSOs selected on the basis of IR excess may be incomplete because the circumstellar-disks
in young stars may disappear on time-scales of just a few Myr \citep[see][]{2007prpl.conf..345B}.
Recent studies of a few SFRs reveal that disk fraction decreases with age of the region, e.g., 
Sh 2-311: 35\% disk fraction, age=4 Myr, distance = 5 kpc \citep[][]{2016MNRAS.461.2502Y};
NGC 2282: 58\% disk fraction, age= 2-5 Myr, distance=1.65 kpc \citep[][]{2015MNRAS.454.3597D};
W3-AFGL333: 50-60\% disk fraction, age=2 Myr, distance=2 kpc \citep[][]{2016ApJ...822...49J}.
Since the average age of the YSOs in NGC 7538 region is 1.4 Myr (cf. Section 4.1), 
we can expect less than 40\% of YSOs do not show excess emission in NIR/MIR
\citep{2001ApJ...553L.153H,2008ApJ...686.1195H}.

Since X-ray detection method is sensitive also to young stars that have already 
dispersed their circumstellar disks \citep{2011AA...530A..34P},
we have  paid attention to X-ray emitting stars and 
classified them  according to their
evolutionary stages using the classical NIR TCDs 
\citep{2008MNRAS.384.1675J, 2008MNRAS.383.1241P, 2012PASJ...64..107S, 2014A&A...567A.109K}.

In Fig.~\ref{nirccd}, we have plotted the NIR TCD  for the X-ray counterparts.
Since none of the previous studies on the NGC 7538 region \citep[cf.][]{2004ApJ...616.1042O,2014MNRAS.443.3218M,2014MNRAS.439.3719C} 
have published their deep photometric catalogs online, we have used 2MASS data to find 
the NIR counterparts of the X-ray sources.
The {\it Chandra} on-axis point-spread function (PSF) is $0.5^{\prime\prime}$ and it degrades at
large off-axis angles \citep[see e.g.][]{2005ApJS..160..319G, 0004-637X-714-2-1582}. So we used
an optimal matching radius of 1 arcsec to determine the 2MASS NIR counterparts of the
X-ray sources. This size of the matching radius is well established in other studies as well
\citep[see e.g.][]{2002ApJ...574..258F, 2007ApJS..168..100W}. 
Ninety of the X-ray emitting sources have 2MASS NIR counterparts falling 
in the our selected region ($15\times15$ arcmin$^2$ FOV) of NGC 7538.
All the 2MASS magnitudes and colours have been converted into the California Institute of
Technology (CIT) system\footnote{http://www.astro.caltech.edu/~jmc/2mass/v3/transformations/}.
All the curves and lines are also in the CIT system. 
The sources falling in the `F' region and above the
extension of the intrinsic CTTS locus as well as the sources having $(J-H) \geq 0.6$ and
lying to the left of the upper reddening line  
are assigned as WTTSs/Class III sources (45 sources) \citep[see, e.g.,][]{2008MNRAS.384.1675J, 2008MNRAS.383.1241P, 2012PASJ...64..107S, 2014A&A...567A.109K}.
The X-ray counterparts  falling in the `T' (12 sources) and `P' (1 sources) regions are
classified as Class II and Class I sources, respectively.

Sixteen X-ray sources do not have $J$ band data.  The classification of these sources was done
by comparing the dereddened  $K_s{_0}/(H-K_s){_0}$ CMD of the studied region with that of the reference field, 
having the equal area (cf. Fig.~\ref{nirband}).
This scheme has been explained in detail by \citet{2014A&A...567A.109K}.
The amounts of reddening and extinction were estimated on the extinction map obtained by using the  NIR TCD.
Briefly, the stars having $(J-H)$ colour $ \ge 0.6$ mag and lying above the
CTTS locus or its extension were traced back to the CTTS locus or its extension to get their 
amounts of reddening.
Once we have the amount of reddening for individual stars, we can generate 
extinction maps for the target and field regions. The extinction maps were then used to deredden the remaining stars.
The red solid curve in Fig.~\ref{nirband} represents the outer boundary of the field star distribution 
in the CMD taking into account the scatter due to photometric errors and variations in the $A_V$  
values \citep[see for details,][]{2014A&A...567A.109K}.
All the stars having a colour `$(H-K)_0 - \sigma_{(H-K)_0}$' larger than the $red$ cut-off curve
in the studied region, are believed to have excess emission in the $K_s$ band and can be presumed as 
Class I YSOs \citep[see also][]{2004ApJ...616.1042O,2012ApJ...759...48M, 2014A&A...567A.109K}. 
 We have classified six sources without $J$ band detections  as Class I on the basis of the above criterion.

In total 64 sources with X-ray emission were classified as YSOs (45 as Class III, 12 as Class II and  
7 as Class I) on the basis of their positions in the NIR TCD/CMD. 

\subsubsection{The YSOs sample}

We have cross identified the YSOs detected in the present survey based on H$\alpha$ (21 sources, cf. 3.1.1) 
and X-ray emission (64 sources, cf 3.1.3) with those which were detected on the basis of excess IR emission 
and are  available in the literature 
\citep[890 sources, cf. Section 3.1.2 and Table \ref{data1_yso},][]{2004ApJ...616.1042O,2014MNRAS.443.3218M,2014MNRAS.439.3719C},
using a search radius of 1 arcsec. 
 We found 6 H$\alpha$ and 26 X-ray emitting sources are also showing excess IR emission, 
thus we confirmed their identification.
The remaining 53 sources (cf. Table \ref{data2_yso}) are new additions and have been added to make a catalog of altogether 943 YSOs 
in the 15$\times$15 arcmin$^2$ field around NGC 7538.
Optical counterparts for 74 of these YSOs were also identified by using a search radius of 1 arcsec.
The position, magnitude, colour and classification of these YSOs are given in Table \ref{data3_yso}.

Since the aim of this work is to study the star formation activities in the NGC 7538 region, 
the information regarding individual properties of the YSOs is vital, 
which are derived by using the SED fitting analysis (cf. Section 3.2.1).
The SEDs of YSOs can be generated by using the multiwavelength data (i.e. optical to MIR) under the 
condition that a minimum of 5 data points should be available.
Out of 943 YSOs, 463 satisfy this criterion and therefore are used in the further analysis.
The evolutionary classes of most of these YSOs were defined on the basis  of available data sets.
First they were classified on the basis of $\alpha_{IRAC}$ \citep[cf.][]{2014MNRAS.439.3719C}. 
Those which do not have $\alpha_{IRAC}$, were then classified on the basis of MIR+NIR TCD.
And if IRAC  data are not available, then the NIR TCD classification scheme was used (cf. Section 3.1.2).
Finally, the newly identified YSOs on the basis of their X-ray 
and H$\alpha$ emission (cf. Section 3.1.1 and Section 3.1.3) 
were classified based on their IR excess properties and EWs of H$\alpha$ line, respectively.

\subsubsection{Source contamination and  the completeness of the YSOs sample}

We have a sample of 419 YSOs for which SEDs can be generated and are associated with the NGC 7538 SFR. 
Our candidate YSOs may be contaminated by IR excess sources such as star forming galaxies, 
broad-line active galactic nuclei, unresolved shock emission knots, objects that suffer from polycyclic aromatic hydrocarbon (PAH) emissions etc.,
that mimic the colours of YSOs. Since we are observing through the Galactic plane, contamination due
to galaxies should be negligible \citep{2015A&A...573A..95M}.  In order to have a statistical estimate of possible
galaxy contamination in our YSOs sample, we used the Spitzer Wide-area Infrared Extragalactic (SWIRE) catalog 
obtained from the observations of the ELAIS N1 field \citep{2013MNRAS.428.1958R}. SWIRE is a survey of the
extragalactic field using the Spitzer-IRAC and MIPS bands and can be used to predict the number of galaxies 
in a sample of YSOs \citep{2009ApJS..181..321E}.  
The SWIRE catalog is resampled for the spatial extent
as well as the sensitivity limits of the photometric data in the NGC 7538 region and is also corrected by the average extinction
(i.e. $A_V$ = 11 mag, cf. Section 4.1). 
17 sources in the SWIRE catalog could satisfy these criteria.
Similarly, out of the 397 YSOs having IRAC photometry, 7 can be approximately categorized as
obscured AGB stars as they have very bright MIR flux, i.e., [4.5]$\leq$7.8 mag \citep[cf.][]{2008AJ....136.2413R}.
Therefore, the contribution of various contaminants in
our YSO sample should be $\sim$6\%, which is a small fraction of the total number of YSOs. 
We also applied the colour/magnitude criteria by \citet{2009ApJS..184...18G} to the YSOs classified by  \citet{2014MNRAS.439.3719C}
on the basis of their IRAC SEDs, and $\sim$28\% of them are matched with the colours 
of PAH contaminated apertures, shock emissions, PAH emitting galaxies and AGNs. 
Recently, \citet{2016ApJ...822...49J} have also checked the same scheme applied to their sample of YSOs in the W3-AFGL333 region 
and found that $\sim$38\% of Class I sources and $\sim$15\% of Class II sources in their sample turned out to be  contaminants.
However,  using statistical approach as used by us, they have found only $<$5\% contaminants in their sample of YSOs. 
Several studies \citep[e.g.,][]{2008ApJ...688.1142K,2011ApJ...743...39R,2013ApJ...778...96W}
have noticed that the use of the \citet{2009ApJS..184...18G}  criteria for a region at $\sim$2 kpc
would likely provide an overestimation of the contaminations.

\begin{figure}
\centering\includegraphics[height=3cm,width=4cm,angle=0]{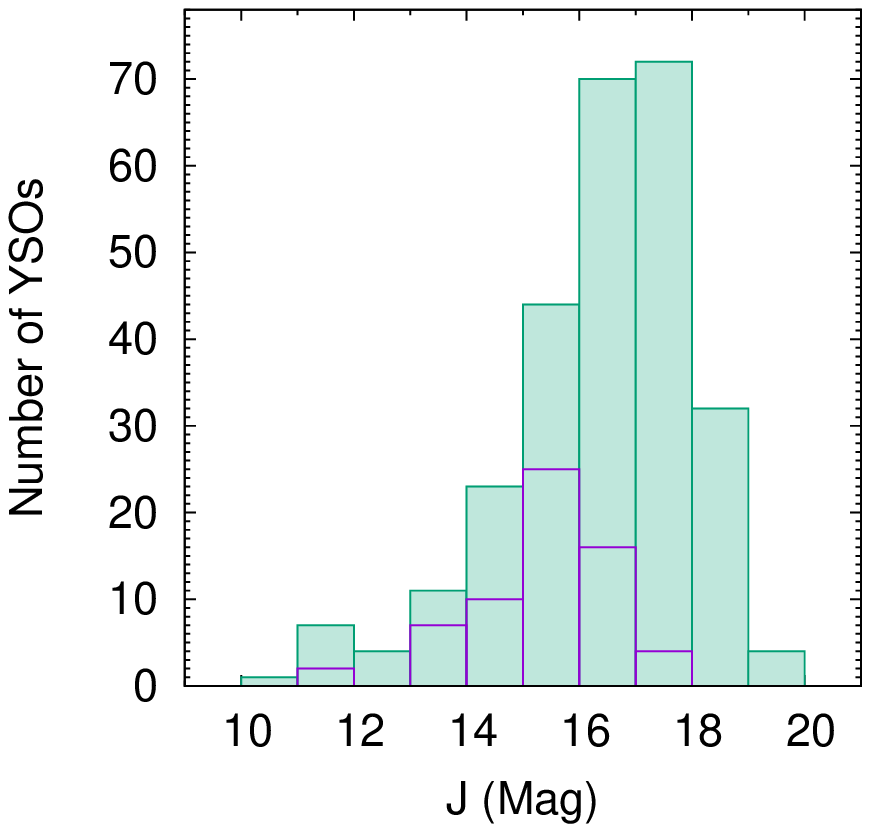}
\centering\includegraphics[height=3cm,width=4cm,angle=0]{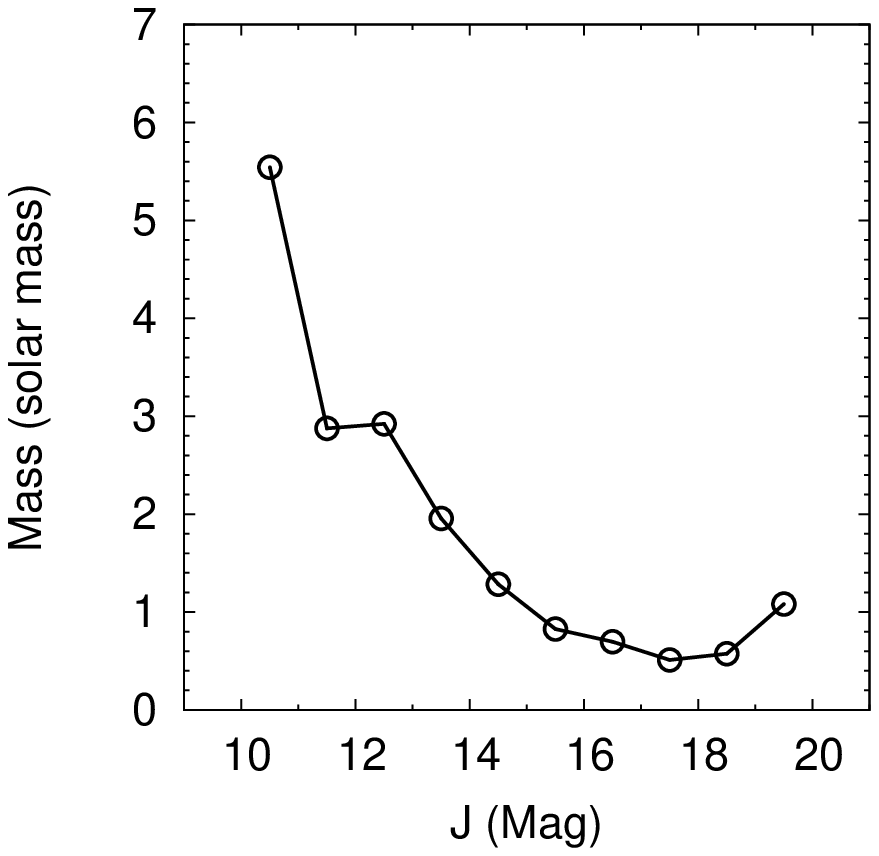}
\caption{\label{FcftJ} Histogram (green) showing the distribution of the number of YSOs in different $J$ magnitude (left-hand panel) and the
corresponding mass limit of the YSOs (right-hand panel).
The blue histogram represents the sample of X-ray emitting YSOs (64) having 2MASS detections.
}
\end{figure}

For the above mentioned sample of YSOs having data taken from various surveys, knowledge of its
completeness in terms of mass is necessary, but we cannot perform the $ADDSTAR$ 
routine to derive it.
Therefore, we have instead drawn histograms for the numbers of YSOs in different magnitude bins and checked for their peaks.
This peak will roughly represent the completeness limit of the data in terms of magnitude \citep{2013MNRAS.432.3445J,2013ApJ...778...96W,2016ApJ...822...49J,2016AJ....151..126S}. 
Once we know this, the corresponding mass is taken as that of the lowest mass YSO.
For example, in Fig. \ref{FcftJ} we show the distribution of the numbers of YSOs in different $J$ magnitudes and the
corresponding mass limit of the YSOs.
The peak of this histogram is in between 17-18 mag and for this magnitude bin 
the corresponding mass is 0.5 M$_\odot$.
In this way we found that our data are complete down to 0.5, 0.8, 0.8, 0.8 and 0.6 M$_\odot$ in the $J$, $H$, $K_s$, 
3.6 $\mu$m and 4.8 $\mu$m bands, respectively.
Since the sources undetected at 5.8 $\mu$m and 8.0 $\mu$m were classified on the 
basis of the 3.6 $\mu$m, 4.8 $\mu$m,  $H$ and $K_s$ band photometry, 
the completeness limit for them is not included here.
The mass completeness for the YSOs having X-ray and H$\alpha$ emission  will depend on the 
completeness limit of the photometric data from which they were identified.
The completeness limit for the X-ray emitting YSOs (64) comes out to be
0.8 M$_\odot$ as inferred from the peak of the blue histogram in Fig. \ref{FcftJ}. 
For the H$\alpha$ emitting sources, their completeness can be approximated as 0.8 M$_\odot$, which is 
equivalent to that of the optical data. 
To conclude, we have taken the highest of these, i.e., 0.8 M$_\odot$, as the completeness limit 
for the detection of the current YSO sample.

\begin{table}
\caption{\label{data1_yso}  A sample table containing information for 890 cataloged YSOs by  
 \citet{2004ApJ...616.1042O}, \citet{2014MNRAS.439.3719C} and  \citet{2014MNRAS.443.3218M} in the NGC 7538 region. 
The complete table is available in an electronic form only. }
\begin{tabular}{@{}lccc@{}c@{}c@{}c@{}c@{}}
\hline
ID$^*$& $\alpha_{(2000)}$  & $\delta_{(2000)}$  & Class$^a$   & Class$^b$ & Class$^c$ & Class$^d$\\
&  ($^h$:$^m$:$^s$)  & ($^\circ$:$^\prime$:$^\prime$$^\prime$) & &                &              &                \\
\hline
$C_{2361}$ & 23:12:34.40 &+61:27:11.3 & II&      -  &      - &   -  \\
$C_{2362}$ & 23:12:37.42 &+61:36:16.2 & -&      -   &      - &   -  \\
$C_{2363}$ & 23:12:40.88 &+61:24:56.5 & II&     -   &      - &   -  \\
$C_{2364}$ & 23:12:40.88 &+61:31:38.8 & II&      -  &      - &   -  \\
$C_{2365}$ & 23:12:41.18 &+61:28:02.1 & II&      -  &      - &   -  \\
\hline
\end{tabular}
$^*$: O, C and M are YSOs cataloged in  \citet{2004ApJ...616.1042O}, \citet{2014MNRAS.439.3719C} and  \citet{2014MNRAS.443.3218M}, respectively.\\
$^a$: Classification by \citet{2014MNRAS.439.3719C} based on IRAC SED slope.\\
$^b$: Classification by IRAC TCDs.\\  
$^c$: Classification by \citet{2004ApJ...616.1042O} and \citet{2014MNRAS.443.3218M} through NIR TCDs.\\
$^d$: Classification by H$\alpha$ and X-ray emission (i.e., I/II/III(Comment), if Comment=7: H$\alpha$ emission star (spectroscopy), Comment=8: H$\alpha$ emission star (photometry), Comment=9: X-ray emitting star.\\

\end{table}

\begin{table*}
\caption{\label{data2_yso} A sample table containing information for 53 newly identified YSOs. The $J$, $H$ and $K_s$ magnitudes and their errors are from the 2MASS PSC. The complete table is available in an electronic form only.}
\begin{tabular}{@{}lccccccccc@{}}
\hline  
ID& $\alpha_{(2000)}$  & $\delta_{(2000)}$ & $J\pm \sigma$  & $H\pm \sigma$  &  $K_s\pm \sigma$ & Class$^d$\\
 & ($^h$:$^m$:$^s$)  &  ($^\circ$:$^\prime$:$^\prime$$^\prime$)                & (mag)& (mag)& (mag)&\\
\hline  
$S_1$  &  23:13:49.12 &+61:30:15.4&$11.966\pm0.024$&$11.120\pm0.030$&$10.783\pm0.019$& II(8) \\
$S_2$  &  23:13:29.47 &+61:30:38.9&$12.447\pm0.034$&$11.945\pm0.039$&$11.649\pm0.039$& II(8) \\
$S_3$  &  23:13:56.29 &+61:25:07.3&$13.455\pm0.031$&$12.843\pm0.035$&$12.553\pm0.032$& III(7) \\
$S_4$  &  23:14:15.91 &+61:31:12.1&$13.608\pm0.027$&$12.843\pm0.028$&$12.666\pm0.023$& III(9)  \\
$S_5$  &  23:13:31.96 &+61:27:47.3&$13.697\pm0.030$&$12.223\pm0.032$&$11.495\pm0.019$& III(9)  \\
\hline
\end{tabular}

$^d$: Classification by H$\alpha$ and X-ray emission (i.e., I/II/III(Comment), if Comment=7: H$\alpha$ emission star (spectroscopy), Comment=8: H$\alpha$ emission star (photometry), Comment=9: X-ray emitting star.\\
\end{table*}

\begin{figure*}
\centering\includegraphics[height=6cm,width=7cm,angle=0]{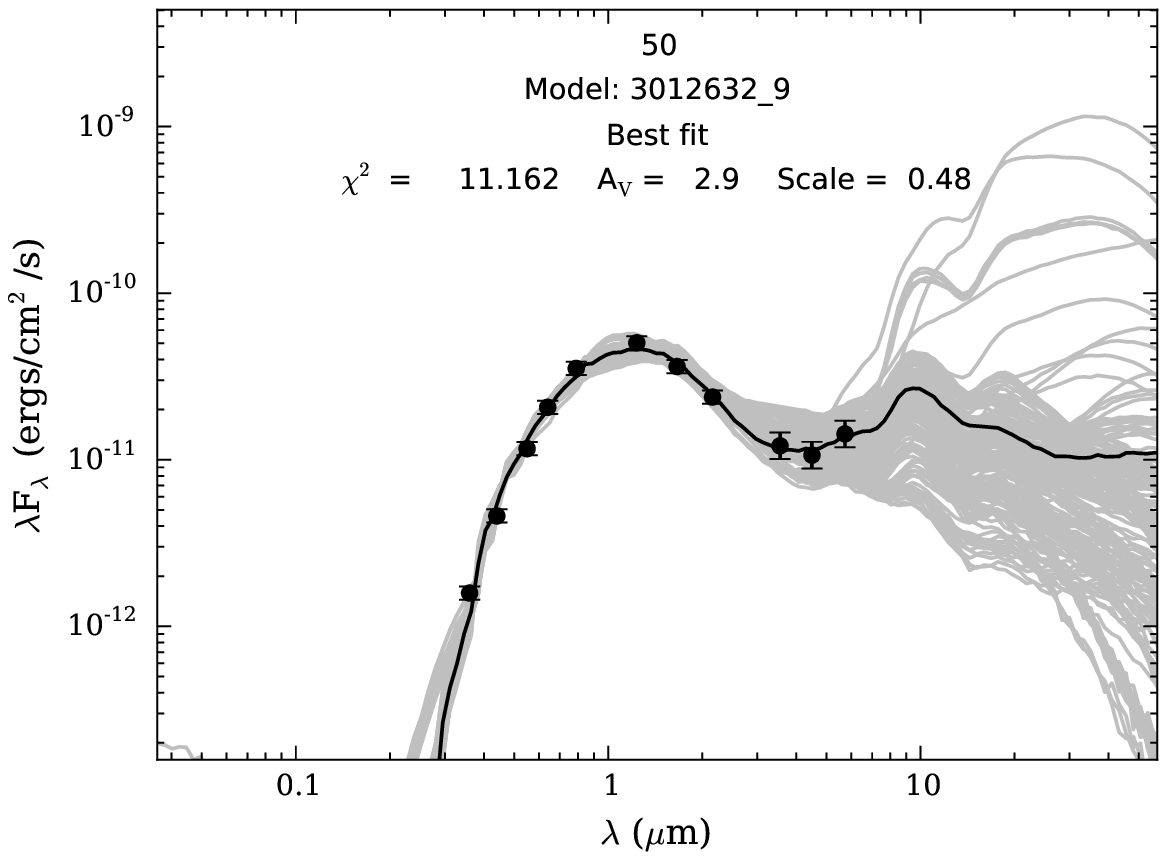}
\centering\includegraphics[height=6cm,width=7cm,angle=0]{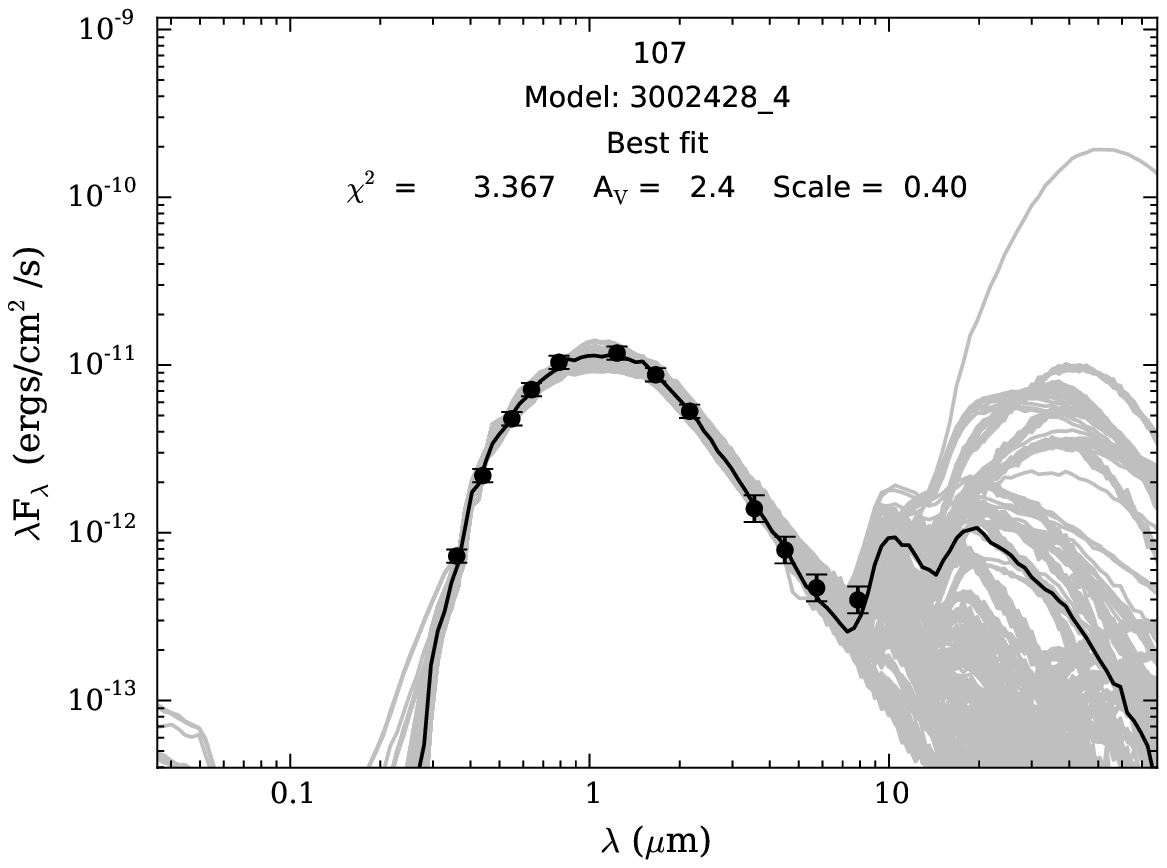}
\caption{\label{sed} Sample SEDs for Class I (left-hand panel) and Class II (right-hand panel)  sources 
created by the SED fitting tools of \citet{2007ApJS..169..328R}. 
The black curve shows the best fit and the gray curves show the subsequent well fits. 
The filled circles with error bars denote the input flux values.}
\end{figure*}

\subsection{YSOs parameters }

\subsubsection{From spectral energy distribution analysis}

We constructed SEDs of the YSOs using the grid models and fitting tools of
\citet{2006ApJS..167..256R,2007ApJS..169..328R} for characterizing and understanding their nature. 
The models were computed by using a Monte-Carlo based 20000 2-D radiation transfer calculations 
from \citet{2003ApJ...598.1079W,2003ApJ...591.1049W, 2004ApJ...617.1177W}
and by adopting several combinations of a central star, a disk, an in-falling envelope, 
and a bipolar cavity in a reasonably large parameter space and with 10 viewing angles (inclinations). 
The SED fitting tools provide the evolutionary stage and physical parameters such as mass, age, disk mass, 
disk accretion rate and stellar temperature of YSOs and hence are an ideal tool to study the
evolutionary status of YSOs.

We constructed the SEDs of  463 YSOs using the multiwavelength data 
(optical to MIR wavelengths, i.e. 0.37, 0.44, 0.55, 0.65, 0.80, 1.2, 1.6, 2.2, 3.6, 4.5, 5.8 and 8.0 $\mu m$) and 
with a condition that a minimum of 5 data points should be available (cf. Section 3.1.4).
The SED fitting tool fits each of the models to the data allowing the distance
and extinction as free parameters. 
The distance of the NGC 7538 region is taken as 2.65$^{+0.12}_{ -0.11}$ kpc (cf. Appendix A), but 
the input value ranges to the fitting tool is given with three times the 
error of the adopted distance, i.e., 2.65-0.12$\times$3 to 2.65+0.12$\times$3 = 2.3 to 3.0 kpc. 
Since, this region is highly nebulous and \citet[][Fig. 6]{2004ApJ...616.1042O} detected YSOs up to $A_V$ = 25 mag, 
we varied $A_V$ in a broader range (i.e., from 0 to 30 mag) with three times the
errors associated with the foreground $A_V$ value ($\pm$2 mag, cf. Appendix A) and 
keeping in mind the nebulosity associated with this SFR
\citep[see also,][]{2012ApJ...755...20S,2013MNRAS.432.3445J,2014MNRAS.443.1614P}. 

We further set photometric uncertainties of 10\% for
optical and 20\% for both NIR and MIR
data. These values are adopted instead of the formal errors
in the catalog in order to fit without any possible bias caused
by underestimating the flux uncertainties. We obtained the
physical parameters of the YSOs using the relative probability
distribution for the stages of all the `well-fit' models. The
well-fit models for each source are defined by

$\chi^2 - \chi^2_{min} \leq 2 N_{data}$

where $\chi^2_{min}$ is the goodness-of-fit parameter for the
best-fit model and $N_{data}$ is the number of input data points.
In Fig. \ref{sed}, we show example SEDs of Class I and Class II
sources, where the solid black curves represent 
the best-fit and the gray curves are the subsequent well-fits. 
As can be seen, the SED of the Class I source rises substantially in the MIR in comparison to that of the Class II source
due to its optically thick disk.
The sheer number of gray curves in the SED indicates how difficult it was to select a single best-fitting model to
these data, which is not surprising given that there are no anchors at long wavelengths. 
Here it is worthwhile to take note that the Class I source
shown in Fig. \ref {sed} is optically detected, which is not usual,
but there are also several other cases where such Class I sources are not detected optically,
e.g., \citet[][NGC 1931]{2013ApJ...764..172P}, \citet[][Sh 2-252]{2013MNRAS.432.3445J}, 
 \citet[][W5-East]{2011MNRAS.415.1202C}, \citet[][Sh 2-255-257]{2011ApJ...738..156O}
and \citet[][Sh-2 311]{2016MNRAS.461.2502Y}. 
In the present study, five Class I sources have optical detections.
Optical detection of Class I sources can be attributed to their viewing angle 
\citep{2005A&A...429..543N,2007prpl.conf..117W,2011ARA&A..49...67W}.
Also, some Class I and Class II sources can exhibit similar SEDs in the $Spitzer$ bands 
\citep{2005ApJ...629..881H,2011ARA&A..49...67W,2006ApJS..167..256R}. As an example, the models of 
\citet{2003ApJ...591.1049W} show that mid-latitude ($\sim40^\circ$)-viewed Class I stars have optical
and NIR/MIR characteristics similar to those of more edge-on disk ($\sim75^\circ$) Class II stars. 
The effects of the edge-on disk orientation are most severe for evolutionary diagnostics 
determined in the NIR/MIR such as the 2-25 $\mu$m spectral 
index \citep{2007prpl.conf..117W}. The definitive identification of Class I 
sources requires other observations that better constrain the presence of an envelope, 
such as MIR spectroscopy, far-IR and millimeter photometry, 
and high-resolution images \citep{2010ApJS..186..111L}.

From the well-fit models for each source derived
from the SED fitting tool, we calculated the $\chi^2$ weighted
model parameters such as the $A_V$, distance, stellar mass and stellar age of each  YSO
and they are given in Table \ref {data4_yso} along with their
adopted evolutionary classes (cf. Section 3.1.4). 
The error in each parameter is calculated from the standard deviation of all well-fit parameters. 

\begin{table*}
\caption{\label{data3_yso}  A sample table containing information for 74 optically identified YSOs. The age and mass of the YSOs are derived from the optical CMD. IDs are the same as in Table \ref{data1_yso} and Table \ref{data2_yso}. 
The complete table is available in an electronic form only.
 }
\begin{tabular}{@{}rcccccccc@{}}
\hline
ID &  $V\pm \sigma$  & $(U-B) \pm \sigma$ &  $(B-V) \pm \sigma$ &  $(V-R_c)\pm \sigma$& $(V-I_c)\pm \sigma$ & Age$\pm \sigma$ & Mass$\pm \sigma$ \\
  &  (mag)& (mag) &  (mag)& (mag)&  (mag) & (Myr) & (M$_\odot$)  \\
\hline
$C_{2833}$&$12.369\pm0.004$&$-0.047\pm0.005$& $0.611\pm0.005$& $0.401\pm0.007$& $0.806\pm0.008$&  $0.4\pm0.1$ & $5.8\pm0.1$\\
$C_{2864}$&$13.903\pm0.005$&$ 0.676\pm0.007$& $0.970\pm0.006$& $0.576\pm0.011$& $1.077\pm0.009$&  $1.2\pm0.1$ & $3.9\pm0.3$\\
$C_{2523}$&$14.259\pm0.007$&$ 0.081\pm0.006$& $1.402\pm0.008$& $1.050\pm0.012$& $2.251\pm0.013$&  $0.1\pm0.1$ & $5.4\pm0.1$\\
$C_{2411}$&$14.360\pm0.006$&$ 0.016\pm0.006$& $0.742\pm0.006$& $0.460\pm0.009$& $0.950\pm0.009$&  $2.5\pm1.9$ & $4.9\pm0.3$\\
\hline
\end{tabular}
\end{table*}

\begin{table}
\caption{\label{data4_yso}  A sample table containing stellar parameters of selected 419 YSOs derived by using the SED fitting analysis. IDs are the same as in Table \ref{data1_yso} and Table \ref{data2_yso}. The complete table is available in an electronic form only. }
\begin{tabular}{@{}r@{ }c@{ }c@{ }c@{ }c@{ }c@{ }c@{ }c@{}}
\hline
 ID &   $N_{data}$ & Class  & $\chi^2_{min}$ &  Distance$\pm \sigma$   &  $A_V$$\pm \sigma$     &  Age$\pm \sigma$ &  Mass$\pm \sigma$ \\
    &     &     &     &   (kpc)         &   (mag)       & (Myr)      &($M_{\odot}$) \\
\hline
$C_{2361}$ &  6 & II &  1.7 &$2.5\pm0.2$& $19.8\pm3.7$&$0.8\pm0.9$&$1.7\pm1.0$\\
$C_{2363}$ &  6 & II &  3.1 &$2.6\pm0.3$& $18.0\pm2.6$&$1.2\pm1.7$&$1.1\pm0.8$\\
$C_{2364}$ &  7 & II &  0.7 &$2.6\pm0.3$& $ 4.4\pm1.8$&$1.5\pm1.5$&$1.2\pm0.7$\\
$C_{2365}$ &  6 & II &  0.6 &$2.6\pm0.3$& $21.2\pm3.6$&$1.0\pm1.4$&$1.5\pm0.9$\\
\hline
\end{tabular}
The evolutionary class of the YSOs are from Table \ref{data1_yso} and Table \ref{data2_yso} (cf. Section 3.1.4).
\end{table}

\subsubsection{From optical colour-magnitude diagram}

\begin{figure}
\centering\includegraphics[height=7cm,width=7cm,angle=0]{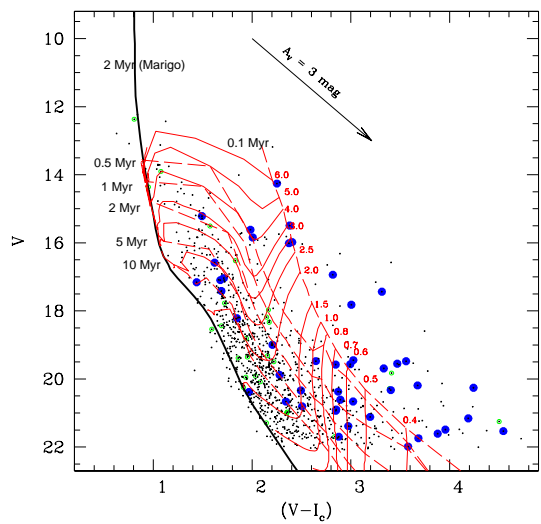}
\caption{\label{vi} $V/(V-I_c)$ CMD for all the optically detected sources in the NGC 7538 region.
Blue dots are identified YSOs and green circles are the sources categorized as 
non-members of the region  (cf. Section 4.1 for detail).
The isochrone for 2 Myr by \citet{2008AA...482..883M}  (thick black curve) along with the PMS isochrones
of 0.1,0.5,1,2,5,10 Myr (red dashed curves) and the evolutionary tracks of different masses 
(red curves)  by \citet{2000AA...358..593S} are also shown.
All the isochrones and evolutionary tracks are corrected for the distance of 2.65 kpc and reddening $E(B-V)=0.75$ mag.
}
\end{figure}

By comparing the positions of YSOs in the optical CMD with theoretical isochrones, 
we can determine their age and mass.
In Fig.~\ref{vi}, the $V /(V - I_c)$ CMD has been plotted for all the 
sources in the NGC 7538 region along with optically detected 74 YSOs.
The post-main-sequence isochrone for 2 Myr calculated by \citet{2008AA...482..883M} (thick black curve) along with the PMS isochrones
of 0.1,0.5,1,2,5,10 Myr (red dashed curves) and evolutionary tracks of different masses (red curves)  by \citet{2000AA...358..593S} are also shown.
These isochrones are corrected for the distance (2.65$^{+0.12}_{ -0.11}$ kpc, cf. Appendix A) 
and the foreground reddening
($E(B - V )_{min} = 0.75\pm0.2$ mag, cf. Appendix A) of NGC 7538 by using the reddening law $R_V$=2.82 (cf. Appendix B).
Here, it is highly likely that not all the YSOs are situated  on the surface of the associated cloud 
but embedded in it where the reddening law is anomalous ($R_V$=3.85, cf. Appendix B).
Change of the reddening law can affect the derivation of the physical parameters of the YSOs.
Therefore, we checked the change in the position of a YSO in the dereddened optical CMD, 
corrected for the extinction $A_V$= 3 mag (i.e. the average extinction value
of the optically detected YSOs, cf. Section 3.2.1, Table \ref{data4_yso}) by applying
different $R_V$ values.
After removing the foreground contribution,
the  intra-cluster extinction in $V$ and $(V-I_c)$  will be 0.9 mag and 0.34 mag, respectively for $R_V$=3.85
and,  0.7 mag and 0.28 mag, respectively for $R_V$ = 2.82.
Therefore, the change in reddening law will have a marginal effect to the derived physical parameters. 
But the amount of intra-cluster reddening can affect the derivation of ages/masses of YSOs, if not corrected individually. 
As we can see from Fig. \ref{vi}, the reddening vector is almost parallel to the isochrones, 
but nearly perpendicular to the evolutionary tracks for various masses, 
this effect will be nominal in deriving the ages, but can be substantial in the mass estimation. 

The age and mass of the YSOs have been derived from the optical CMD (cf. Fig. \ref{vi}) by applying the following procedure.
We created an error box around each observed data point using the errors associated with photometry 
as well as errors associated with the estimation of reddening and distance as given in previous paragraph. 
Five hundred random data points were generated by using Monte-Carlo simulations in this box.
The age and mass of each generated point were estimated from the nearest passing isochrone and evolutionary track by \citet{2000AA...358..593S}.
For accuracy, the isochrones and  evolutionary tracks  were used in a bin size of 0.1 Myr and were interpolated by 2000 points.
At the end we have taken their mean and standard deviation as the final derived values and errors and 
are given in  Table~\ref{data3_yso}.

\begin{figure}
\centering\includegraphics[height=5.5cm,width=7cm,angle=0]{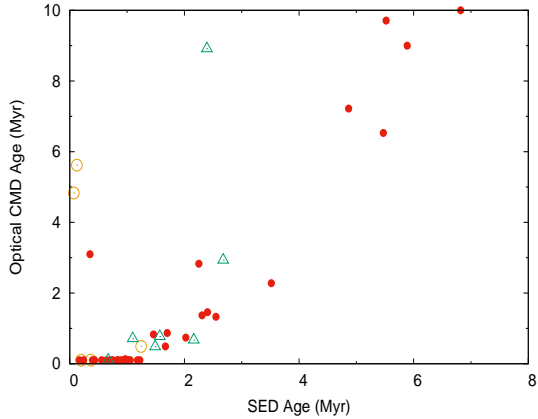}
\caption{\label{cmp} Comparison of the age estimates obtained from the 
CMD analysis with those from the SED fitting.
Yellow open circles, red filled circles and green open triangles 
represent Class I, Class II  and Class III sources, respectively.}
\end{figure}

In Fig. \ref{cmp}, we have compared the age estimates obtained from the CMD and the SED fitting. 
The distribution indicates in general a reasonable agreement between them with a large scatter.
We have calculated the linear Pearson correlation coefficient (r=0.71) for this distribution
and found that the probability of having no correlation is negligible (i.e., 10$^{-6}$). 
The scatter in Fig. \ref{cmp} may be due to the difference in reddening/extinction corrections to 
the magnitudes/colours of the YSOs in these two techniques.
In the optical CMD, we have used a fixed foreground reddening value for all the YSOs,
while in the SED fitting, $A_V$ was given as a free parameter in a broad range to deredden the individual YSOs. Therefore, by the CMDs the very young YSOs 
which tend to be deeply embedded in the cloud can be assigned lower ages than their actual ones.
ここ

\begin{figure}
\centering\includegraphics[height=7cm,width=7cm,angle=-90]{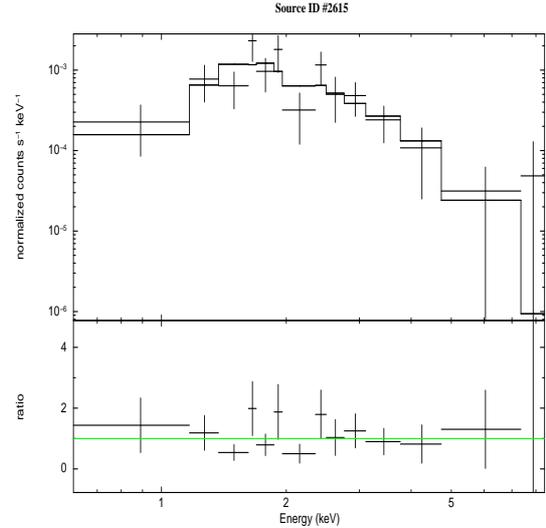}
\caption{\label{spectra} An example X-ray spectrum of a YSO (source ID 2615) fitted by the spectral model 
 {\sc{phabs}} $\times$ {\sc{Apec}}. 
}
\end{figure}

\subsection{X-ray spectral analysis}

Present data contains 190 X-ray emitting sources, of which 91 are classified as YSOs, 
either on the basis of their location on the NIR TCD/CMD (64, cf. Section 3.1.3),
or having their counterpart in already published list of YSOs (27, cf. Section 3.1.2).
For X-ray spectral analysis, we have selected 21 of them, having more than 35  counts
(SNR $>$ 5) in the energy band 0.2-8.0 keV
to ensure a minimum quality of spectral fits.
The spectra of these sources and the background were extracted by
using the  $specextract$ task. The radii of the extracted regions of the sources
varied between 5 arcsec and 15 arcsec depending on the position
of the source detected by the $PWDetect$ task (see Section 2.3.1) on the 
detector as well as on its angular separation 
with respect to the neighboring X-ray sources.
For each source, the background spectrum was obtained from multiple source-free regions
chosen according to the source location on the same CCD.
The spectra were binned to have a minimum of 5 counts per spectral bin by using $grppha$ task 
included in {\sc ftools}.
Finally, a spectral analysis was performed based on the global fitting by using
the Astrophysical Plasma Emission Code ({\sc{APEC}}) version 1.10 modeled by \citet{smi+01} and
implemented in the {\sc xspec} version 12.3.0. The plasma model
{\sc{APEC}} calculates both line and continuum emissivities for a hot,
optically thin plasma that is in collisional ionization equilibrium.
The photoelectric absorption model ($photoelectric~absorption~screens$; {\sc phabs}) by  
\citet{bal+92} was used to account for the Galactic absorption. 
The simplest isothermal gas model was considered for the fitting and expressed as {\sc{phabs}} $\times$ {\sc{Apec}} by using
Cash maximum-likelihood scheme (cstat) in {\sc xspec}. 
Plasma abundances were fixed at 0.3 times the solar abundances, as this value is routinely found in X-ray spectral fitting of young stars  \citep[e.g.,][]{fei+02, cur+09, bha+13}.
The hydrogen column density ($\rm{N_H}$) and plasma temperature (kT) determined from the fitting were used to
calculate the X-ray flux of the above 21 YSOs by using the cflux model in xspec.
For reference, we have shown a sample spectrum in Fig. \ref{spectra}.
The X-ray fluxes of the remaining YSOs have been  derived from their X-ray count rates.
The count conversion factor (CCF, i.e.,  $\rm{2.6\times10^{-11}}$ ${\rm erg~s^{-1}~cm^{-2}}$) to convert count rates into un-absorbed X-ray fluxes has been estimated 
from WebPIMMS\footnote{http://heasarc.gsfc.nasa.gov/cgi-bin/Tools/w3pimms/pim\_adv} 
by using the {\sc 1T APEC} plasma model. The input parameters, kT ($\sim$2.7 keV) and $\rm{N_H}$ ($\sim$$\rm{0.75\times10^{22}~cm^{-2}}$) in WebPIMMS 
were calculated as the mean of the kT  derived from the spectral fitting of the 21 X-ray sources which were
having counts greater than 35, and by using the LAB model\footnote{http://heasarc.gsfc.nasa.gov/cgi-bin/Tools/w3nh/w3nh.pl} \citep{kal+05}, respectively.

\begin{table}
\caption{\label{xray-param}  A sample table containing the spectral parameters of the X-ray sources having NIR counterparts and identified as YSOs. 
IDs are the same as in Table \ref{data1_yso} and Table \ref{data2_yso}.
The complete table is available in an electronic form only.
}
\begin{tabular}{lcccc}
\hline
ID     &  $\rm{N_H (10^{22}~cm^{-2})}$    &  kT (keV)             &   $\rm log({{L_X} (erg~s^{-1})})$ \\
\hline
$S_3$   & $      0.45^{+0.18}_{-0.18} $&$  0.59^{+0.33}_{-0.20}    $&$  31.67_{-0.06}^{+0.07}$\\
$S_4$   & $      0.68^{+0.17}_{-0.16} $&$  0.65^{+0.18}_{-0.22}    $&$  31.85_{-0.07}^{+0.06}$\\
$S_5$   & $       -                   $&$  -                       $&$  30.59_{-0.19}^{+0.34}$\\
$S_7$   & $     <0.39                 $&$  4.56^{+7.90}_{-2.16}    $&$  31.28_{-0.08}^{+0.08}$\\
\hline
\end{tabular}
\end{table} 

\begin{figure*}
\centering\includegraphics[height=5cm,width=5.5cm,angle=0]{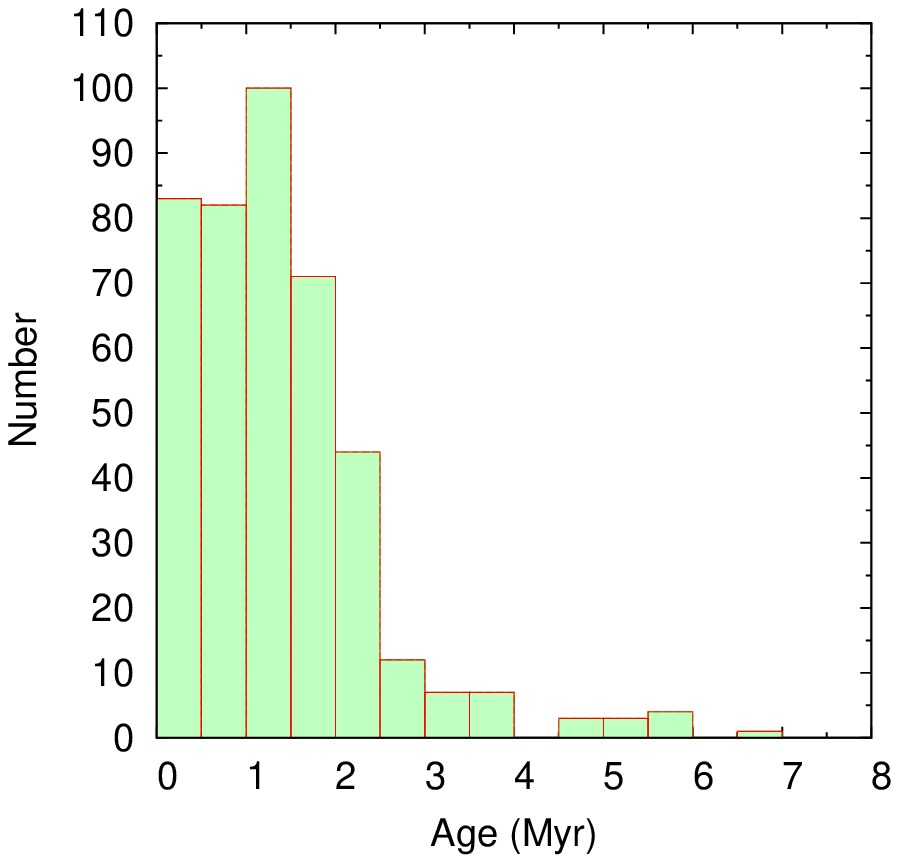}
\centering\includegraphics[height=5cm,width=5.5cm,angle=0]{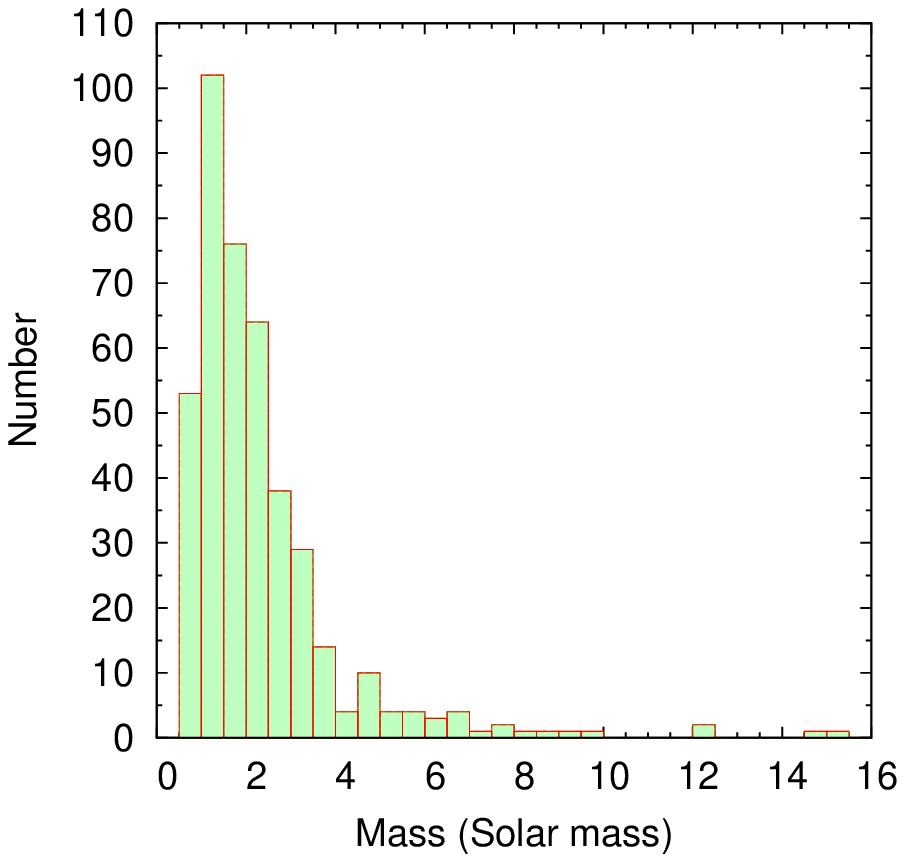}
\centering\includegraphics[height=5cm,width=5.5cm,angle=0]{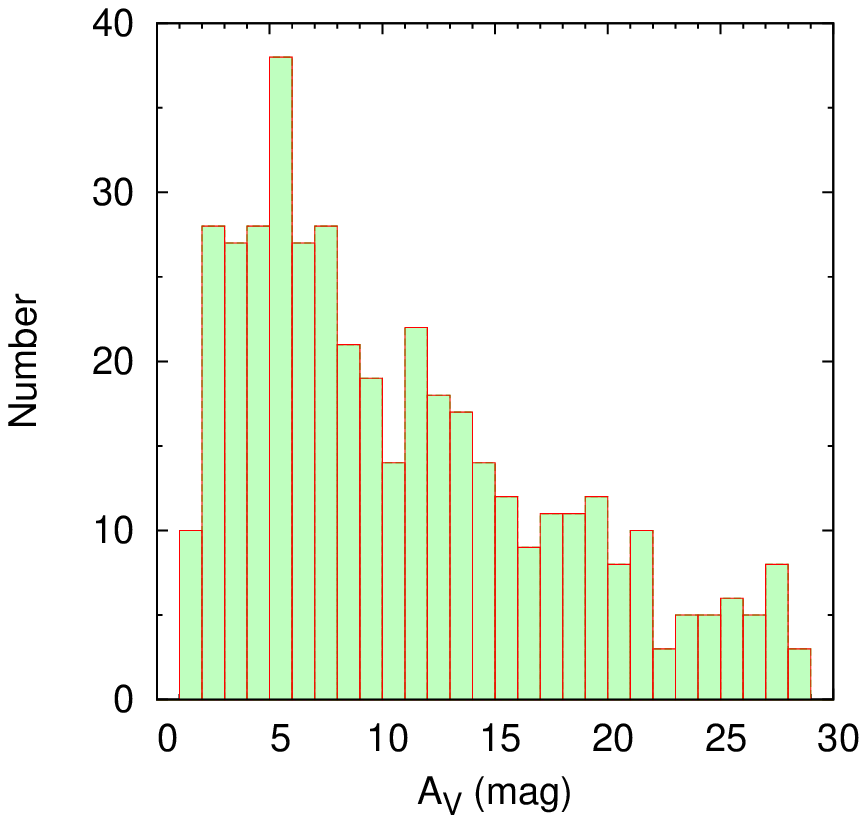}
\caption{\label{histogram} Histograms showing the distribution of the ages (left-hand panel),
masses (center panel) and extinction values `$A_V$'  (right-hand panel) of the YSOs in the NGC 7538 region.
The age, mass and $A_V$ are derived from the SED fitting analysis (cf. Section 3.4.1).
}
\end{figure*}

Finally, the X-ray luminosities ($\rm{L_X}$) for all 91 sources were estimated from the X-ray fluxes
by using the distance of NGC 7538, i.e., 2.65 kpc.
The spectral parameters of these YSOs are given in Table \ref{xray-param}.
The N$\rm _H$ value for these  X-ray sources are in the range of $0.4 - 4.6 \times 10^{22}~ \rm cm^{-2}$,
which corresponds to an $A_V$ range of 2 - 24 mag \citep[$\rm N_H$ = 1.87 $\times$ 10$^{21} A_V$,][]{1978ApJ...224..132B},
which is comparable to the $A_V$ range of the YSOs in the present study (cf. Section 4.1).
Therefore, we can safely assume  that these sources are the members of the NGC 7538 SFR.

\begin{figure}
\centering\includegraphics[height=3cm,width=4cm,angle=0]{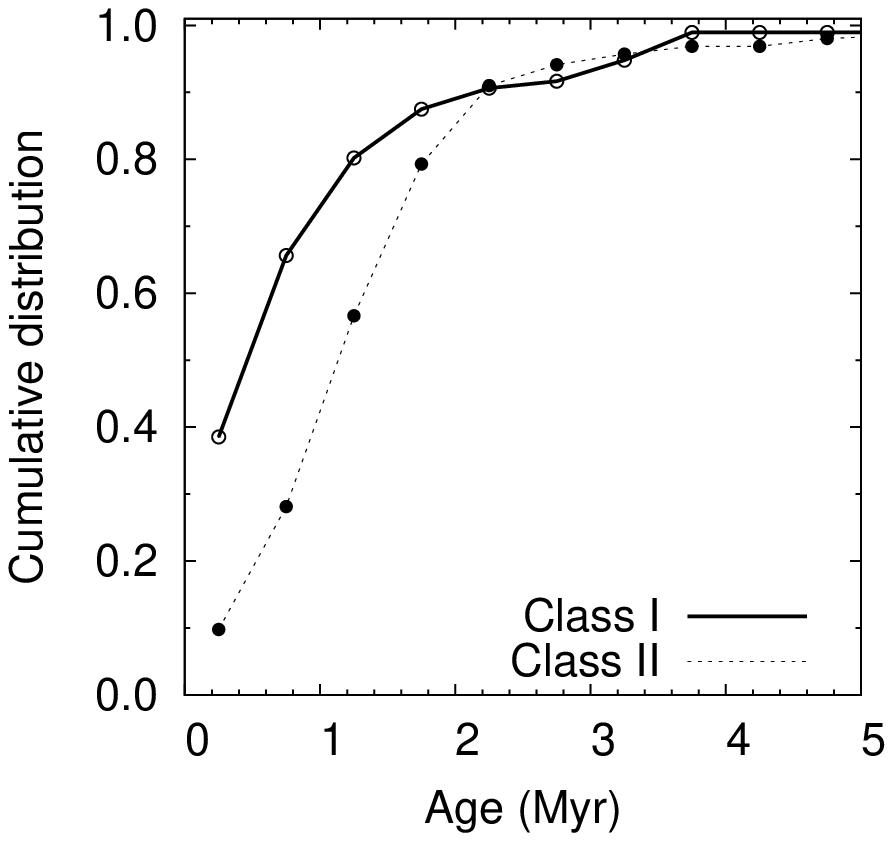}
\centering\includegraphics[height=3cm,width=4cm,angle=0]{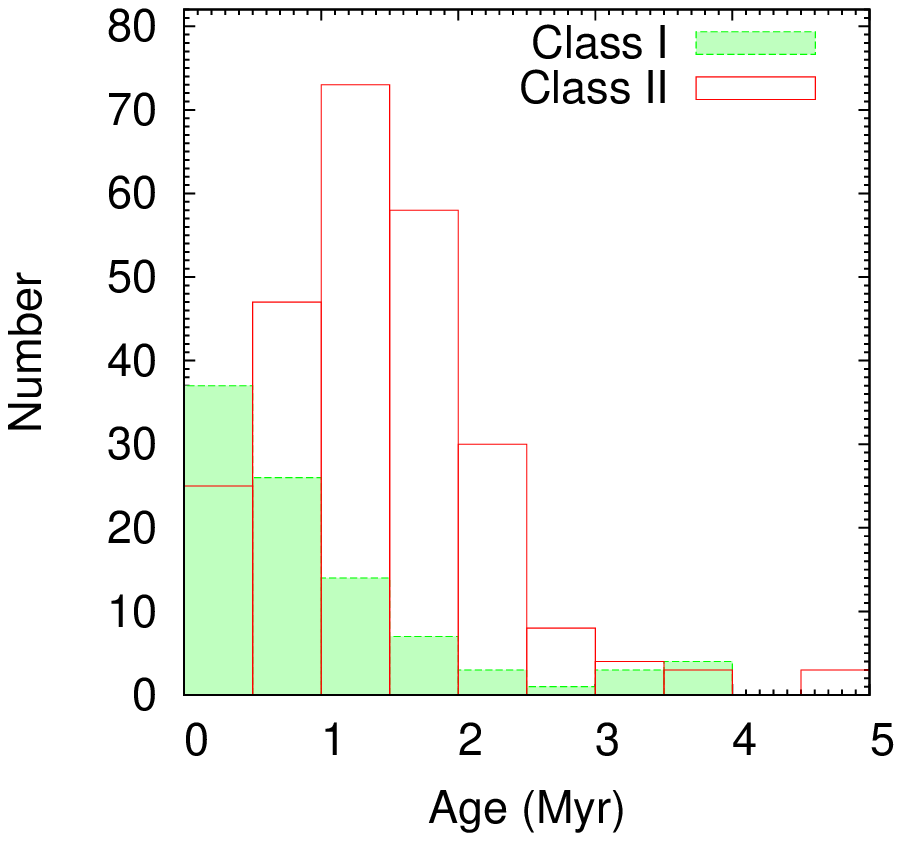}
\caption{\label{cum} (Left-hand panel): Cumulative distribution of Class I (solid line) and Class II (dotted line)
sources as a function of their age.
(Right-hand panel): Age distribution of Class I (filled histogram) and Class II (un-filled histogram) sources.}
\end{figure}

\section{Discussion}

\subsection{Stellar parameters and X-ray properties of the candidate YSOs}

In section 3.2.1, we have determined stellar parameters of 463 YSOs using the SED fitting method.
The YSOs which fall outside the distance range between 2.53 and 2.77
[i.e., $\rm (D_{s} + \sigma_{s}) < (2.65-0.12) ~kpc $ or $\rm  (D_{s} - \sigma_{s}) > (2.65+0.12) ~kpc $]
as well as those which have very low $A_V$ value [i.e., $A_{V_{s}} + \sigma_{A_{V_{s}}}<$ 2.8-0.2 mag were considered as non-members.
In the above D$_s$ and $\sigma_{s}$ are the distance and its error and 2.65 and 0.12 are the corresponding values in kpc and 2.8 and 0.2 are the foreground extinction and its error in magnitude of the  NGC 7538 region (cf. Appendix A).
A total of 44 sources have been found to be foreground/background populations and mistaken as associated YSOs 
and these are not used in further analyses.   Therefore, we have selected the remaining 419 YSOs  
as probable members of  the NGC 7538 SFR (cf. Table \ref {data4_yso}). 
Histograms of the age, mass and $A_V$ of these YSOs are shown in Fig.~\ref{histogram}.  
It is found that $\sim$91\% (380/419) of the sources have ages between 0.1 to 2.5 Myr.
Similar results have been reported by \citet{2010A&A...517A...2P} and \citet{2004AJ....128.2942B}. 
The masses of the YSOs are between 0.5 to 15.2 M$_\odot$, a majority ($\sim$86\%) of them
being between 0.5 to 3.5 M$_\odot$. 
These age and mass are comparable with those of TTSs.
Here it is worthwhile to note that the derived masses of four YSOs 
are significantly higher than the cut-off limit of 8 M$_\odot$
for which one cannot separate observationally the luminosity due to accretion
from the intrinsic luminosity of the protostar \citep{2011isf..book.....W}. 
The $A_V$ distribution shows a long tail indicating its large spread from $A_V$=1 - 30 mag,
which is consistent with the nebulous nature of this region.
The average age, mass and extinction ($A_V$) for this sample of YSOs are 1.4 Myr, 2.3 M$_\odot$ and 11 mag, respectively.

\begin{figure*}
\includegraphics[width=6.0cm,angle=270]{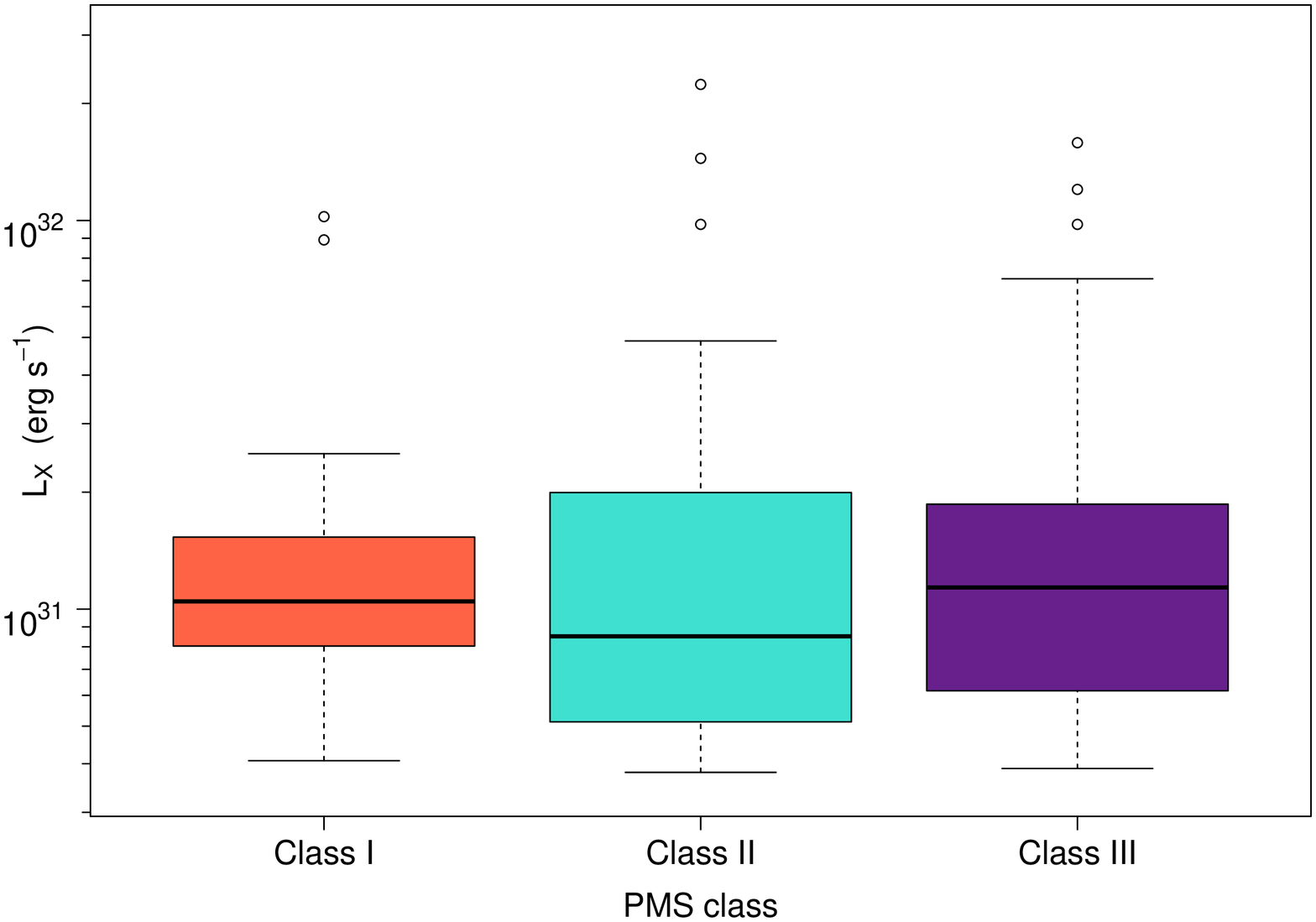}
\includegraphics[width=6.0cm,angle=270]{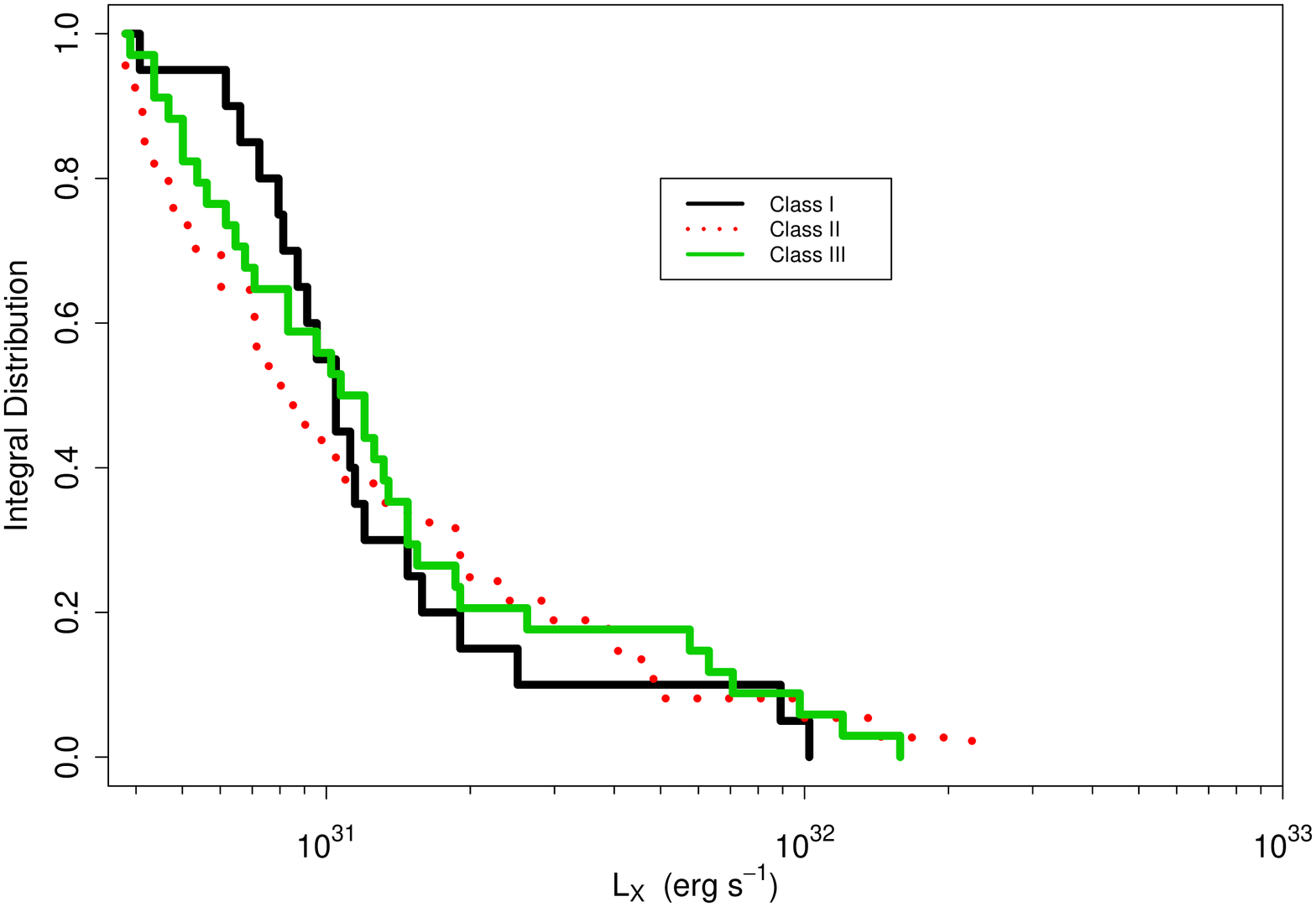}
\caption{(Left-hand panel): Boxplots of X-ray activity of Class I, Class II and Class III objects.
Data inferred from the upper and lower quartile is shown with a box and its range, denoting any points that fall outside this range as open circles. 
Open circles are the points outside 5 times the interquartile range above the upper quartile and below the lower quartile.
(Right-hand panel): Comparison of the XLFs of Class I, Class II and Class III objects. 
}
\label{fig:compare1}
\end{figure*}

The evolutionary class of the  selected 419 YSOs given in the Table \ref {data4_yso} reveals that
$\sim$24\% (99), $\sim$62\% (258) and $\sim$2\% (10) sources are Class I, Class II and Class III YSOs, respectively.
Remaining 52 of them could not be classified in the present study (cf. Section 3.1.2). 
The high percentage of Class I/II YSOs indicates the youth of this region.
In Fig. \ref{cum} (left-hand panel), we have shown the cumulative distribution of Class I and Class II YSOs as a function
of their ages, which manifests that Class I sources are relatively younger than Class II sources as expected. We have performed
a  Kolmogorov-Smirnov (KS) test for this age distribution. The test indicates 
that the chance of the two populations having been drawn from the same distribution is $\sim$2\%. 
The right-hand panel of Fig. \ref{cum}  plots the distribution of ages for the Class I and Class II sources.
The distribution of the Class I sources shows a peak at a very young age, i.e., $\lesssim$0.5 Myr, whereas that
of the Class II sources peaks at $\sim$1-1.5 Myr. 
Both of these figures were generated for the YSOs having masses greater than the completeness limit 
of this sample (i.e. $>$0.8 M$_\odot$) and show 
an age difference of $\sim$1 Myr between the Class I and Class II sources.
 Here it is worthwhile to take note that \citet{2009ApJS..181..321E} through c2d $Spitzer$ Legacy projects studied YSOs
associated with five nearby molecular clouds and concluded that the life time of the Class I phase is 0.54 Myr.
The peak in the histogram of Class I sources agrees well with them.

\begin{figure*}
\includegraphics[width=16cm,angle=0]{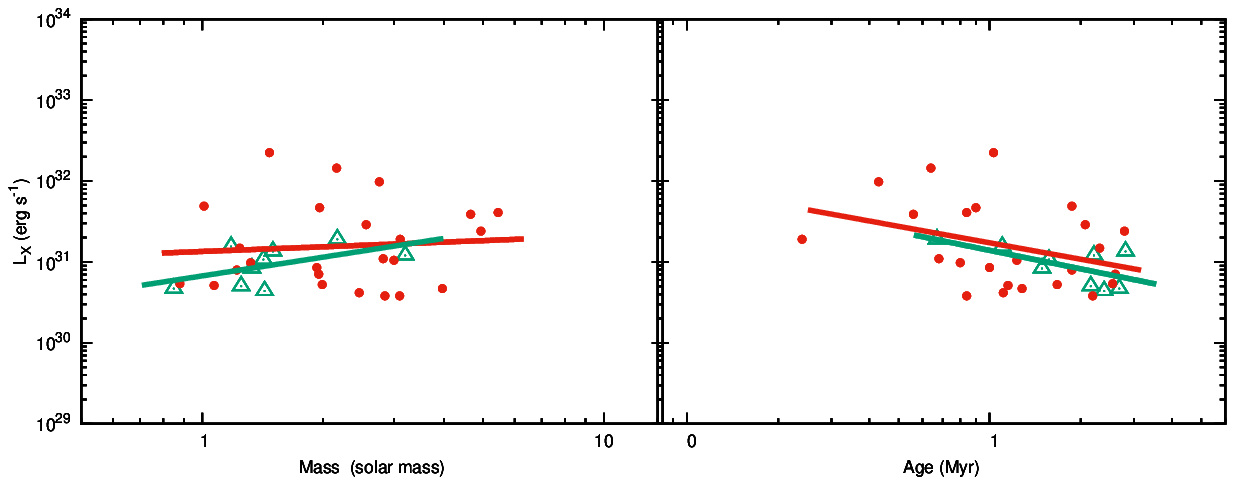}
\caption{
$L_X$  versus mass/age in a logarithmic scale for the identified TTSs.
Red dots and green triangles represent Class II  and Class III YSOs, respectively.
Red and green lines are the least square fit to the data for Class II and Class III sources, respectively.
}
\label{fig:compare2}
\end{figure*}

\begin{table}
\caption{Results of two sample tests. Columns 2, 3 and 4 represent the probability that the
samples have a common parent $\rm{L_X}$ distribution }             
\label{tab:two_sample}      
\begin{tabular}{lccc}       
\hline
Test                                & Class I    &  Class II    &  Class I    \\
                                    &   versus         &   versus          &     versus        \\
                                     & Class II   &   Class III   &  Class III   \\
\hline
Wilcoxon Rank Sum                    &    0.35 &    0.44 &    0.91  \\
Peto and Peto Generalized            &    0.36 &    0.44 &    0.91  \\
Wilcoxon                            &         &         &           \\
Kolmogorov-Smirnov                   &    0.19 &    0.91 &    0.68  \\
Anderson-Darling                     &    0.16 &    0.44 &    0.43  \\
\hline
\end{tabular}
\end{table}

The stellar parameters of the X-ray emitting sources identified as YSOs 
can be used to study the possible physical origin of the X-ray emission in PMS stars. 
The coronal activity which is primarily responsible for the generation of X-ray in low mass stars may be affected by the
composition of X-ray emitting plasma and the disk fraction during the PMS evolution in different classes of YSOs.
A number of studies have been done in various SFRs, but no firm conclusion has been drawn so far on account of contradictory results.
In some cases, such as Chameleon I \citep{fei+93}, $\rho$ Ophiuchus \citep{cas+95},  IC 348
\citep{pre+02} and NGC 1893 \citep{2014NewA...29...18P}, no significant difference has been found in the  X-ray  luminosity between Class  II  and  Class III  YSOs.
However, there are several examples \citep{ste+01, fla+03, sta+04, pre+05, fla+06, tel+07, gua+12}
which show that Class  II YSOs have lower X-ray  luminosities  than  Class  III  YSOs.

We have derived $L_X$ for 91 YSOs in the NGC 7538 region (cf. Section 3.3).
A comparison of the X-ray activity of the YSOs with different classes 
[Class I: 20, Class II: 34 and Class III: 37]
are represented in Fig. \ref{fig:compare1}  with boxplots and  X-ray luminosity functions (XLFs) by using the Kaplan Meier estimator
of integral distribution functions in  the R package (ver 3.2.0). 
The distribution shows that the X-ray activity is nearly similar in all YSO classes 
with a mean value of $\rm{log (L_X)}$ around 31.1 $\rm{erg~s^{-1}}$.
The distribution of  the Class II YSOs shows a relatively larger scatter in
comparison to those of  the Class I and Class III sources.
To derive the statistical significance of the comparison,
we performed the two sample tests for estimating the probability of having  common parent distributions
and the results are given in Table~\ref{tab:two_sample}.
They 
show that the X-ray activity in  the Class I, Class II and Class III objects is not significantly different from each other.
It is thought that the X-ray activity in low mass stars is associated mainly with the rotation rate and depth of the convection zone.
Our results may imply that the increase of the X-ray surface activity with an increase of the 
rotation rate may be compensated by the decrease of the stellar surface 
area during  the PMS evolution \citep{1997A&A...324..690P, bha+13}.

Out of  the 91 YSOs having  $L_X$ estimation (cf. Section 3.3), the age and mass have been
estimated for 47 YSOs on the basis of SED fitting. The masses of these YSOs is $>$0.8 M$_\odot$.
It is worthwhile to mention that the completeness limit for the X-ray emitting YSOs (64) is 0.8 M$_\odot$ (cf. Section 3.1.5).

To study the effect of  circumstellar disks on the X-ray emission,
in Fig. \ref{fig:compare2} we have tried to see  if any correlation of $L_X$ exists with the age and mass. 
When concluding any results from these distributions, we have to be careful about the large errors in the 
estimated values of the age/mass of the TTSs.

\begin{table}
\scriptsize
\caption{\label{coefficients}  Coefficients of the straight lines fitted to the $L_X$ versus mass/age distribution of  the YSOs of different classes.}
\begin{tabular}{@{}lcccccr@{}}
\hline
Class~&Mass   & a    &  b  &  Age  & a & b \\
&(M$_\odot$)  &      &     &  (Myr) &   &   \\
\hline
II &0.9 -  5.5  &$  31.1\pm 0.2$&$ 0.2\pm 0.5$&$ 0.2 - 2.8   $&$   31.2\pm 0.1$&$ -0.7\pm 0.4$\\
III &0.9 -  3.2  &$  30.8\pm 0.1$&$ 0.8\pm 0.5$&$ 0.7 - 2.8   $&$   31.1\pm 0.1$&$ -0.8\pm 0.3$\\
\hline
\end{tabular}
\end{table}

The $L_X$ versus mass distribution show a large scatter but with  an indication of increasing  
$L_X$ with mass for the Class III sources.
The values of coefficients `a'  and `b'  of the linear regression fit log($L_X$) = a + b$\times$log(M$_\odot$) 
for the Class II and Class III sources are given in Table \ref{coefficients}.
The intercepts `a' for the sources of different classes have comparable
values to each other, indicating that the presence
of circumstellar disks has practically no influence on the X-ray emission.
This result is in agreement with that reported by \citet{2002ApJ...574..258F} and  \citet{2014NewA...29...18P}, 
and is in contradiction with  those by \citet{2001A&A...377..538S}, \citet{2005ApJS..160..401P} and \citet{2007A&A...468..425T}. 
There is  an indication for a higher value of `b' for Class III as compared to Class II sources.
Recently, \citet{2014NewA...29...18P} have found, for their sample of Class II and Class III sources in the mass range 0.2-2.0 M$_\odot$ in the NGC 1893 region,
the `b' values to be $0.51\pm0.20$ and $1.13\pm0.13$, 
respectively, whereas the `a' values to be  $30.71\pm0.07$ and $30.74\pm0.05$, 
respectively.
\citet{2007A&A...468..425T} have found a =  $30.13\pm0.09$ and  $30.57\pm0.09$ and b = $1.70\pm0.20$ and  $1.78\pm0.17$ 
for  Class II and Class III sources, respectively, in the Taurus molecular cloud. 
These slope values are higher than that obtained in the present study.
One possible reason for the lower value for the NGC 7538 region may be a bias in our
sample toward X-ray luminous, low mass stars. 
In the case of NGC 2264, the linear fit for the detected sources yields a =$30.6\pm0.3$ and b = $0.8\pm0.1$ \citep{2007AJ....134..999D}. However,
after taking into account the X-ray non-detections, the slope is found to be steeper (b = $1.5\pm0.1$).
The mass range of YSOs used for the fitting also plays a crucial role as already demonstrated by \citet{2012ApJ...753..117G}.
They have found that for Class III sources having
mass $\leq$0.8 M$_\odot$ in NGC 6611 gave a value of slope b = $1.1\pm0.3$, whereas the distribution of stars more massive
than 0.8 M$_\odot$ is flatter with a slope of $0.4\pm0.2$.

As for the $L_X$ versus age distribution, $L_X$ for both Class II and Class III sources can be seen to decrease systematically with
age (in the age range of $\sim$0.4 to 2.8 Myr). The values of the coefficients `a'  and `b' of a 
linear regression fit log($L_X$) = a + b$\times$log(age) for the sources of different classes  are given in Table \ref{coefficients}.
The nearly similar values of intercept `a'  and slope `b' for Class II and Class III indicate that
circumstellar disks have practically no effect on the X-ray emission.
Recently, \citet{2014NewA...29...18P} have found  a = $35.06 \pm0.63$ and $35.29\pm0.65$ and b = $-0.78\pm0.10$ and $-0.81\pm0.11$, 
respectively, for their sample of Class II and Class III sources in NGC 1893. 
In NGC 7538 we found that the values of `a' for both Class II and Class III are much lower than those of NGC 1893,
but the `b' values are almost similar to theirs. 
These are slightly steeper in comparison to those (-0.2 to -0.5) reported by
\citet{2005ApJS..160..390P} and  \citet[$-0.36\pm011$;][]{2007A&A...468..425T}.
This difference in the slope could be more significant 
in the case of an unbiased sample where we expect more less luminous, low mass X-ray sources which would give
a higher value for the slope. If X-ray luminosities of accreting PMS stars are systematically lower than non-accreting PMS
stars \citep[e.g.,][]{2003ApJ...582..382F,2005ApJS..160..401P} and if
Class II sources evolve to Class III sources, one might expect that in the case
of  the Class II source, $L_X$ should increase with age rather than decrease.
However, the evolution of the Class II sources up to $\sim$2.8 Myr does
not show any sign of increase in X-ray luminosity.

\begin{figure*}
\centering
\centering\includegraphics[height=8.7cm,width=8.7cm]{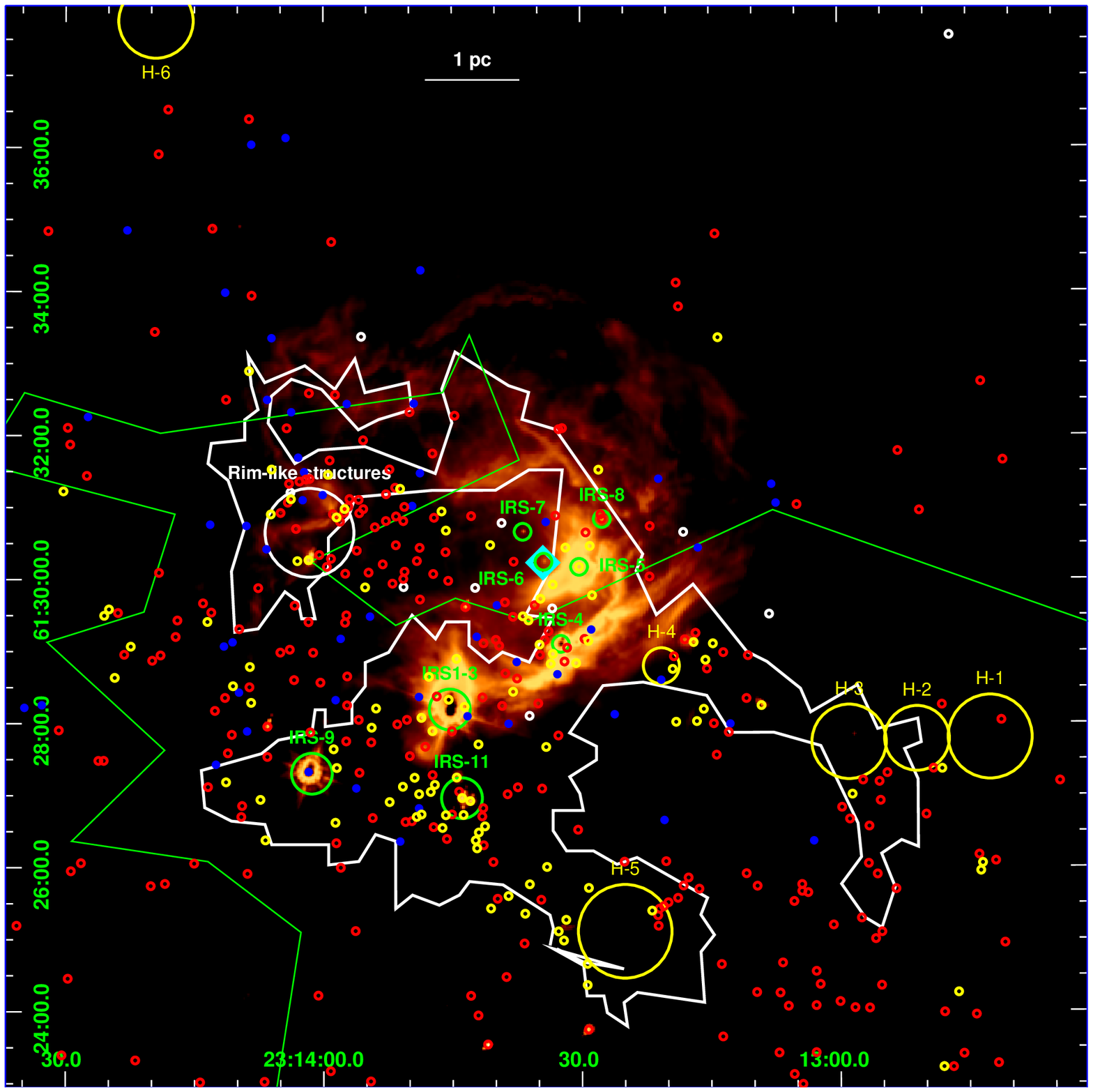}
\centering\includegraphics[height=8.7cm,width=8.7cm]{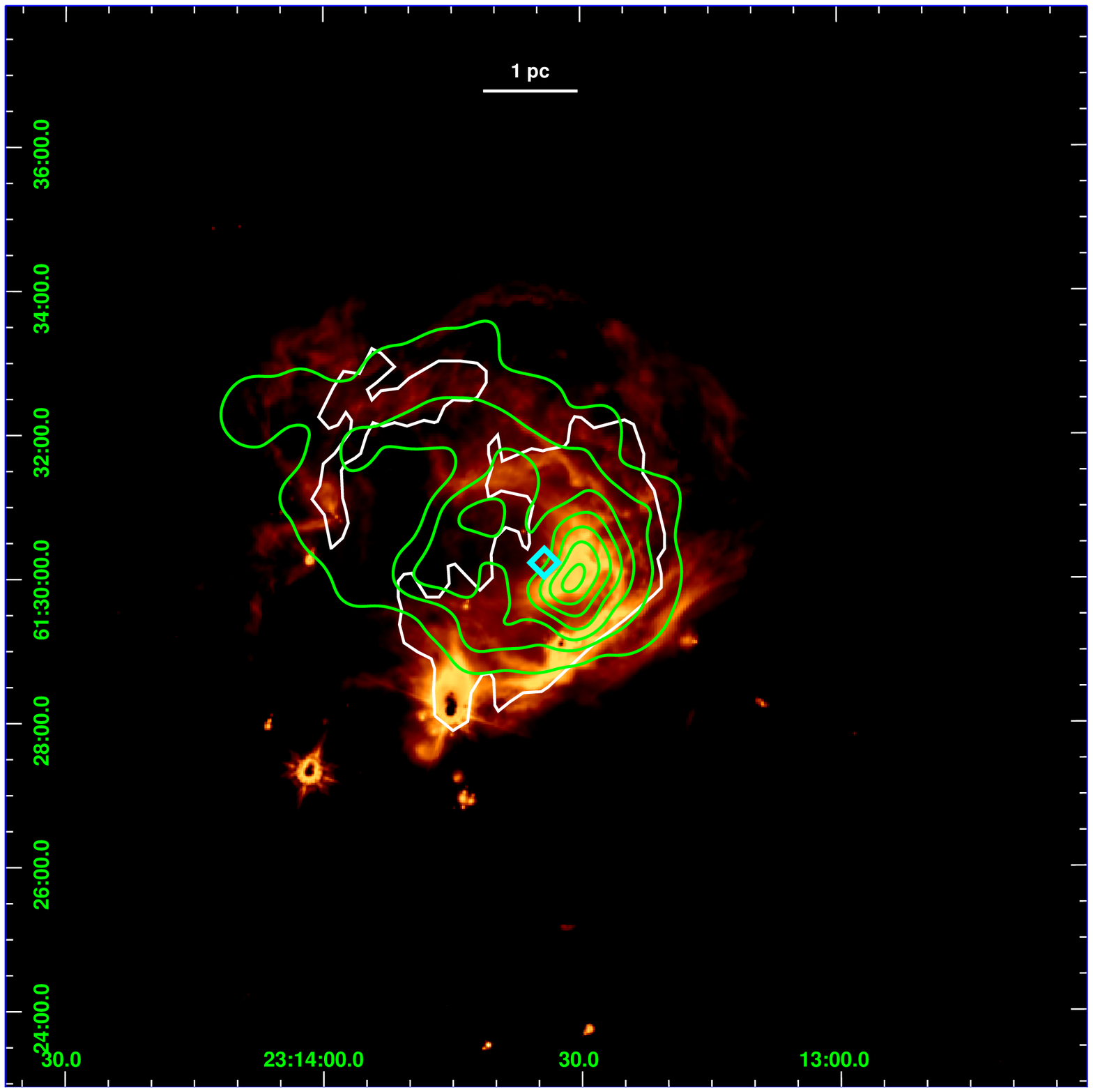}
\caption{\label{spa} (Left-hand panel): Spatial distribution of the YSOs
superimposed on the $15^\prime\times15^\prime$ IRAC  8.0 $\mu$m  image of the NGC 7538 region.
The location of Class I (yellow circles), Class II (red dots), Class III (white dots) and unclassified (blue dots) 
sources are shown along with the $J = 1-0$ line of $^{13}$CO (green contour) and 
850 $\mu$m continuum emission (white contour) taken from \citet{2014MNRAS.439.3719C}.
Also shown are the positions of high-mass dense clumps \citep{2013ApJ...773..102F}  along with the IR sources.
(Right-hand panel): Distributions of the ionized gas as traced by 
radio \citep[1280 MHz, green contours,][]{2004ApJ...616.1042O}  
and by H$\alpha$ emission (white contours) are shown overlaid on the IRAC  8.0 $\mu$m image of the same region.
}. 
\end{figure*}

\subsection{Triggered star formation}

\citet{1991MmSAI..62..715M} have suggested the presence of YSOs
of various evolutionary stages in the vicinity of NGC 7538 with considerable 
substructures (cf. Fig. \ref{color}). 
Brief description of each of them is summarized as follows:

\noindent
{\bf IRS 1-3:} This active region has three massive IR sources, each associated with its own compact H\,{\sevensize II}  region \citep{1974ApJ...187..473W}.
IRS 1 has been identified as a high-mass ($\sim$30 M$_\odot$) protostar with a CO outflow, 
and is considered as the source injecting energy to the UC H\,{\sevensize II}  region NGC 7538 A \citep{2010A&A...517A...2P}.
IRS 2, situated $\sim$10 arcsec north of IRS 1, is inferred to be an 
O9.5V star and possesses the most extended H\,{\sevensize II}  region. 
IRS 3 is situated about $\sim$15 arcsec west of IRS 1 \citep{2004ApJ...616.1042O}
and is the least luminous of the three.

\noindent
{\bf IRS 4-8:} The composition of this active region is a group of young stars located at the southern 
rim of the optical H\,{\sevensize II}  region IRS 4,
a bright NIR reflection nebula containing an O9V star IRS 5, 
and the  main ionizing source  IRS 6 (O3V type) of the H\,{\sevensize II}  region NGC 7538 \citep{2010A&A...517A...2P}. 
IRS 7 and IRS 8 have been inferred as foreground field stars \citep{2005prpl.conf.8307T,2010A&A...517A...2P}.

\noindent
{\bf IRS 9:} This bright reflection nebula  located at the south-eastern tip of NGC 7538 harbors 
massive protostars \citep{1979MNRAS.188..463W,2006A&A...448L..57P, 2010A&A...517A...2P}.

\noindent
{\bf IRS 11:} This may be a large contracting or rotating filament that is
fragmenting at scales of 0.1 pc to 0.01 pc to form multiple high-mass stars  
($\sim$10 M$_\odot$) having disks and envelopes as well as shedding outflows 
\citep{2006A&A...448L..57P,2012ApJ...757...58N}. 

\noindent
{\bf Rim-like structures (globules):} This nebular region is seen on the north-eastern side of 
NGC 7538 consisting two cone-shaped rim-like structures.
Both point toward IRS 6 and have faint stars on their tip.

\noindent
Apart from these active regions, \citet{2013ApJ...773..102F} have recently detected 13 more candidates for 
high-mass dense clump in a one square degree field of NGC 7538.
The positions and sizes of some of them are shown in Fig. \ref{color} as H1 to H6, located mainly in the south-western periphery of NGC 7538.
These are potential sites of intermediate- to high-mass star formation to be identified 
through far infra-red (FIR) observations by $Herschel$ space based telescope.

\begin{figure*}
\centering
\centering\includegraphics[height=12cm,width=12cm]{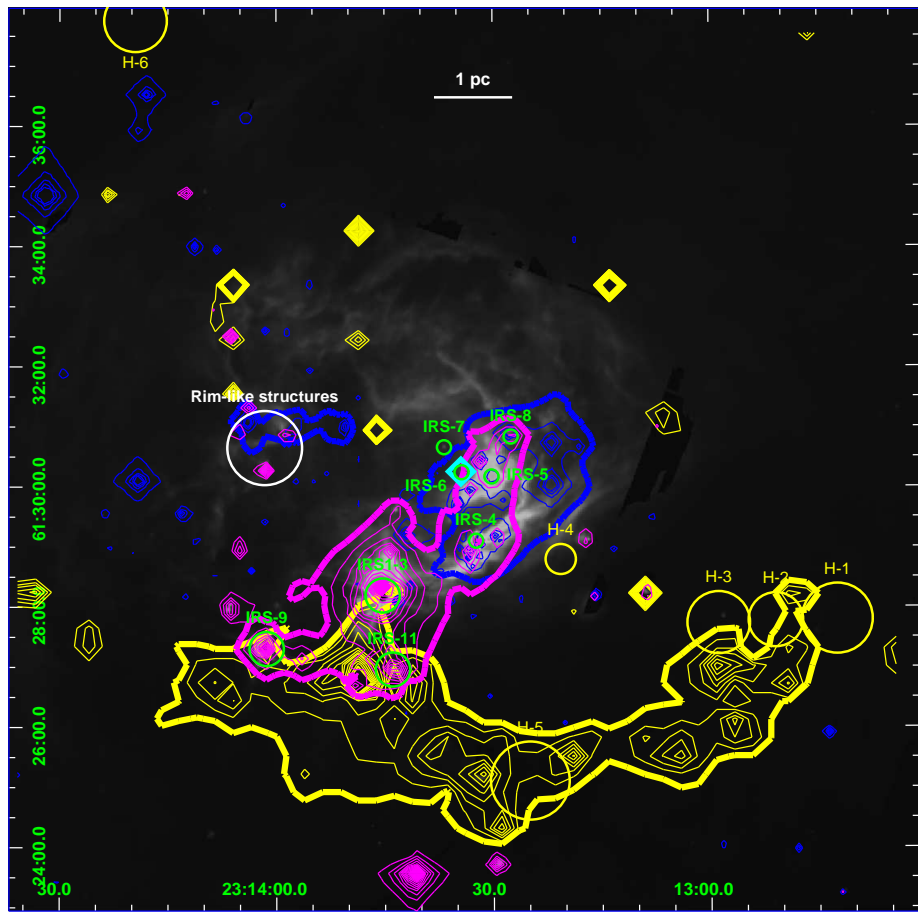}
\caption{\label{spaage} Distribution of younger (yellow contours), older (blue contours) and massive (purple contours) 
populations overlaid on the IRAC 8.0 $\mu$m image. 
The location of the IR sources, globules and cold clumps are also shown.
}
\end{figure*}

Fig. \ref{spa} (left-hand panel) shows the spatial distribution of
member YSOs (419 sources, cf. Section 4.1)
superimposed on the IRAC 8.0 $\mu$m  image of $15\times15$ arcmin$^2$ of the NGC 7538 region.
This sample of YSOs is obtained from 
multiwavelength data  taken from  various surveys having 
different completeness limits. 
In section 3.1.5, we have discussed  the completeness of these surveys
and found that approximately they can be assumed complete for mass $\geq$0.8 M$_\odot$.
A majority (94\%) of the YSOs selected here have masses $\geq$0.8 M$_\odot$,
therefore, incompleteness in this sample will have minimal effect on the over-all spatial distribution of the YSOs.

The  $^{13}$CO $J = 1-0$ line (green contour) and 850 $\mu$m continuum (white contour) emission 
maps taken from \citet{2014MNRAS.439.3719C} are shown in Fig. \ref{spa} (left-hand panel). 
For simplicity we have shown only the outer most contours, representing the extent of gas and dust 
in this region.
The resolution of the $^{13}$CO and 850 $\mu$m observations are 46 arcsec and 14 arcsec, respectively.
\citet{2014MNRAS.439.3719C} have studied the groupings of YSOs in this region and found several of them. 
They  estimated physical parameters of these groups and found that younger sources  are located in  
regions having higher YSOs surface density and are correlated with densest molecular clouds. 
Since they have not discussed the effect of high mass stars on the recent star formation through the distribution of YSOs, gas and dust,
we further used these distributions to trace star formation activities in this region.
The positions of the high-mass dense clumps \citep{2013ApJ...773..102F}  along with the IR sources are also shown in the figure.
In the right-hand panel of Fig. \ref{spa}, the distributions of ionized gas as seen in the 
1280 MHz radio continuum emission \citep[green contours,][]{2004ApJ...616.1042O} and in H$\alpha$ emission (white contours) are 
shown superposed on the IRAC  8.0 $\mu$m image of the same region.
The YSOs are distributed either on the nebulosity or towards the southern regions.
The distribution of the $^{13}$CO, 850 $\mu$m and H$\alpha$ emission indicates a bubble-like feature around the ionizing
source IRS 6, likely created due to the expansion of the H\,{\sevensize II}  region. 
The correlation of the PAH emission  
as seen in IRAC 8.0 $\mu$m image \citep{2009A&A...494..987P} with the $^{13}$CO emission indicates that the ionized gas
is confined inside the molecular cloud. 
The lack of  diffuse emission at 8.0 $\mu$m can be noticed towards the south-east of IRS 6. However, as can be seen 
from the radio continuum contours  in Fig. \ref{spa} (right-hand panel), 
the ionized gas is bounded more sharply to the south-western region.
The distribution of  Class I YSOs shows a nice correlation with that
of molecular gas and the PAH emission/H\,{\sevensize II}  region boundaries. 
Very few Class I sources are located towards the central region near IRS 6 as compared to the outer, southern regions.
The strong positional coincidence between the YSOs and the molecular cloud
suggests an enhanced star formation activity towards this region, as
often observed in other SFRs \citep[e.g.][]{2009ApJS..181..321E,2009A&A...504..461F}.
There is a concentration of very young YSOs (mainly Class I) towards the southern region outside of the dust rim (IRS 11).
Five high-mass dense clump candidates \citep{2013ApJ...773..102F} are located just outside the NGC 7538 H\,{\sevensize II}  region towards southwest.
There are many separate groups of younger YSOs located near these cold molecular clumps.
In summary, there are two main concentrations of YSOs, one  
in the  H\,{\sevensize II}  region and the other in the southern region. 

\begin{figure}
\centering
\includegraphics[height=4.5cm,width=7cm]{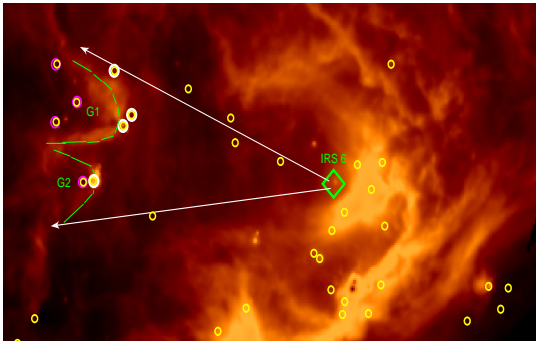}
\caption{\label {brc} Location of the two globules along with the distribution of Class I YSOs 
(circles) and `IRS 6', the ionizing source of NGC 7538. 
}
\end{figure}

By analyzing the distribution of YSOs of various ages and masses
in relation to the molecular cloud structure, 
we can study the mode of star formation in this region.
For this, we have  generated  contour maps of the age/mass of YSOs
smoothened to the  resolution of 9 arcsec grid size. 
In Fig. \ref{spaage}, we have over-plotted the age contours on the IRAC 8.0 $\mu$m image 
representing younger (yellow contours) and comparatively older (blue contours) YSOs.
The outermost age contours for them correspond to 0.7 Myr and 1.8 Myr, respectively.
In the same figure, we have also plotted the mass contours (purple contours) of the YSOs representing the massive ones. 
The  outer-most contour for it corresponds to 4.2 M$_\odot$. 
The step size are 0.2 Myr and 0.2 M$_\odot$ for the age and mass contours, respectively.
The older population is mainly associated with the  H\,{\sevensize II}  region enclosed by thick blue contours, 
whereas the younger one is located outside the H\,{\sevensize II}  region mainly in the south-western part enclosed
in a thick yellow contour.
The mass distribution shows that the massive population, enclosed in the thick purple contours, is sandwiched between these two populations
and is associated mainly with IRS 1-3, IRS 9 and IRS 11.
The mean values of ages and masses of the YSOs associated with these subregions are given in Table \ref{stats}.

\begin{table}
\caption{\label{stats} Mean values of ages and masses of YSOs in different regions.}
\begin{tabular}{@{}lcccc@{}}
\hline
Region & N &  Mean mass & Mean age  \\
      &   &  (M$_\odot$) & (Myr)  \\    
\hline
&&&&\\
Whole  & 419 & $2.3\pm0.1$ & $1.4\pm0.1$ \\
&&&&\\
South-west (young)   &91 &$1.8\pm 0.1$&$ 1.0\pm 0.1$\\
&&&&\\
Central (old)     &73 &$2.7\pm 0.2$&$ 2.0\pm 0.2$\\
"~~~~~  (old-globules) &32 &$2.5\pm 0.2$&$ 1.9\pm 0.2$\\
"~~~~~  (old-IRS 6)  &41 &$2.9\pm 0.2$&$ 2.2\pm 0.2$\\
&&&&\\
South-central (massive) &35 &$4.7\pm 0.6$&$ 1.2\pm 0.2$\\
\hline
\end{tabular}
\end{table}

The location of the IR sources, rim-like structures and the cold clumps are also 
shown in Fig. \ref{spaage}.
The oldest region (thick blue contours) is made up of two separate groups, i.e., one around IRS 6 and the other near the rim-like structures (globules).
The masses and ages of the YSOs near the rim-like structures are lower as compared to those in the central region near IRS 6.  
The two rim-like structures with sharp edges (maybe due to the PAH emission) 
pointing towards the central star IRS 6 indicate that the ionization front (IF) interacts with the molecular ridge as seen in the $^{13}$CO emission.
They morphologically resemble bright-rimmed clouds that result from the pre-existing dense molecular clouds impacted by the ultra-violet 
photons from nearby OB stars \citep{1994A&A...289..559L}.
The low mass YSOs on their tips are generally believed to be formed as a result of triggering effect of the expanding H\,{\sevensize II} region 
\citep{2005A&A...433..565D,2006A&A...446..171Z,2008ApJ...688.1142K,2009A&A...496..177D}.
Their elongated distribution and age difference with respect to the location of
the ionization source can be used to check whether the RDI \citep[][]{1994A&A...289..559L,2006MNRAS.369..143M} 
mode of triggered star formation was effective in the region.

For this purpose, we have compared the time elapsed during the formation of the Class I YSOs  in the globule 
regions with the age/lifetime of the O3 ionization source.
In Fig. \ref{brc}, we show the location of the two globules, the ionizing source IRS 6 and the Class I sources.
The presence of extremely young Class I sources presumably represents the 
very recent star formation event in the region,
thus can be taken as a proxy to trace the triggered star formation.
The position of the tips of the globules is situated at a 
projected distance of $\sim$2.4 pc from IRS 6. 
We estimated the time needed for the IF to travel there 
as $\sim0.26$ Myr, assuming that it expanded at
the speed of $\sim$9 km s$^{-1}$ \citep[see, e.g.,][]{1976RMxAA...1..373P}.
The age of the Class I sources (white circles in Fig. \ref{brc}) on this rim is  $1.9\pm0.6$ Myr,
which is $\sim$0.3 Myr younger than the estimated age of the O3V star \citep[$\sim$2.2 Myr, cf.][]{2010A&A...517A...2P}.
The mean age of the Class I sources inside the rim (magenta circles in Fig. \ref{brc}) is $1.5\pm0.6$ Myr. 
We have evaluated the shock crossing time in the globules to see whether the star formation 
there initiated by the propagation of the shock or whether it had already taken place prior to the arrival of the shock.         
Assuming a typical shock propagation velocity of 1-2 km s$^{-1}$, as found in
the case of bright-rimmed clouds \citep[see, e.g.,][]{1999A&A...342..233W,2004A&A...414.1017T}, 
the shock travel time to the YSOs, which are projected at distances
$\sim$0.6 pc from the head, is $\sim0.4$ Myr.
This time-scale is comparable to the difference in the ages of the YSOs on the rims and inside them.
Although the sample is small and the errors are large, these results seems to support the notion that
the formation of the YSOs in the globules could be due to the RDI mechanism.
The above analysis is not statistically significant to conclude  the triggered star formation, 
but  \citet{2015MNRAS.450.1199D} have stated that the system where many indicators can be satisfied 
simultaneously can be a genuine site of triggering.

Therefore, to investigate further, we have calculated the dynamical age of the NGC 7538 H\,{\sevensize II} 
region from its radius using the equation given by \citet{1978ppim.book.....S}:

\begin{figure}
\centering\includegraphics[height=5cm,width=7cm,angle=0]{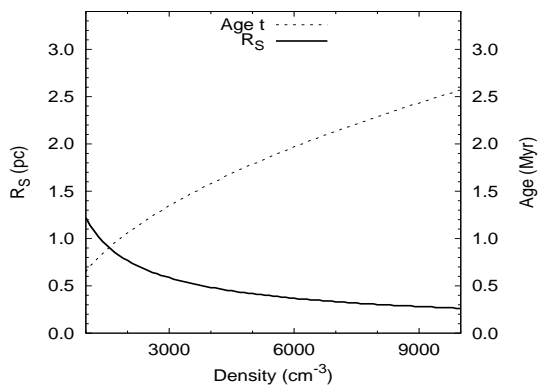}
\caption{\label{rs} Plots showing the variation of the Str\"omgren radius and dynamical age with the ambient density.
}
\end{figure}

\begin{figure*}
\centering\includegraphics[height=4.5cm,width=5.5cm,angle=0]{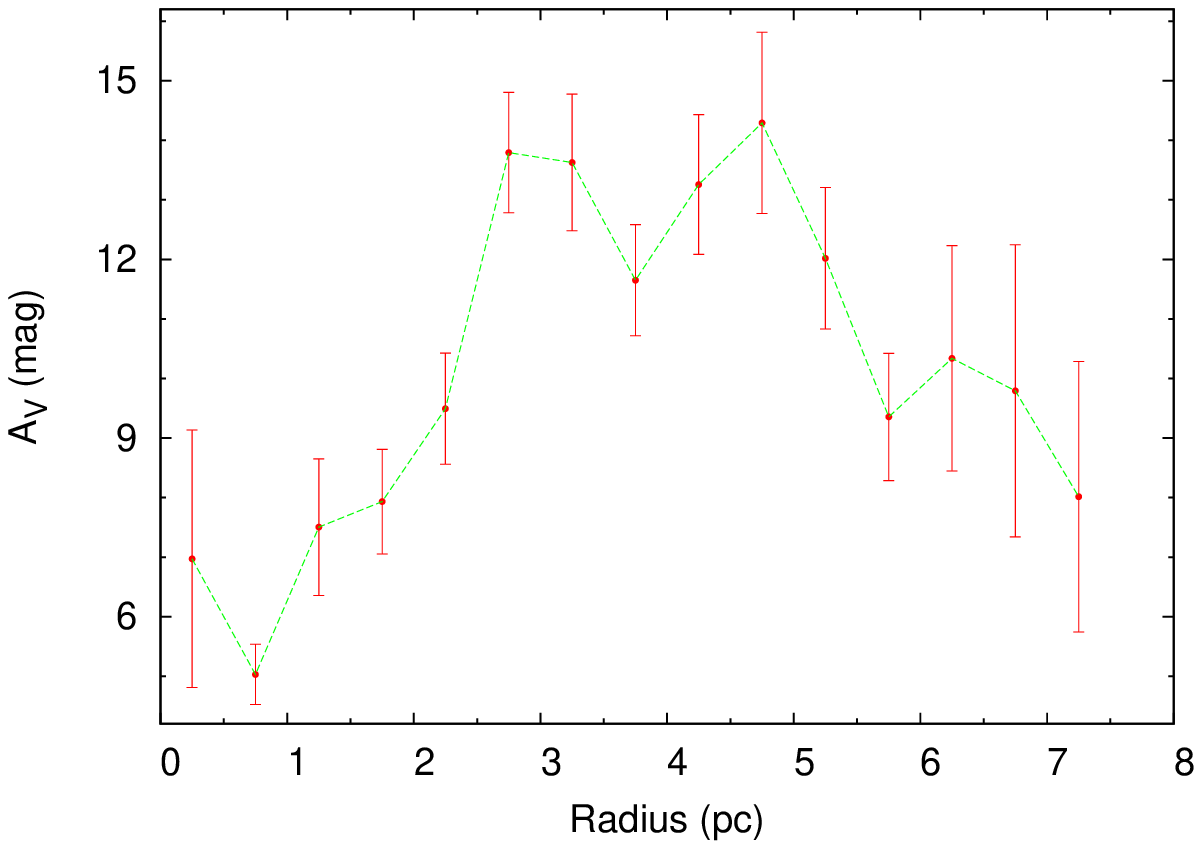}
\centering\includegraphics[height=4.5cm,width=5.5cm,angle=0]{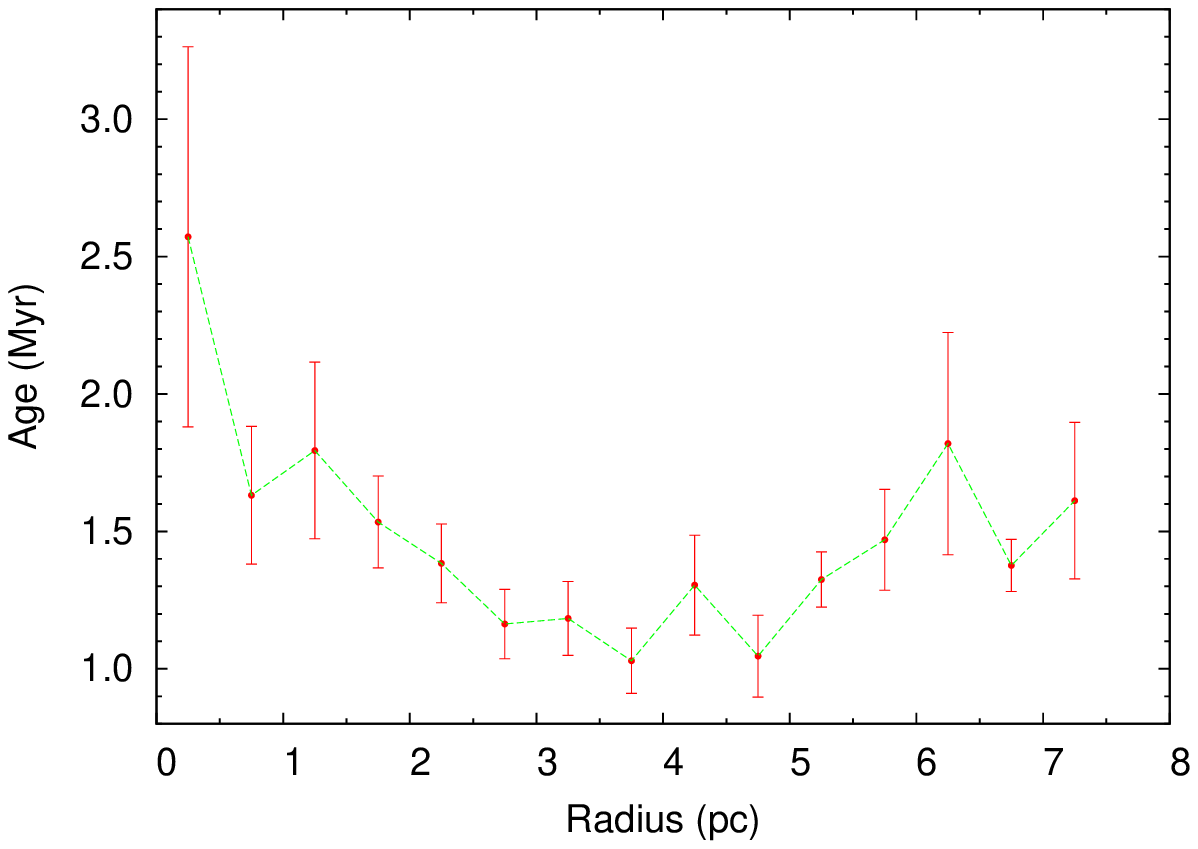}
\centering\includegraphics[height=4.5cm,width=5.5cm,angle=0]{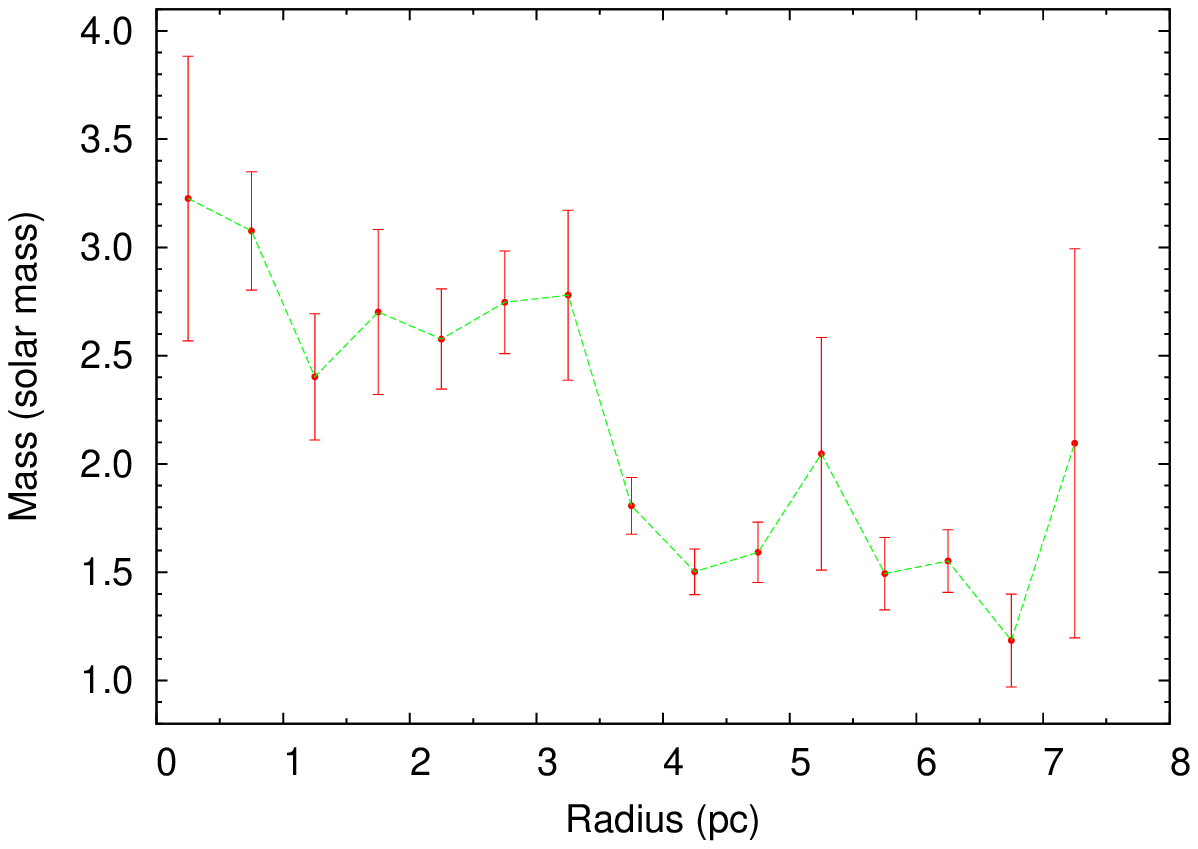}
\caption{\label{rage} Distribution of $A_V$, age and mass of the YSOs as a function of radial distance from the ionizing source IRS 6. }
\end{figure*}

$R(t) = R_S (1+ {{7ct}\over{4R_s}})^{4/7} $,

\noindent
where $R(t)$ denotes the radius of the H\,{\sevensize II}  region at time $t$, and $c$ is the sound speed.
The latter was assumed as $\sim$9 km s$^{-1}$ \citep{1976RMxAA...1..373P,2005fost.book.....S}, and 
the former was taken to be 3 pc as is derived from the radio and H$\alpha$ maps (cf. Fig. \ref{spa}).
Then the dynamical age $(t)$ can be calculated if we know Str\"omgren radius $R_S$, which 
is estimated by using the relation given in \citet{2011isf..book.....W} and \citet{2005fost.book.....S}
and by assuming an O3V star as the  ionizing source emitting 7.4 $\times 10^{49}$ UV photons per second \citep{1996ApJ...460..914V}. 
However, the information on the initial ambient density is needed, which we don't know. 
So we have left it as a free parameter and calculated the Str\"omgren radius ($R_S$) and the 
corresponding dynamical age $(t)$ of the H\,{\sevensize II} region for a range of 
ambient density from $10^3$ to 
$10^4$ cm$^{-3}$. Fig. \ref{rs} shows the results, where $R_S$ and $t$ are plotted as a function of the initial ambient density. The former varies from $\sim$ 1.3 to 0.25 pc, while the latter varies from $\sim$ 0.7 to 2.6 Myr. 
This upper limit of the dynamical age is comparable to the age of the central 
O3 star and corresponds to  a higher ambient  density for this region.
The mean age of the YSOs in this region is 1.4 Myr, which is less than the dynamical age of the region
and, as such, their formation could have been influenced by the  expanding H\,{\sevensize II} region.

We have also looked for the radial distance dependence of $A_V$/age/mass of YSOs with respect 
to the ionizing source IRS 6 as shown in Fig. \ref{rage}.
As expected, the $A_V$ distribution reveals less extinction near IRS 6 as compared to the outer region.
There is a broad peak starting at the projected distance of $\sim$3 pc and we see a decreasing trend after 5 pc. 
This seems to indicate the presence of a shell-like layer of collected medium just outside the H\,{\sevensize II}  region.
It is very interesting to note that the age distribution shows a clear decrease from the center to a distance of 3 pc.
Also 
the masses of the YSOs within 3 pc is higher as compared to those outside indicating a
difference in the  physical properties of the YSOs  within 3 pc.
These trends are indicative of triggered star formation in the inner region (within 3 pc), 
where the O3 star played an active role.
From 3 pc out, we see a completely different distribution 
with a large number of young and low mass YSOs (cf. Table \ref{stats}) located in the southwest region of NGC 7538 
(cf. Fig. \ref{spaage}, thick yellow contour).
It seems that the YSOs in this group might have formed spontaneously in the absence of any triggering mechanism 
due to the O3 star. The distribution of the cold clumps also suggests that spontaneous low-mass star formation is under way there.
However, disentangling triggered star formation from spontaneous star formation accurately requires 
precise determination of the proper motions and ages of individual sources \citep{2015MNRAS.450.1199D}.

\subsection{Mass function}

The distribution of stellar masses that form in one star-formation event in a given
volume of space is called IMF. Together with star formation rate,
it is one of the important issues of star-formation studies.
Since, environment effects due to the presence of high mass stars 
may be more revealing at the low-mass end of the present day MF,
we will try to study it in the NGC 7538 SFR.

The MF is often expressed by a power law,
$N (\log m) \propto m^{\Gamma}$ and  the slope of the MF is given as:

   $$ \Gamma = d \log N (\log m)/d \log m  $$

\noindent
where $N (\log m)$ is the number of stars per unit logarithmic mass interval.
The first empirical determination of MF was by \citet{1955ApJ...121..161S}, which gave $\Gamma = -1.35$ for
the field stars in the Galaxy in the mass range $0.4 \leq$ m/M$_\odot \leq 10$. However, subsequent works
\citep[eg.][]{1979ApJS...41..513M, 1986FCPh...11....1S, 1991ARAA..29..129R, 2002Sci...295...82K}
suggest that the MF in the Galaxy often deviates from  the pure power law.
It has been shown \citep[see e.g.][]{1986FCPh...11....1S, 1998ASPC..142..201S, 
2002Sci...295...82K, 2003PASP..115..763C, 2005ASSL..327.....C} that, for masses above $\sim$1 M$_\odot$, the MF
can generally be approximated by a declining power law with a slope similar to
that found by \citet{1955ApJ...121..161S}. However, it is now clear that this power law does
not extend to masses below $\sim$1 M$_\odot$. The distribution becomes flatter
below 1 M$_\odot$ and turns down at the lowest stellar masses. \citet{2002Sci...295...82K} divided
the MF slopes for four different mass intervals.
It was also often claimed that some (very) massive SFRs have 
truncated MFs, i.e., contain much smaller numbers of low-mass
stars than expected from the field MF.
However, most of the recent and sensitive studies of massive SFRs
\citep[see, e.g.][]{2009MNRAS.396.1665L, 2009A&A...501..563E} found large numbers
of low-mass stars in agreement with the expectation from
the ``normal" field star MF.
\citet{2011AA...530A..34P} confirmed these results for the Carina Nebula
and supported the assumption of a universal IMF (at least in
our Galaxy). In consequence, this result also supports the notion that
OB associations and  massive star clusters are the
dominant supply sources for the Galactic field star population, as
already suggested by \citet{1978PASP...90..506M}.

\begin{figure}
\centering
\includegraphics[height=6cm,width=8cm]{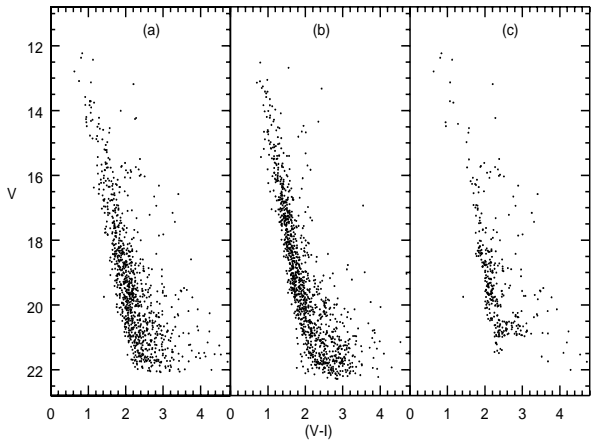}
\caption{\label {band} $V/(V-I_c)$ CMD for (a) stars in the NGC 7538 region and (b) stars in the reference region. (c) is a statistically cleaned CMD.  }
\end{figure}

\begin{figure}
\centering\includegraphics[height=5cm,width=7cm,angle=0]{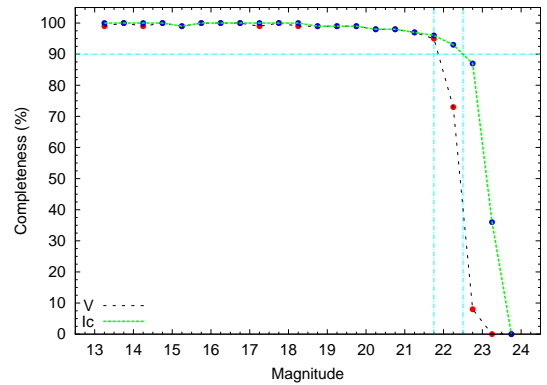}
\caption{\label{cft} Completeness levels for $V$ and $I_c$ (off-setted by 2 mag) bands as a function of magnitude
derived from the artificial star experiments ({\it ADDSTAR}, see Section 4.3).}
\end{figure}

\begin{figure}
\centering
\includegraphics[height=8cm,width=8cm]{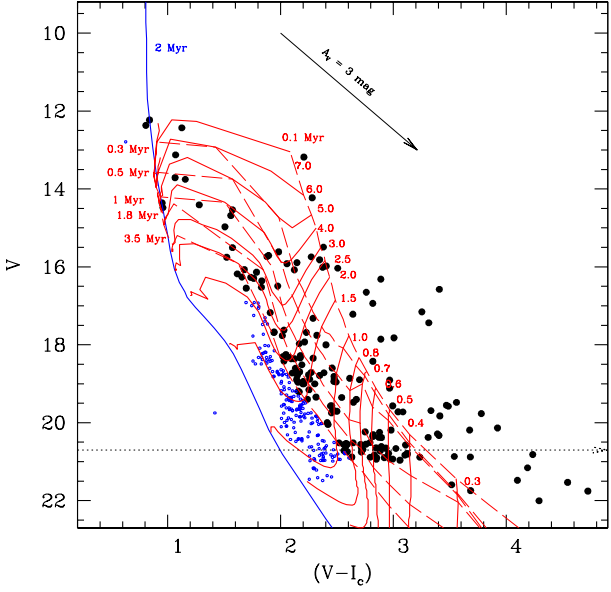}
\caption{\label{cleaned} Statistically cleaned $V/(V-I_c)$ CMD  for stars lying in the NGC 7538 region.
Filled circles (Ages $\leq$ 3.5 Myr) are used to estimate the MF of the region.
The isochrone of 2 Myr by \citet{2008AA...482..883M} and the PMS isochrones
of 0.1,0.3,0.5,1,1.8,3.5 Myr along with the evolutionary tracks for different masses by \citet{2000AA...358..593S} are also shown. 
All the curves are corrected for the  distance of 2.65 kpc and the foreground extinction $A_V$= 2.1 mag.
The dashed horizontal line represents the completeness limit of the data after taking into account the average extinction of the YSOs.}
\end{figure}

As discussed earlier the sample of YSOs used is compiled from various surveys having different 
completeness limits. So we have tried to use only our deep and homogeneous optical data to generate the MF of
the NGC 7538 region \citep[cf.][]{2007MNRAS.380.1141S,2008MNRAS.383.1241P,2011MNRAS.415.1202C,2013ApJ...764..172P,2013MNRAS.432.3445J}.
For this, we have utilized the optical $V$ versus $(V-I_c)$ CMD of all the 
sources in the NGC 7538 FOV and that of the nearby field region of equal area
and decontaminated the former sources from foreground/background
stars using a statistical subtraction method.
In  Fig. \ref{band}, we have shown the $V/(V-I_c)$ CMDs for the stars lying within the NGC 7538  FOV  (left panel) and for those in the reference field region (middle panel).
To statistically subtract the latter from the former, the both CMDs were divided
into grids of $\Delta V=1$ mag by $\Delta (V-I_c) = 0.4$  mag. The number of stars in each grid of the both CMDs were then counted
and the probable number of cluster members in each grid were estimated from the difference.
The estimated numbers of contaminating field stars
(the numbers in the bin - the probable numbers of cluster members) were removed from
the cluster CMD one by one that is the nearest to the randomly selected star in the CMD of the reference region of that bin.
The both CMDs were also corrected for the  incompleteness of the data.
The photometric data may be incomplete due to various reasons, e.g., nebulosity, crowding of the stars, 
detection limit etc. In particular it is very important to know the completeness limits  in terms of mass.        
The $IRAF$ routine $ADDSTAR$ of $DAOPHOT II$ was used to determine the 
completeness factor (CF) \citep[for detail, see][]{2008AJ....135.1934S}.
Briefly, in this method artificial stars of known magnitudes and positions 
are randomly added in the original frames and then these artificially generated frames are
re-reduced by the same procedure as used in the original reduction. The ratio of the
number of stars recovered to those added in each magnitude gives the CF as a function of magnitude.
To determine the completeness  of the $V$ versus $(V-I_c)$ CMD, 
we followed the procedure given by \citet{1991A&A...250..324S} by adding artificial 
stars to both $V$ and $I$ images in such a way that they have similar geometrical 
locations but differ in $I$ brightness according to the mean $(V-I_c)$ colours of the 
MS stars. Since the mean $(V-I_c)$ colour of the  MS stars is $\sim$2 mag in the NGC 7538 region 
(cf. Fig. \ref{band}), the $I$ band magnitude is offset-ed to the $V$ band  magnitude by adding a correction 
of $\sim$2 mag in Fig.~\ref{cft} showing the CF as a function of magnitudes. As expected
the CF decreases with fainter magnitudes. Our photometry is more than 90\% complete up to V$\simeq$21.5 mag,
which corresponds to the detection limit of a 0.8 M$_\odot$ (cf. Fig. \ref{cleaned}) PMS star of 
$\simeq$1.8 Myr age embedded in the nebulosity of $A_V\simeq$3.0 mag 
(i.e., the average values for the optically detected YSOs, cf. Table \ref{data4_yso}).

In Fig. \ref{cleaned}, we have plotted the statistically cleaned $V/(V-I_c)$ CMD for the NGC 7538 region showing
 the presence of PMS stars in the region.
We have also plotted the ZAMS by \citet{2000AA...358..593S} and  the PMS isochrones by \citet{2000AA...358..593S}.  
The evolutionary tracks by \citet{2000AA...358..593S} for various masses have also been plotted.
The dashed horizontal line represents the completeness limit of the data at $(V-I_c)$ = 2 mag
 after taking into account the average extinction of the YSOs corrected for the distance.

\begin{figure}
\includegraphics[height=6cm,width=8cm,angle=0]{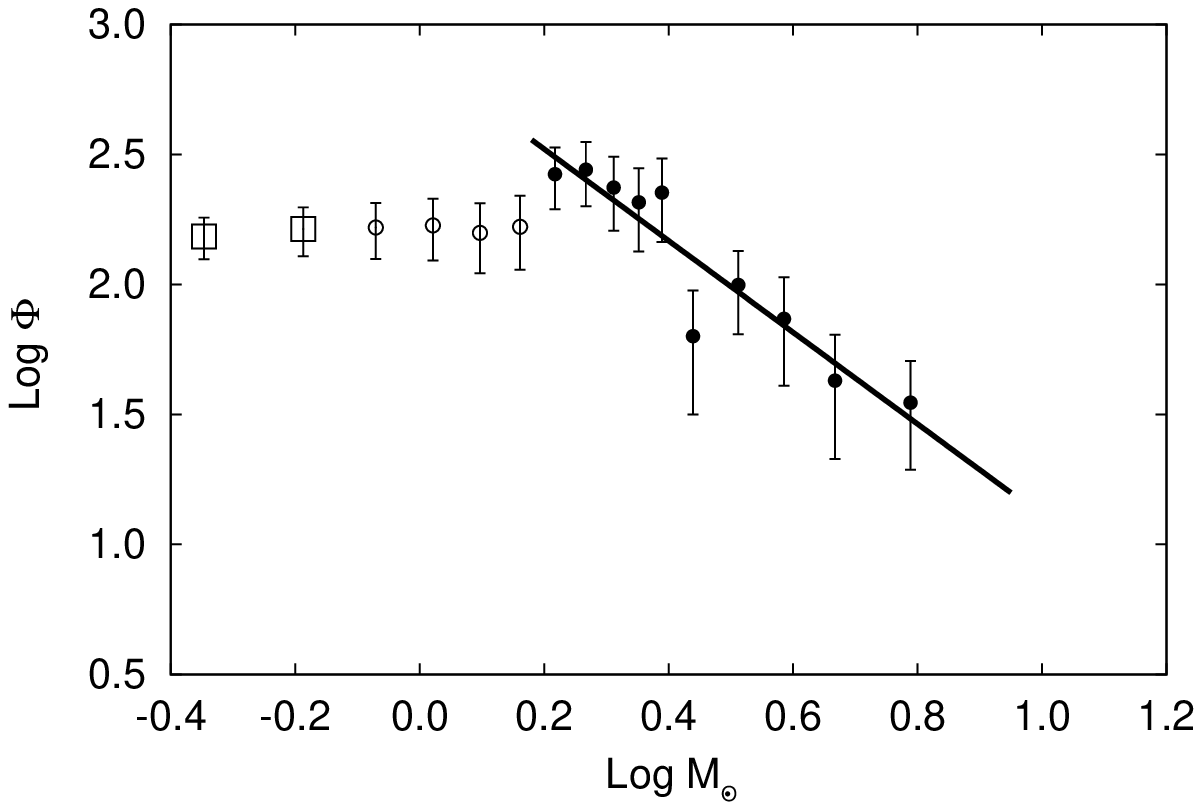}
\includegraphics[height=6cm,width=8cm,angle=0]{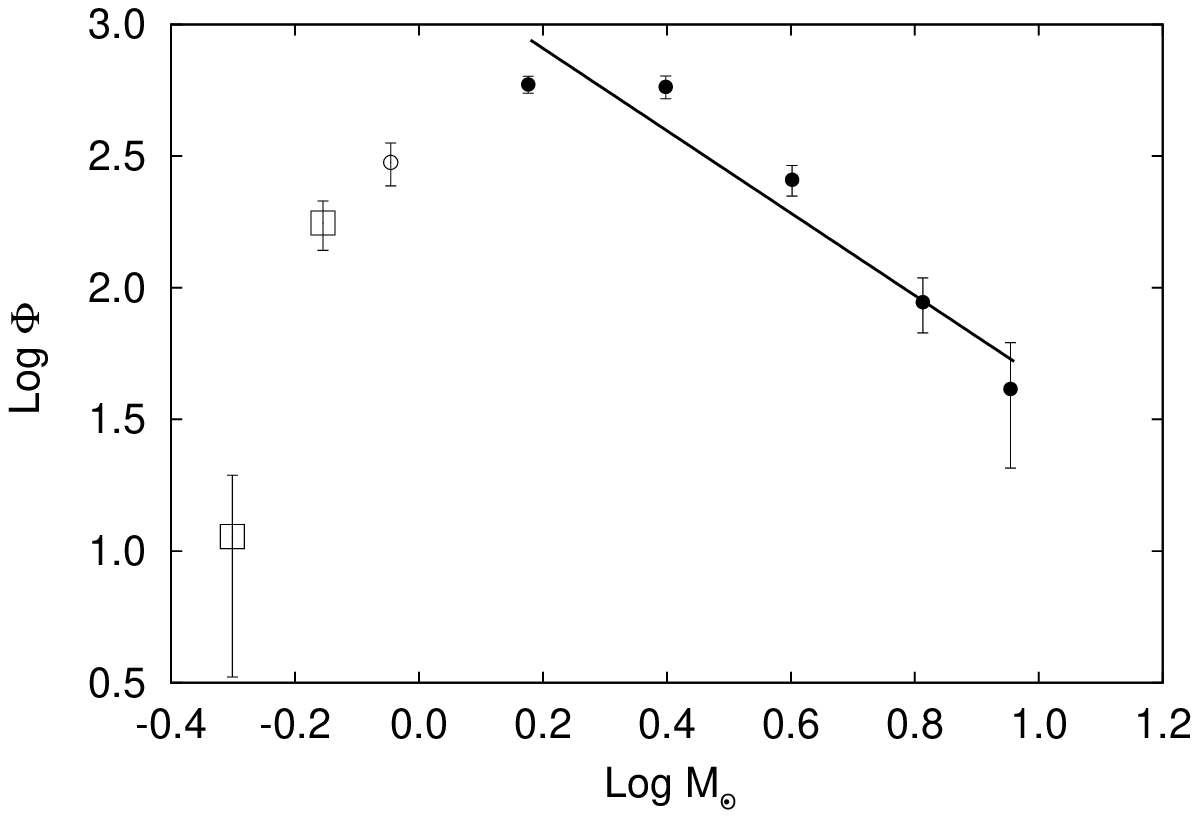}
\caption{\label{mf} A plot of the MF for the statistically cleaned CMD (top panel) and
the identified YSOs (bottom panel) in the NGC 7538 region.
Log $\phi$ represents log($N$/dlog $m$). The error bars represent $\pm\sqrt N$ errors. The solid line shows a least
squares fit to the MF distribution (Black dots).
Open squares are the data points falling below the completeness limit of 0.8 M$_\odot$.
Open circles are the data points near turn-off point in the MF distribution and are not used in the fitting.
}
\end{figure}

The masses of individual stars were then estimated by the same technique mentioned in 
Section 3.2.2, and the corresponding MF has been plotted in Fig. \ref{mf} (upper panel). 
For this, we have used only those sources which have ages equivalent to the average age 
of the optically identified YSOs combined with error (i.e., $\leq$3.5 Myr, cf. Table \ref{data4_yso}).
There is a change of slope from the high mass to low mass end  with a turn-of
at around 1.5 M$_\odot$, as has often been noticed in other regions
\citep{2007MNRAS.380.1141S, 2008MNRAS.383.1241P, 2008MNRAS.384.1675J}.
The slope of the MF $\Gamma$ for this sample in the mass range $\sim1.5<$M/M$_\odot < 6$ comes out to be $-1.76\pm0.24$,
which is steeper than the value -1.35 given by \citet{1955ApJ...121..161S}.
We have optical photometry of only two IR sources associated with the NGC 7538 region,
i.e., IRS 5 (O9, 20 M$_\odot$) and IRS 6 (O3, 60 M$_\odot$) 
as others could not be resolved or are foreground sources. 
For the generation of the MF, we have not used these two stars as they will introduce 
large gaps between the points in the MF distribution and corresponding  errors will be quite large.
Also, a lower mass range is required to compare this MF distribution to that of YSOs.

Since the detection limit for the optical sample is upto $A_V\simeq$7.7 mag (cf. Appendix A), the optical sample
will represent YSOs which are on the surface or partially embedded in the nebulosity of the NGC 7538 SFR.
We need NIR/MIR data to cover the deeply embedded YSOs. In Section 3.2.1, with the help of the SED fitting, 
we have estimated the age and mass  of 419 YSOs by using mostly NIR/MIR data. Therefore, we have used this sample
to trace further deep in the NGC 7538 region. 
The completeness limit for this sample, where the data of individual YSOs are taken from various surveys,
is discussed in section 3.1.5 and found to be 0.8 M$_\odot$.
The masses of the YSOs in this sample were used to generate the MF and is shown in the lower-panel of Fig. \ref{mf}
\citep[see also,][]{2014A&A...567A.109K,2016MNRAS.461.2502Y}.
We have not used here the four high mass YSOs (cf. Section 4.1).
The slope of the MF $\Gamma$ down to a similar mass limit (i.e. 1.5 M$_\odot$) for this sample of YSOs
comes out to be $-1.56\pm0.27$, which is similar within the error with that derived by using optical CMD. 
From this, we can infer that the optical and SED fitting samples represent the same population of YSOs.

\section{Conclusion}

Although the NGC 7538 region has already been studied extensively in IR and radio wavelengths,
it is rather neglected in the optical and X-ray. 
To the best of our knowledge, the present study is the first detailed multiwavelength 
study  (including the optical) of this region. 
We have added some more YSOs based on their H$\alpha$ or X-ray emission, thus complemented
the previous IR excess based  catalog of the YSOs in the region. 
We have checked the association of these YSOs with NGC 7538 and derived their individual physical 
parameters based on the SED fitting.
The spatial distribution of these YSOs along with those of the MIR and radio emission have been used to constrain the star-formation history  in the region.
The XLF and MF for the YSOs in this region have also been derived.
The main results of this study can be summarized as follows:

\begin{itemize}

\item
Analysis of the $Chandra$ X-ray data  in a $\approx 17\times17$ arcmin$^2$ field of the NGC 7538 region reveals 190 X-ray emitting sources.

\item
On the basis of H$\alpha$ grism spectroscopy,  
H$\alpha$ (photometry) and X-ray emission, we have identified 6, 15 and 64 YSOs, respectively, in the region.
We have compiled a catalog of 943 YSOs by combining those detected in the present study
with those previously cataloged. 53 YSOs are new additions from this study.
This new catalog is complete down to 0.8 M$_\odot$.

\item
We have derived the age/mass of 463 YSOs based on the SED fitting analysis.
419 of them are confirmed to belong to the NGC 7538 SFR and
$\sim$91\% (380/419) of them have ages between 0.1 to 2.5 Myr.
A majority ($\sim$86\%) of the 419 YSOs have masses between 0.5 to 3.5 M$_\odot$
as derived by SED fitting analysis.
These ages and masses are comparable with those of TTSs.
The $A_V$ value for these YSOs shows a spread from 1 to 30 mag.
The average age, mass and extinction ($A_V$) for this sample of YSOs are 1.4 Myr, 2.3 M$_\odot$ and 11 mag, respectively.

\item
Of the above 419 YSOs, around 24\% (99), 62\% (258) and 2\% (10) are found to be Class I, Class II and Class III sources, respectively.
The high percentage ($\sim$85\%) of Class I and Class II YSOs indicates the youth of this region.
A KS test for the age distribution of the Class I and Class II sources
suggests that it is  different.
The ages of Class I sources are $\lesssim$0.5 Myr, whereas the Class II sources have ages 
$\sim$1-1.5 Myr mostly, indicating an age difference of $\sim$1 Myr between them.

\item
The mean value of $\rm{log(L_X)}$ of identified YSOs in the region is found to be around 31.1 $\rm{erg~s^{-1}}$.
It is found that the X-ray activity of the Class I, Class II and Class III sources is not significantly different from each other.
The $L_X$ seems to increase with the mass of the Class III sources, whereas it decreases with the age of the Class II and Class III sources. 
However, we find no statistically significant difference in the slopes and intercepts for the
$L_X$ versus age distribution for the Class II and Class III sources, which indicates that the presence
of circumstellar disks has practically no influence on the X-ray emission.
This suggests that the  increase of the X-ray surface activity with the increase of the 
rotation rate may be compensated by the decrease of the stellar surface 
area during the PMS evolution.

\item
The spatial distribution of the YSOs along with those of MIR/radio/CO emission has been used to understand the 
star formation process in the region.
The YSOs in the inner region (within 3 pc from IRS 6, containing the bright H\,{\sevensize II}  region) may have been formed by a triggering mechanism caused by the central high mass star IRS 6.

\item
The MF changes its slopes from high mass to low mass sides with a turn-off at around 1.5 M$_\odot$.
The slope of the MF $\Gamma$ in the mass range $1.5 <$M/M$_\odot < 6$ comes out to be $-1.76\pm0.24$, which is steeper than
the \citet{1955ApJ...121..161S} value -1.35.

\end{itemize}

\section*{Acknowledgments}

Authors are thankful to the anonymous referee for his/her useful comments.
HB acknowledges the Inspire Faculty Grant Support (IFA11-PH02) by Department of Science \& Technology, India.
The observations reported in this paper were obtained by using the 1 m Sampurnanand telescope, Nainital, India and  the 2 m HCT at IAO, Hanle, the High Altitude Station of Indian Institute of Astrophysics, Bangalore, India. This publication makes use of data from the Two Micron All Sky Survey, which is a joint project of the University of Massachusetts and the Infrared Processing and Analysis Center/California Institute of Technology, funded by the National Aeronautics and Space Administration and the National Science Foundation.

\bibliography{7538}{}

\begin{thebibliography}{}
\makeatletter
\relax
\def\mn@urlcharsother{\let\do\@makeother \do\$\do\&\do\#\do\^\do\_\do\%\do\~}
\def\mn@doi{\begingroup\mn@urlcharsother \@ifnextchar [ {\mn@doi@}
  {\mn@doi@[]}}
\def\mn@doi@[#1]#2{\def\@tempa{#1}\ifx\@tempa\@empty \href
  {http://dx.doi.org/#2} {doi:#2}\else \href {http://dx.doi.org/#2} {#1}\fi
  \endgroup}
\def\mn@eprint#1#2{\mn@eprint@#1:#2::\@nil}
\def\mn@eprint@arXiv#1{\href {http://arxiv.org/abs/#1} {{\tt arXiv:#1}}}
\def\mn@eprint@dblp#1{\href {http://dblp.uni-trier.de/rec/bibtex/#1.xml}
  {dblp:#1}}
\def\mn@eprint@#1:#2:#3:#4\@nil{\def\@tempa {#1}\def\@tempb {#2}\def\@tempc
  {#3}\ifx \@tempc \@empty \let \@tempc \@tempb \let \@tempb \@tempa \fi \ifx
  \@tempb \@empty \def\@tempb {arXiv}\fi \@ifundefined
  {mn@eprint@\@tempb}{\@tempb:\@tempc}{\expandafter \expandafter \csname
  mn@eprint@\@tempb\endcsname \expandafter{\@tempc}}}

\bibitem[\protect\citeauthoryear{{Balog}, {Kenyon}, {Lada}, {Barsony},
  {Vink{\'o}}  \& {G{\'a}spa{\'r}}}{{Balog} et~al.}{2004}]{2004AJ....128.2942B}
{Balog} Z.,  {Kenyon} S.~J.,  {Lada} E.~A.,  {Barsony} M.,  {Vink{\'o}} J.,
  {G{\'a}spa{\'r}} A.,  2004, \mn@doi [\aj] {10.1086/425548}, \href
  {http://adsabs.harvard.edu/abs/2004AJ....128.2942B} {128, 2942}

\bibitem[\protect\citeauthoryear{{Balucinska-Church} \&
  {McCammon}}{{Balucinska-Church} \& {McCammon}}{1992}]{bal+92}
{Balucinska-Church} M.,  {McCammon} D.,  1992, \mn@doi [\apj] {10.1086/172032},
  \href {http://adsabs.harvard.edu/abs/1992ApJ...400..699B} {400, 699}

\bibitem[\protect\citeauthoryear{{Bertoldi}}{{Bertoldi}}{1989}]{1989ApJ...346..735B}
{Bertoldi} F.,  1989, \mn@doi [\apj] {10.1086/168055}, \href
  {http://adsabs.harvard.edu/abs/1989ApJ...346..735B} {346, 735}

\bibitem[\protect\citeauthoryear{{Bessell} \& {Brett}}{{Bessell} \&
  {Brett}}{1988}]{1988PASP..100.1134B}
{Bessell} M.~S.,  {Brett} J.~M.,  1988, \mn@doi [\pasp] {10.1086/132281}, \href
  {http://adsabs.harvard.edu/abs/1988PASP..100.1134B} {100, 1134}

\bibitem[\protect\citeauthoryear{{Bhatt}, {Pandey}, {Singh}, {Sagar}  \&
  {Kumar}}{{Bhatt} et~al.}{2013}]{bha+13}
{Bhatt} H.,  {Pandey} J.~C.,  {Singh} K.~P.,  {Sagar} R.,   {Kumar} B.,  2013,
  \mn@doi [Journal of Astrophysics and Astronomy] {10.1007/s12036-013-9190-8},
  \href {http://adsabs.harvard.edu/abs/2013JApA...34..393B} {34, 393}

\bibitem[\protect\citeauthoryear{{Bica}, {Dutra}  \& {Barbuy}}{{Bica}
  et~al.}{2003}]{2003A&A...397..177B}
{Bica} E.,  {Dutra} C.~M.,   {Barbuy} B.,  2003, \mn@doi [\aap]
  {10.1051/0004-6361:20021479}, \href
  {http://adsabs.harvard.edu/abs/2003A%26A...397..177B} {397, 177}

\bibitem[\protect\citeauthoryear{{Bohlin}, {Savage}  \& {Drake}}{{Bohlin}
  et~al.}{1978}]{1978ApJ...224..132B}
{Bohlin} R.~C.,  {Savage} B.~D.,   {Drake} J.~F.,  1978, \mn@doi [\apj]
  {10.1086/156357}, \href {http://adsabs.harvard.edu/abs/1978ApJ...224..132B}
  {224, 132}

\bibitem[\protect\citeauthoryear{{Brice{\~n}o}, {Preibisch}, {Sherry},
  {Mamajek}, {Mathieu}, {Walter}  \& {Zinnecker}}{{Brice{\~n}o}
  et~al.}{2007}]{2007prpl.conf..345B}
{Brice{\~n}o} C.,  {Preibisch} T.,  {Sherry} W.~H.,  {Mamajek} E.~A.,
  {Mathieu} R.~D.,  {Walter} F.~M.,   {Zinnecker} H.,  2007, Protostars and
  Planets V, \href {http://adsabs.harvard.edu/abs/2007prpl.conf..345B} {pp
  345--360}

\bibitem[\protect\citeauthoryear{Broos, Townsley, Feigelson, Getman, Bauer  \&
  Garmire}{Broos et~al.}{2010}]{0004-637X-714-2-1582}
Broos P.~S.,  Townsley L.~K.,  Feigelson E.~D.,  Getman K.~V.,  Bauer F.~E.,
  Garmire G.~P.,  2010, The Astrophysical Journal, 714, 1582

\bibitem[\protect\citeauthoryear{{Caramazza} et~al.,}{{Caramazza}
  et~al.}{2012}]{2012A&A...539A..74C}
{Caramazza} M.,  et~al., 2012, \mn@doi [\aap] {10.1051/0004-6361/201117256},
  \href {http://adsabs.harvard.edu/abs/2012A\%26A...539A..74C} {539, A74}

\bibitem[\protect\citeauthoryear{{Cardelli}, {Clayton}  \& {Mathis}}{{Cardelli}
  et~al.}{1989}]{1989ApJ...345..245C}
{Cardelli} J.~A.,  {Clayton} G.~C.,   {Mathis} J.~S.,  1989, \mn@doi [\apj]
  {10.1086/167900}, \href {http://adsabs.harvard.edu/abs/1989ApJ...345..245C}
  {345, 245}

\bibitem[\protect\citeauthoryear{{Carpenter}, {Snell}, {Schloerb}  \&
  {Skrutskie}}{{Carpenter} et~al.}{1993}]{1993ApJ...407..657C}
{Carpenter} J.~M.,  {Snell} R.~L.,  {Schloerb} F.~P.,   {Skrutskie} M.~F.,
  1993, \mn@doi [\apj] {10.1086/172548}, \href
  {http://adsabs.harvard.edu/abs/1993ApJ...407..657C} {407, 657}

\bibitem[\protect\citeauthoryear{{Casanova}, {Montmerle}, {Feigelson}  \&
  {Andre}}{{Casanova} et~al.}{1995}]{cas+95}
{Casanova} S.,  {Montmerle} T.,  {Feigelson} E.~D.,   {Andre} P.,  1995,
  \mn@doi [\apj] {10.1086/175214}, \href
  {http://adsabs.harvard.edu/abs/1995ApJ...439..752C} {439, 752}

\bibitem[\protect\citeauthoryear{{Chabrier}}{{Chabrier}}{2003}]{2003PASP..115..763C}
{Chabrier} G.,  2003, \mn@doi [\pasp] {10.1086/376392}, \href
  {http://adsabs.harvard.edu/abs/2003PASP..115..763C} {115, 763}

\bibitem[\protect\citeauthoryear{{Chauhan}, {Pandey}, {Ogura}, {Ojha}, {Bhatt},
  {Ghosh}  \& {Rawat}}{{Chauhan} et~al.}{2009}]{2009MNRAS.396..964C}
{Chauhan} N.,  {Pandey} A.~K.,  {Ogura} K.,  {Ojha} D.~K.,  {Bhatt} B.~C.,
  {Ghosh} S.~K.,   {Rawat} P.~S.,  2009, \mn@doi [\mnras]
  {10.1111/j.1365-2966.2009.14756.x}, \href
  {http://adsabs.harvard.edu/abs/2009MNRAS.396..964C} {396, 964}

\bibitem[\protect\citeauthoryear{{Chauhan}, {Pandey}, {Ogura}, {Jose}, {Ojha},
  {Samal}  \& {Mito}}{{Chauhan} et~al.}{2011}]{2011MNRAS.415.1202C}
{Chauhan} N.,  {Pandey} A.~K.,  {Ogura} K.,  {Jose} J.,  {Ojha} D.~K.,  {Samal}
  M.~R.,   {Mito} H.,  2011, \mn@doi [\mnras]
  {10.1111/j.1365-2966.2011.18742.x}, \href
  {http://adsabs.harvard.edu/abs/2011MNRAS.415.1202C} {415, 1202}

\bibitem[\protect\citeauthoryear{{Chavarr{\'{\i}}a}, {Allen}, {Brunt}, {Hora},
  {Muench}  \& {Fazio}}{{Chavarr{\'{\i}}a} et~al.}{2014}]{2014MNRAS.439.3719C}
{Chavarr{\'{\i}}a} L.,  {Allen} L.,  {Brunt} C.,  {Hora} J.~L.,  {Muench} A.,
  {Fazio} G.,  2014, \mn@doi [\mnras] {10.1093/mnras/stu224}, \href
  {http://adsabs.harvard.edu/abs/2014MNRAS.439.3719C} {439, 3719}

\bibitem[\protect\citeauthoryear{{Chini}, {Kruegel}  \& {Kreysa}}{{Chini}
  et~al.}{1990}]{1990A&A...227L...5C}
{Chini} R.,  {Kruegel} E.,   {Kreysa} E.,  1990, \aap, \href
  {http://adsabs.harvard.edu/abs/1990A\%26A...227L...5C} {227, L5}

\bibitem[\protect\citeauthoryear{{Cohen}, {Persson}, {Elias}  \&
  {Frogel}}{{Cohen} et~al.}{1981}]{1981ApJ...249..481C}
{Cohen} J.~G.,  {Persson} S.~E.,  {Elias} J.~H.,   {Frogel} J.~A.,  1981,
  \mn@doi [\apj] {10.1086/159308}, \href
  {http://adsabs.harvard.edu/abs/1981ApJ...249..481C} {249, 481}

\bibitem[\protect\citeauthoryear{{Corbelli}, {Palla}  \&
  {Zinnecker}}{{Corbelli} et~al.}{2005}]{2005ASSL..327.....C}
{Corbelli} E.,  {Palla} F.,   {Zinnecker} H.,  eds, 2005, {The Initial Mass
  Function 50 years later}  Astrophysics and Space Science Library Vol. 327

\bibitem[\protect\citeauthoryear{{Crampton}, {Georgelin}  \&
  {Georgelin}}{{Crampton} et~al.}{1978}]{1978A&A....66....1C}
{Crampton} D.,  {Georgelin} Y.~M.,   {Georgelin} Y.~P.,  1978, \aap, \href
  {http://adsabs.harvard.edu/abs/1978A\%26A....66....1C} {66, 1}

\bibitem[\protect\citeauthoryear{{Currie}, {Evans}, {Spitzbart}, {Irwin},
  {Wolk}, {Hernandez}, {Kenyon}  \& {Pasachoff}}{{Currie}
  et~al.}{2009}]{cur+09}
{Currie} T.,  {Evans} N.~R.,  {Spitzbart} B.~D.,  {Irwin} J.,  {Wolk} S.~J.,
  {Hernandez} J.,  {Kenyon} S.~J.,   {Pasachoff} J.~M.,  2009, \mn@doi [\aj]
  {10.1088/0004-6256/137/2/3210}, \href
  {http://adsabs.harvard.edu/abs/2009AJ....137.3210C} {137, 3210}

\bibitem[\protect\citeauthoryear{{Cutri} et~al.,}{{Cutri}
  et~al.}{2003}]{2003yCat.2246....0C}
{Cutri} R.~M.,  et~al., 2003, VizieR Online Data Catalog, \href
  {http://adsabs.harvard.edu/abs/2003yCat.2246....0C} {2246, 0}

\bibitem[\protect\citeauthoryear{{Dahm}, {Simon}, {Proszkow}  \&
  {Patten}}{{Dahm} et~al.}{2007}]{2007AJ....134..999D}
{Dahm} S.~E.,  {Simon} T.,  {Proszkow} E.~M.,   {Patten} B.~M.,  2007, \mn@doi
  [\aj] {10.1086/519954}, \href
  {http://adsabs.harvard.edu/abs/2007AJ....134..999D} {134, 999}

\bibitem[\protect\citeauthoryear{{Dale}, {Haworth}  \& {Bressert}}{{Dale}
  et~al.}{2015}]{2015MNRAS.450.1199D}
{Dale} J.~E.,  {Haworth} T.~J.,   {Bressert} E.,  2015, \mn@doi [\mnras]
  {10.1093/mnras/stv396}, \href
  {http://adsabs.harvard.edu/abs/2015MNRAS.450.1199D} {450, 1199}

\bibitem[\protect\citeauthoryear{{Damiani}, {Maggio}, {Micela}  \&
  {Sciortino}}{{Damiani} et~al.}{1997}]{1997ApJ...483..350D}
{Damiani} F.,  {Maggio} A.,  {Micela} G.,   {Sciortino} S.,  1997, \mn@doi
  [\apj] {10.1086/304217}, \href
  {http://adsabs.harvard.edu/abs/1997ApJ...483..350D} {483, 350}

\bibitem[\protect\citeauthoryear{{Deharveng}, {Zavagno}  \&
  {Caplan}}{{Deharveng} et~al.}{2005}]{2005A&A...433..565D}
{Deharveng} L.,  {Zavagno} A.,   {Caplan} J.,  2005, \mn@doi [\aap]
  {10.1051/0004-6361:20041946}, \href
  {http://adsabs.harvard.edu/abs/2005A\%26A...433..565D} {433, 565}

\bibitem[\protect\citeauthoryear{{Deharveng}, {Zavagno}, {Schuller}, {Caplan},
  {Pomar{\`e}s}  \& {De Breuck}}{{Deharveng}
  et~al.}{2009}]{2009A&A...496..177D}
{Deharveng} L.,  {Zavagno} A.,  {Schuller} F.,  {Caplan} J.,  {Pomar{\`e}s} M.,
    {De Breuck} C.,  2009, \mn@doi [\aap] {10.1051/0004-6361/200811337}, \href
  {http://adsabs.harvard.edu/abs/2009A\%26A...496..177D} {496, 177}

\bibitem[\protect\citeauthoryear{{Deharveng} et~al.,}{{Deharveng}
  et~al.}{2010}]{2010A&A...523A...6D}
{Deharveng} L.,  et~al., 2010, \mn@doi [\aap] {10.1051/0004-6361/201014422},
  \href {http://adsabs.harvard.edu/abs/2010A%26A...523A...6D} {523, A6}

\bibitem[\protect\citeauthoryear{{Dutta}, {Mondal}, {Jose}, {Das}, {Samal}  \&
  {Ghosh}}{{Dutta} et~al.}{2015}]{2015MNRAS.454.3597D}
{Dutta} S.,  {Mondal} S.,  {Jose} J.,  {Das} R.~K.,  {Samal} M.~R.,   {Ghosh}
  S.,  2015, \mn@doi [\mnras] {10.1093/mnras/stv2190}, \href
  {http://adsabs.harvard.edu/abs/2015MNRAS.454.3597D} {454, 3597}

\bibitem[\protect\citeauthoryear{{Elmegreen}}{{Elmegreen}}{1998}]{1998ASPC..148..150E}
{Elmegreen} B.~G.,  1998, in {Woodward} C.~E.,  {Shull} J.~M.,   {Thronson} Jr.
  H.~A.,  eds,  Astronomical Society of the Pacific Conference Series Vol. 148,
  Origins. p.~150 (\mn@eprint {} {astro-ph/9712352})

\bibitem[\protect\citeauthoryear{{Elmegreen} \& {Lada}}{{Elmegreen} \&
  {Lada}}{1977}]{1977ApJ...214..725E}
{Elmegreen} B.~G.,  {Lada} C.~J.,  1977, \mn@doi [\apj] {10.1086/155302}, \href
  {http://adsabs.harvard.edu/abs/1977ApJ...214..725E} {214, 725}

\bibitem[\protect\citeauthoryear{{Espinoza}, {Selman}  \& {Melnick}}{{Espinoza}
  et~al.}{2009}]{2009A&A...501..563E}
{Espinoza} P.,  {Selman} F.~J.,   {Melnick} J.,  2009, \mn@doi [\aap]
  {10.1051/0004-6361/20078597}, \href
  {http://adsabs.harvard.edu/abs/2009A\%26A...501..563E} {501, 563}

\bibitem[\protect\citeauthoryear{{Eswaraiah}, {Pandey}, {Maheswar}, {Chen},
  {Ojha}  \& {Chandola}}{{Eswaraiah} et~al.}{2012}]{2012MNRAS.419.2587E}
{Eswaraiah} C.,  {Pandey} A.~K.,  {Maheswar} G.,  {Chen} W.~P.,  {Ojha} D.~K.,
   {Chandola} H.~C.,  2012, \mn@doi [\mnras]
  {10.1111/j.1365-2966.2011.19908.x}, \href
  {http://adsabs.harvard.edu/abs/2012MNRAS.419.2587E} {419, 2587}

\bibitem[\protect\citeauthoryear{{Evans} II et~al.,}{{Evans}
  et~al.}{2009}]{2009ApJS..181..321E}
{Evans} II N.~J.,  et~al., 2009, \mn@doi [\apjs] {10.1088/0067-0049/181/2/321},
  \href {http://adsabs.harvard.edu/abs/2009ApJS..181..321E} {181, 321}

\bibitem[\protect\citeauthoryear{{Fallscheer} et~al.,}{{Fallscheer}
  et~al.}{2013}]{2013ApJ...773..102F}
{Fallscheer} C.,  et~al., 2013, \mn@doi [\apj] {10.1088/0004-637X/773/2/102},
  \href {http://adsabs.harvard.edu/abs/2013ApJ...773..102F} {773, 102}

\bibitem[\protect\citeauthoryear{{Fang}, {van Boekel}, {Wang}, {Carmona},
  {Sicilia-Aguilar}  \& {Henning}}{{Fang} et~al.}{2009}]{2009A&A...504..461F}
{Fang} M.,  {van Boekel} R.,  {Wang} W.,  {Carmona} A.,  {Sicilia-Aguilar} A.,
   {Henning} T.,  2009, \mn@doi [\aap] {10.1051/0004-6361/200912468}, \href
  {http://adsabs.harvard.edu/abs/2009A\%26A...504..461F} {504, 461}

\bibitem[\protect\citeauthoryear{{Favata} \& {Micela}}{{Favata} \&
  {Micela}}{2003}]{2003SSRv..108..577F}
{Favata} F.,  {Micela} G.,  2003, \mn@doi [\ssr]
  {10.1023/B:SPAC.0000007491.80144.21}, \href
  {http://adsabs.harvard.edu/abs/2003SSRv..108..577F} {108, 577}

\bibitem[\protect\citeauthoryear{{Feigelson} \& {Montmerle}}{{Feigelson} \&
  {Montmerle}}{1999a}]{1999ARAA..37..363F}
{Feigelson} E.~D.,  {Montmerle} T.,  1999a, \mn@doi [\araa]
  {10.1146/annurev.astro.37.1.363}, \href
  {http://adsabs.harvard.edu/abs/1999ARA\%26A..37..363F} {37, 363}

\bibitem[\protect\citeauthoryear{{Feigelson} \& {Montmerle}}{{Feigelson} \&
  {Montmerle}}{1999b}]{1999ARA&A..37..363F}
{Feigelson} E.~D.,  {Montmerle} T.,  1999b, \mn@doi [\araa]
  {10.1146/annurev.astro.37.1.363}, \href
  {http://adsabs.harvard.edu/abs/1999ARA\%26A..37..363F} {37, 363}

\bibitem[\protect\citeauthoryear{{Feigelson}, {Casanova}, {Montmerle}  \&
  {Guibert}}{{Feigelson} et~al.}{1993}]{fei+93}
{Feigelson} E.~D.,  {Casanova} S.,  {Montmerle} T.,   {Guibert} J.,  1993,
  \mn@doi [\apj] {10.1086/173264}, \href
  {http://adsabs.harvard.edu/abs/1993ApJ...416..623F} {416, 623}

\bibitem[\protect\citeauthoryear{{Feigelson}, {Broos}, {Gaffney}, {Garmire},
  {Hillenbrand}, {Pravdo}, {Townsley}  \& {Tsuboi}}{{Feigelson}
  et~al.}{2002a}]{2002ApJ...574..258F}
{Feigelson} E.~D.,  {Broos} P.,  {Gaffney} III J.~A.,  {Garmire} G.,
  {Hillenbrand} L.~A.,  {Pravdo} S.~H.,  {Townsley} L.,   {Tsuboi} Y.,  2002a,
  \mn@doi [\apj] {10.1086/340936}, \href
  {http://adsabs.harvard.edu/abs/2002ApJ...574..258F} {574, 258}

\bibitem[\protect\citeauthoryear{{Feigelson}, {Broos}, {Gaffney}, {Garmire},
  {Hillenbrand}, {Pravdo}, {Townsley}  \& {Tsuboi}}{{Feigelson}
  et~al.}{2002b}]{fei+02}
{Feigelson} E.~D.,  {Broos} P.,  {Gaffney} III J.~A.,  {Garmire} G.,
  {Hillenbrand} L.~A.,  {Pravdo} S.~H.,  {Townsley} L.,   {Tsuboi} Y.,  2002b,
  \mn@doi [\apj] {10.1086/340936}, \href
  {http://adsabs.harvard.edu/abs/2002ApJ...574..258F} {574, 258}

\bibitem[\protect\citeauthoryear{{Flaccomio}, {Damiani}, {Micela}, {Sciortino},
  {Harnden}, {Murray}  \& {Wolk}}{{Flaccomio}
  et~al.}{2003a}]{2003ApJ...582..382F}
{Flaccomio} E.,  {Damiani} F.,  {Micela} G.,  {Sciortino} S.,  {Harnden} Jr.
  F.~R.,  {Murray} S.~S.,   {Wolk} S.~J.,  2003a, \mn@doi [\apj]
  {10.1086/344535}, \href {http://adsabs.harvard.edu/abs/2003ApJ...582..382F}
  {582, 382}

\bibitem[\protect\citeauthoryear{{Flaccomio}, {Damiani}, {Micela}, {Sciortino},
  {Harnden}, {Murray}  \& {Wolk}}{{Flaccomio} et~al.}{2003b}]{fla+03}
{Flaccomio} E.,  {Damiani} F.,  {Micela} G.,  {Sciortino} S.,  {Harnden} Jr.
  F.~R.,  {Murray} S.~S.,   {Wolk} S.~J.,  2003b, \mn@doi [\apj]
  {10.1086/344536}, \href {http://adsabs.harvard.edu/abs/2003ApJ...582..398F}
  {582, 398}

\bibitem[\protect\citeauthoryear{{Flaccomio}, {Micela}  \&
  {Sciortino}}{{Flaccomio} et~al.}{2006}]{fla+06}
{Flaccomio} E.,  {Micela} G.,   {Sciortino} S.,  2006, \mn@doi [\aap]
  {10.1051/0004-6361:20065084}, \href
  {http://adsabs.harvard.edu/abs/2006A\%26A...455..903F} {455, 903}

\bibitem[\protect\citeauthoryear{{Frieswijk}, {Spaans}, {Shipman}, {Teyssier},
  {Carey}  \& {Tielens}}{{Frieswijk} et~al.}{2008}]{2008ApJ...685L..51F}
{Frieswijk} W.~F.,  {Spaans} M.,  {Shipman} R.~F.,  {Teyssier} D.,  {Carey}
  S.~J.,   {Tielens} A.~G.~G.~M.,  2008, \mn@doi [\apjl] {10.1086/592382},
  \href {http://adsabs.harvard.edu/abs/2008ApJ...685L..51F} {685, L51}

\bibitem[\protect\citeauthoryear{{Getman} et~al.,}{{Getman}
  et~al.}{2005}]{2005ApJS..160..319G}
{Getman} K.~V.,  et~al., 2005, \mn@doi [\apjs] {10.1086/432092}, \href
  {http://adsabs.harvard.edu/abs/2005ApJS..160..319G} {160, 319}

\bibitem[\protect\citeauthoryear{{Golay}}{{Golay}}{1974}]{1974ASSL...41.....G}
{Golay} M.,  ed. 1974, {Introduction to astronomical photometry}  Astrophysics
  and Space Science Library Vol. 41, \mn@doi{10.1007/978-94-010-2169-2.
}

\bibitem[\protect\citeauthoryear{{Guarcello}, {Caramazza}, {Micela},
  {Sciortino}, {Drake}  \& {Prisinzano}}{{Guarcello} et~al.}{2012a}]{gua+12}
{Guarcello} M.~G.,  {Caramazza} M.,  {Micela} G.,  {Sciortino} S.,  {Drake}
  J.~J.,   {Prisinzano} L.,  2012a, \mn@doi [\apj]
  {10.1088/0004-637X/753/2/117}, \href
  {http://adsabs.harvard.edu/abs/2012ApJ...753..117G} {753, 117}

\bibitem[\protect\citeauthoryear{{Guarcello}, {Caramazza}, {Micela},
  {Sciortino}, {Drake}  \& {Prisinzano}}{{Guarcello}
  et~al.}{2012b}]{2012ApJ...753..117G}
{Guarcello} M.~G.,  {Caramazza} M.,  {Micela} G.,  {Sciortino} S.,  {Drake}
  J.~J.,   {Prisinzano} L.,  2012b, \mn@doi [\apj]
  {10.1088/0004-637X/753/2/117}, \href
  {http://adsabs.harvard.edu/abs/2012ApJ...753..117G} {753, 117}

\bibitem[\protect\citeauthoryear{{G{\"u}del}}{{G{\"u}del}}{2004}]{2004A&ARv..12...71G}
{G{\"u}del} M.,  2004, \mn@doi [\aapr] {10.1007/s00159-004-0023-2}, \href
  {http://adsabs.harvard.edu/abs/2004A\%26ARv..12...71G} {12, 71}

\bibitem[\protect\citeauthoryear{{Guetter} \& {Vrba}}{{Guetter} \&
  {Vrba}}{1989}]{1989AJ.....98..611G}
{Guetter} H.~H.,  {Vrba} F.~J.,  1989, \mn@doi [\aj] {10.1086/115161}, \href
  {http://adsabs.harvard.edu/abs/1989AJ.....98..611G} {98, 611}

\bibitem[\protect\citeauthoryear{{Gutermuth}, {Megeath}, {Pipher}, {Williams},
  {Allen}, {Myers}  \& {Raines}}{{Gutermuth}
  et~al.}{2005}]{2005ApJ...632..397G}
{Gutermuth} R.~A.,  {Megeath} S.~T.,  {Pipher} J.~L.,  {Williams} J.~P.,
  {Allen} L.~E.,  {Myers} P.~C.,   {Raines} S.~N.,  2005, \mn@doi [\apj]
  {10.1086/432460}, \href {http://adsabs.harvard.edu/abs/2005ApJ...632..397G}
  {632, 397}

\bibitem[\protect\citeauthoryear{{Gutermuth}, {Megeath}, {Myers}, {Allen},
  {Pipher}  \& {Fazio}}{{Gutermuth} et~al.}{2009}]{2009ApJS..184...18G}
{Gutermuth} R.~A.,  {Megeath} S.~T.,  {Myers} P.~C.,  {Allen} L.~E.,  {Pipher}
  J.~L.,   {Fazio} G.~G.,  2009, \mn@doi [\apjs] {10.1088/0067-0049/184/1/18},
  \href {http://adsabs.harvard.edu/abs/2009ApJS..184...18G} {184, 18}

\bibitem[\protect\citeauthoryear{{Haisch}, {Lada}  \& {Lada}}{{Haisch}
  et~al.}{2001}]{2001ApJ...553L.153H}
{Haisch} Jr. K.~E.,  {Lada} E.~A.,   {Lada} C.~J.,  2001, \mn@doi [\apjl]
  {10.1086/320685}, \href {http://adsabs.harvard.edu/abs/2001ApJ...553L.153H}
  {553, L153}

\bibitem[\protect\citeauthoryear{{Hartmann}, {Megeath}, {Allen}, {Luhman},
  {Calvet}, {D'Alessio}, {Franco-Hernandez}  \& {Fazio}}{{Hartmann}
  et~al.}{2005}]{2005ApJ...629..881H}
{Hartmann} L.,  {Megeath} S.~T.,  {Allen} L.,  {Luhman} K.,  {Calvet} N.,
  {D'Alessio} P.,  {Franco-Hernandez} R.,   {Fazio} G.,  2005, \mn@doi [\apj]
  {10.1086/431472}, \href {http://adsabs.harvard.edu/abs/2005ApJ...629..881H}
  {629, 881}

\bibitem[\protect\citeauthoryear{{He}, {Whittet}, {Kilkenny}  \& {Spencer
  Jones}}{{He} et~al.}{1995}]{1995ApJS..101..335H}
{He} L.,  {Whittet} D.~C.~B.,  {Kilkenny} D.,   {Spencer Jones} J.~H.,  1995,
  \mn@doi [\apjs] {10.1086/192243}, \href
  {http://adsabs.harvard.edu/abs/1995ApJS..101..335H} {101, 335}

\bibitem[\protect\citeauthoryear{{Herbig} \& {Bell}}{{Herbig} \&
  {Bell}}{1988}]{1988cels.book.....H}
{Herbig} G.~H.,  {Bell} K.~R.,  1988, {Third Catalog of Emission-Line Stars of
  the Orion Population : 3 : 1988}

\bibitem[\protect\citeauthoryear{{Hern{\'a}ndez}, {Hartmann}, {Calvet},
  {Jeffries}, {Gutermuth}, {Muzerolle}  \& {Stauffer}}{{Hern{\'a}ndez}
  et~al.}{2008}]{2008ApJ...686.1195H}
{Hern{\'a}ndez} J.,  {Hartmann} L.,  {Calvet} N.,  {Jeffries} R.~D.,
  {Gutermuth} R.,  {Muzerolle} J.,   {Stauffer} J.,  2008, \mn@doi [\apj]
  {10.1086/591224}, \href {http://adsabs.harvard.edu/abs/2008ApJ...686.1195H}
  {686, 1195}

\bibitem[\protect\citeauthoryear{{Hillenbrand}, {Strom}, {Vrba}  \&
  {Keene}}{{Hillenbrand} et~al.}{1992}]{1992ApJ...397..613H}
{Hillenbrand} L.~A.,  {Strom} S.~E.,  {Vrba} F.~J.,   {Keene} J.,  1992,
  \mn@doi [\apj] {10.1086/171819}, \href
  {http://adsabs.harvard.edu/abs/1992ApJ...397..613H} {397, 613}

\bibitem[\protect\citeauthoryear{{Hur}, {Sung}  \& {Bessell}}{{Hur}
  et~al.}{2012}]{2012AJ....143...41H}
{Hur} H.,  {Sung} H.,   {Bessell} M.~S.,  2012, \mn@doi [\aj]
  {10.1088/0004-6256/143/2/41}, \href
  {http://adsabs.harvard.edu/abs/2012AJ....143...41H} {143, 41}

\bibitem[\protect\citeauthoryear{{Jose} et~al.,}{{Jose}
  et~al.}{2008}]{2008MNRAS.384.1675J}
{Jose} J.,  et~al., 2008, \mn@doi [\mnras] {10.1111/j.1365-2966.2007.12825.x},
  \href {http://adsabs.harvard.edu/abs/2008MNRAS.384.1675J} {384, 1675}

\bibitem[\protect\citeauthoryear{{Jose} et~al.,}{{Jose}
  et~al.}{2013}]{2013MNRAS.432.3445J}
{Jose} J.,  et~al., 2013, \mn@doi [\mnras] {10.1093/mnras/stt700}, \href
  {http://adsabs.harvard.edu/abs/2013MNRAS.432.3445J} {432, 3445}

\bibitem[\protect\citeauthoryear{{Jose}, {Kim}, {Herczeg}, {Samal}, {Bieging},
  {Meyer}  \& {Sherry}}{{Jose} et~al.}{2016}]{2016ApJ...822...49J}
{Jose} J.,  {Kim} J.~S.,  {Herczeg} G.~J.,  {Samal} M.~R.,  {Bieging} J.~H.,
  {Meyer} M.~R.,   {Sherry} W.~H.,  2016, \mn@doi [\apj]
  {10.3847/0004-637X/822/1/49}, \href
  {http://adsabs.harvard.edu/abs/2016ApJ...822...49J} {822, 49}

\bibitem[\protect\citeauthoryear{{Kalberla}, {Burton}, {Hartmann}, {Arnal},
  {Bajaja}, {Morras}  \& {P{\"o}ppel}}{{Kalberla} et~al.}{2005}]{kal+05}
{Kalberla} P.~M.~W.,  {Burton} W.~B.,  {Hartmann} D.,  {Arnal} E.~M.,  {Bajaja}
  E.,  {Morras} R.,   {P{\"o}ppel} W.~G.~L.,  2005, \mn@doi [\aap]
  {10.1051/0004-6361:20041864}, \href
  {http://adsabs.harvard.edu/abs/2005A\%26A...440..775K} {440, 775}

\bibitem[\protect\citeauthoryear{{Kendrew} et~al.,}{{Kendrew}
  et~al.}{2012}]{2012ApJ...755...71K}
{Kendrew} S.,  et~al., 2012, \mn@doi [\apj] {10.1088/0004-637X/755/1/71}, \href
  {http://adsabs.harvard.edu/abs/2012ApJ...755...71K} {755, 71}

\bibitem[\protect\citeauthoryear{{Koenig}, {Allen}, {Gutermuth}, {Hora},
  {Brunt}  \& {Muzerolle}}{{Koenig} et~al.}{2008}]{2008ApJ...688.1142K}
{Koenig} X.~P.,  {Allen} L.~E.,  {Gutermuth} R.~A.,  {Hora} J.~L.,  {Brunt}
  C.~M.,   {Muzerolle} J.,  2008, \mn@doi [\apj] {10.1086/592322}, \href
  {http://adsabs.harvard.edu/abs/2008ApJ...688.1142K} {688, 1142}

\bibitem[\protect\citeauthoryear{{Koenig}, {Leisawitz}, {Benford}, {Rebull},
  {Padgett}  \& {Assef}}{{Koenig} et~al.}{2012}]{2012ApJ...744..130K}
{Koenig} X.~P.,  {Leisawitz} D.~T.,  {Benford} D.~J.,  {Rebull} L.~M.,
  {Padgett} D.~L.,   {Assef} R.~J.,  2012, \mn@doi [\apj]
  {10.1088/0004-637X/744/2/130}, \href
  {http://adsabs.harvard.edu/abs/2012ApJ...744..130K} {744, 130}

\bibitem[\protect\citeauthoryear{{Kroupa}}{{Kroupa}}{2002}]{2002Sci...295...82K}
{Kroupa} P.,  2002, \mn@doi [Science] {10.1126/science.1067524}, \href
  {http://adsabs.harvard.edu/abs/2002Sci...295...82K} {295, 82}

\bibitem[\protect\citeauthoryear{{Kumar}, {Sharma}, {Manfroid}, {Gosset},
  {Rauw}, {Naz{\'e}}  \& {Kesh Yadav}}{{Kumar}
  et~al.}{2014}]{2014A&A...567A.109K}
{Kumar} B.,  {Sharma} S.,  {Manfroid} J.,  {Gosset} E.,  {Rauw} G.,  {Naz{\'e}}
  Y.,   {Kesh Yadav} R.,  2014, \mn@doi [\aap] {10.1051/0004-6361/201323027},
  \href {http://adsabs.harvard.edu/abs/2014A\%26A...567A.109K} {567, A109}

\bibitem[\protect\citeauthoryear{{Landolt}}{{Landolt}}{1992}]{1992AJ....104..340L}
{Landolt} A.~U.,  1992, \mn@doi [\aj] {10.1086/116242}, \href
  {http://adsabs.harvard.edu/abs/1992AJ....104..340L} {104, 340}

\bibitem[\protect\citeauthoryear{{Lee}, {Chen}, {Zhang}  \& {Hu}}{{Lee}
  et~al.}{2005}]{2005ApJ...624..808L}
{Lee} H.-T.,  {Chen} W.~P.,  {Zhang} Z.-W.,   {Hu} J.-Y.,  2005, \mn@doi [\apj]
  {10.1086/429122}, \href {http://adsabs.harvard.edu/abs/2005ApJ...624..808L}
  {624, 808}

\bibitem[\protect\citeauthoryear{{Lefloch} \& {Lazareff}}{{Lefloch} \&
  {Lazareff}}{1994}]{1994A&A...289..559L}
{Lefloch} B.,  {Lazareff} B.,  1994, \aap, \href
  {http://adsabs.harvard.edu/abs/1994A\%26A...289..559L} {289, 559}

\bibitem[\protect\citeauthoryear{{Lim}, {Sung}, {Karimov}  \&
  {Ibrahimov}}{{Lim} et~al.}{2011}]{2011JKAS...44...39L}
{Lim} B.,  {Sung} H.~S.,  {Karimov} R.,   {Ibrahimov} M.,  2011, Journal of
  Korean Astronomical Society, \href
  {http://adsabs.harvard.edu/abs/2011JKAS...44...39L} {44, 39}

\bibitem[\protect\citeauthoryear{{Liu}, {de Grijs}, {Deng}, {Hu}, {Baraffe}  \&
  {Beaulieu}}{{Liu} et~al.}{2009}]{2009MNRAS.396.1665L}
{Liu} Q.,  {de Grijs} R.,  {Deng} L.~C.,  {Hu} Y.,  {Baraffe} I.,   {Beaulieu}
  S.~F.,  2009, \mn@doi [\mnras] {10.1111/j.1365-2966.2009.14838.x}, \href
  {http://adsabs.harvard.edu/abs/2009MNRAS.396.1665L} {396, 1665}

\bibitem[\protect\citeauthoryear{{Luhman}, {Allen}, {Espaillat}, {Hartmann}  \&
  {Calvet}}{{Luhman} et~al.}{2010}]{2010ApJS..186..111L}
{Luhman} K.~L.,  {Allen} P.~R.,  {Espaillat} C.,  {Hartmann} L.,   {Calvet} N.,
   2010, \mn@doi [\apjs] {10.1088/0067-0049/186/1/111}, \href
  {http://adsabs.harvard.edu/abs/2010ApJS..186..111L} {186, 111}

\bibitem[\protect\citeauthoryear{{Mallick}, {Ojha}, {Samal}, {Pandey}, {Bhatt},
  {Ghosh}, {Dewangan}  \& {Tamura}}{{Mallick}
  et~al.}{2012}]{2012ApJ...759...48M}
{Mallick} K.~K.,  {Ojha} D.~K.,  {Samal} M.~R.,  {Pandey} A.~K.,  {Bhatt}
  B.~C.,  {Ghosh} S.~K.,  {Dewangan} L.~K.,   {Tamura} M.,  2012, \mn@doi
  [\apj] {10.1088/0004-637X/759/1/48}, \href
  {http://adsabs.harvard.edu/abs/2012ApJ...759...48M} {759, 48}

\bibitem[\protect\citeauthoryear{{Mallick} et~al.,}{{Mallick}
  et~al.}{2014}]{2014MNRAS.443.3218M}
{Mallick} K.~K.,  et~al., 2014, \mn@doi [\mnras] {10.1093/mnras/stu1396}, \href
  {http://adsabs.harvard.edu/abs/2014MNRAS.443.3218M} {443, 3218}

\bibitem[\protect\citeauthoryear{{Marigo}, {Girardi}, {Bressan}, {Groenewegen},
  {Silva}  \& {Granato}}{{Marigo} et~al.}{2008}]{2008AA...482..883M}
{Marigo} P.,  {Girardi} L.,  {Bressan} A.,  {Groenewegen} M.~A.~T.,  {Silva}
  L.,   {Granato} G.~L.,  2008, \mn@doi [\aap] {10.1051/0004-6361:20078467},
  \href {http://adsabs.harvard.edu/abs/2008A\%26A...482..883M} {482, 883}

\bibitem[\protect\citeauthoryear{{Massi}, {Giannetti}, {Di Carlo}, {Brand},
  {Beltr{\'a}n}  \& {Marconi}}{{Massi} et~al.}{2015}]{2015A&A...573A..95M}
{Massi} F.,  {Giannetti} A.,  {Di Carlo} E.,  {Brand} J.,  {Beltr{\'a}n} M.~T.,
    {Marconi} G.,  2015, \mn@doi [\aap] {10.1051/0004-6361/201424388}, \href
  {http://adsabs.harvard.edu/abs/2015A%26A...573A..95M} {573, A95}

\bibitem[\protect\citeauthoryear{{McCaughrean}, {Rayner}  \&
  {Zinnecker}}{{McCaughrean} et~al.}{1991}]{1991MmSAI..62..715M}
{McCaughrean} M.,  {Rayner} J.,   {Zinnecker} H.,  1991, \memsai, \href
  {http://adsabs.harvard.edu/abs/1991MmSAI..62..715M} {62, 715}

\bibitem[\protect\citeauthoryear{{Meyer}, {Calvet}  \& {Hillenbrand}}{{Meyer}
  et~al.}{1997}]{1997AJ....114..288M}
{Meyer} M.~R.,  {Calvet} N.,   {Hillenbrand} L.~A.,  1997, \mn@doi [\aj]
  {10.1086/118474}, \href {http://adsabs.harvard.edu/abs/1997AJ....114..288M}
  {114, 288}

\bibitem[\protect\citeauthoryear{{Miao}, {White}, {Nelson}, {Thompson}  \&
  {Morgan}}{{Miao} et~al.}{2006}]{2006MNRAS.369..143M}
{Miao} J.,  {White} G.~J.,  {Nelson} R.,  {Thompson} M.,   {Morgan} L.,  2006,
  \mn@doi [\mnras] {10.1111/j.1365-2966.2006.10260.x}, \href
  {http://adsabs.harvard.edu/abs/2006MNRAS.369..143M} {369, 143}

\bibitem[\protect\citeauthoryear{{Miller} \& {Scalo}}{{Miller} \&
  {Scalo}}{1978}]{1978PASP...90..506M}
{Miller} G.~E.,  {Scalo} J.~M.,  1978, \mn@doi [\pasp] {10.1086/130373}, \href
  {http://adsabs.harvard.edu/abs/1978PASP...90..506M} {90, 506}

\bibitem[\protect\citeauthoryear{{Miller} \& {Scalo}}{{Miller} \&
  {Scalo}}{1979}]{1979ApJS...41..513M}
{Miller} G.~E.,  {Scalo} J.~M.,  1979, \mn@doi [\apjs] {10.1086/190629}, \href
  {http://adsabs.harvard.edu/abs/1979ApJS...41..513M} {41, 513}

\bibitem[\protect\citeauthoryear{{Moreno} \& {Chavarria-K.}}{{Moreno} \&
  {Chavarria-K.}}{1986}]{1986A&A...161..130M}
{Moreno} M.~A.,  {Chavarria-K.} C.,  1986, \aap, \href
  {http://adsabs.harvard.edu/abs/1986A\%26A...161..130M} {161, 130}

\bibitem[\protect\citeauthoryear{{Moscadelli}, {Reid}, {Menten}, {Brunthaler},
  {Zheng}  \& {Xu}}{{Moscadelli} et~al.}{2009}]{2009ApJ...693..406M}
{Moscadelli} L.,  {Reid} M.~J.,  {Menten} K.~M.,  {Brunthaler} A.,  {Zheng}
  X.~W.,   {Xu} Y.,  2009, \mn@doi [\apj] {10.1088/0004-637X/693/1/406}, \href
  {http://adsabs.harvard.edu/abs/2009ApJ...693..406M} {693, 406}

\bibitem[\protect\citeauthoryear{{Naranjo-Romero}, {Zapata},
  {V{\'a}zquez-Semadeni}, {Takahashi}, {Palau}  \& {Schilke}}{{Naranjo-Romero}
  et~al.}{2012}]{2012ApJ...757...58N}
{Naranjo-Romero} R.,  {Zapata} L.~A.,  {V{\'a}zquez-Semadeni} E.,  {Takahashi}
  S.,  {Palau} A.,   {Schilke} P.,  2012, \mn@doi [\apj]
  {10.1088/0004-637X/757/1/58}, \href
  {http://adsabs.harvard.edu/abs/2012ApJ...757...58N} {757, 58}

\bibitem[\protect\citeauthoryear{{Nisini}, {Antoniucci}, {Giannini}  \&
  {Lorenzetti}}{{Nisini} et~al.}{2005}]{2005A&A...429..543N}
{Nisini} B.,  {Antoniucci} S.,  {Giannini} T.,   {Lorenzetti} D.,  2005,
  \mn@doi [\aap] {10.1051/0004-6361:20041409}, \href
  {http://adsabs.harvard.edu/abs/2005A%26A...429..543N} {429, 543}

\bibitem[\protect\citeauthoryear{{Ogura}, {Sugitani}  \& {Pickles}}{{Ogura}
  et~al.}{2002}]{2002AJ....123.2597O}
{Ogura} K.,  {Sugitani} K.,   {Pickles} A.,  2002, \mn@doi [\aj]
  {10.1086/339976}, \href {http://adsabs.harvard.edu/abs/2002AJ....123.2597O}
  {123, 2597}

\bibitem[\protect\citeauthoryear{{Ojha} et~al.,}{{Ojha}
  et~al.}{2004a}]{2004ApJ...608..797O}
{Ojha} D.~K.,  et~al., 2004a, \mn@doi [\apj] {10.1086/420876}, \href
  {http://adsabs.harvard.edu/abs/2004ApJ...608..797O} {608, 797}

\bibitem[\protect\citeauthoryear{{Ojha} et~al.,}{{Ojha}
  et~al.}{2004b}]{2004ApJ...616.1042O}
{Ojha} D.~K.,  et~al., 2004b, \mn@doi [\apj] {10.1086/425068}, \href
  {http://adsabs.harvard.edu/abs/2004ApJ...616.1042O} {616, 1042}

\bibitem[\protect\citeauthoryear{{Ojha} et~al.,}{{Ojha}
  et~al.}{2011}]{2011ApJ...738..156O}
{Ojha} D.~K.,  et~al., 2011, \mn@doi [\apj] {10.1088/0004-637X/738/2/156},
  \href {http://adsabs.harvard.edu/abs/2011ApJ...738..156O} {738, 156}

\bibitem[\protect\citeauthoryear{{Pandey}, {Ogura}  \& {Sekiguchi}}{{Pandey}
  et~al.}{2000}]{2000PASJ...52..847P}
{Pandey} A.~K.,  {Ogura} K.,   {Sekiguchi} K.,  2000, \pasj, \href
  {http://adsabs.harvard.edu/abs/2000PASJ...52..847P} {52, 847}

\bibitem[\protect\citeauthoryear{{Pandey}, {Upadhyay}, {Nakada}  \&
  {Ogura}}{{Pandey} et~al.}{2003}]{2003AA...397..191P}
{Pandey} A.~K.,  {Upadhyay} K.,  {Nakada} Y.,   {Ogura} K.,  2003, \mn@doi
  [\aap] {10.1051/0004-6361:20021509}, \href
  {http://adsabs.harvard.edu/abs/2003A\%26A...397..191P} {397, 191}

\bibitem[\protect\citeauthoryear{{Pandey}, {Sharma}, {Ogura}, {Ojha}, {Chen},
  {Bhatt}  \& {Ghosh}}{{Pandey} et~al.}{2008}]{2008MNRAS.383.1241P}
{Pandey} A.~K.,  {Sharma} S.,  {Ogura} K.,  {Ojha} D.~K.,  {Chen} W.~P.,
  {Bhatt} B.~C.,   {Ghosh} S.~K.,  2008, \mn@doi [\mnras]
  {10.1111/j.1365-2966.2007.12641.x}, \href
  {http://adsabs.harvard.edu/abs/2008MNRAS.383.1241P} {383, 1241}

\bibitem[\protect\citeauthoryear{{Pandey} et~al.,}{{Pandey}
  et~al.}{2013}]{2013ApJ...764..172P}
{Pandey} A.~K.,  et~al., 2013, \mn@doi [\apj] {10.1088/0004-637X/764/2/172},
  \href {http://adsabs.harvard.edu/abs/2013ApJ...764..172P} {764, 172}

\bibitem[\protect\citeauthoryear{{Pandey}, {Samal}, {Yadav}, {Richichi},
  {Lata}, {Pandey}, {Ojha}  \& {Chen}}{{Pandey}
  et~al.}{2014}]{2014NewA...29...18P}
{Pandey} A.~K.,  {Samal} M.~R.,  {Yadav} R.~K.,  {Richichi} A.,  {Lata} S.,
  {Pandey} J.~C.,  {Ojha} D.~K.,   {Chen} W.~P.,  2014, \mn@doi [\na]
  {10.1016/j.newast.2013.10.007}, \href
  {http://adsabs.harvard.edu/abs/2014NewA...29...18P} {29, 18}

\bibitem[\protect\citeauthoryear{{Panwar}, {Chen}, {Pandey}, {Samal}, {Ogura},
  {Ojha}, {Jose}  \& {Bhatt}}{{Panwar} et~al.}{2014}]{2014MNRAS.443.1614P}
{Panwar} N.,  {Chen} W.~P.,  {Pandey} A.~K.,  {Samal} M.~R.,  {Ogura} K.,
  {Ojha} D.~K.,  {Jose} J.,   {Bhatt} B.~C.,  2014, \mn@doi [\mnras]
  {10.1093/mnras/stu1244}, \href
  {http://adsabs.harvard.edu/abs/2014MNRAS.443.1614P} {443, 1614}

\bibitem[\protect\citeauthoryear{{Pestalozzi}, {Minier}, {Motte}  \&
  {Conway}}{{Pestalozzi} et~al.}{2006}]{2006A&A...448L..57P}
{Pestalozzi} M.~R.,  {Minier} V.,  {Motte} F.,   {Conway} J.~E.,  2006, \mn@doi
  [\aap] {10.1051/0004-6361:200600006}, \href
  {http://adsabs.harvard.edu/abs/2006A\%26A...448L..57P} {448, L57}

\bibitem[\protect\citeauthoryear{{Pestalozzi}, {Elitzur}  \&
  {Conway}}{{Pestalozzi} et~al.}{2009}]{2009A&A...501..999P}
{Pestalozzi} M.~R.,  {Elitzur} M.,   {Conway} J.~E.,  2009, \mn@doi [\aap]
  {10.1051/0004-6361/200811553}, \href
  {http://adsabs.harvard.edu/abs/2009A%26A...501..999P} {501, 999}

\bibitem[\protect\citeauthoryear{{Phelps} \& {Janes}}{{Phelps} \&
  {Janes}}{1994}]{1994ApJS...90...31P}
{Phelps} R.~L.,  {Janes} K.~A.,  1994, \mn@doi [\apjs] {10.1086/191857}, \href
  {http://adsabs.harvard.edu/abs/1994ApJS...90...31P} {90, 31}

\bibitem[\protect\citeauthoryear{{Pismis} \& {Moreno}}{{Pismis} \&
  {Moreno}}{1976}]{1976RMxAA...1..373P}
{Pismis} P.,  {Moreno} M.~A.,  1976, \rmxaa, \href
  {http://adsabs.harvard.edu/abs/1976RMxAA...1..373P} {1, 373}

\bibitem[\protect\citeauthoryear{{Pomar{\`e}s} et~al.,}{{Pomar{\`e}s}
  et~al.}{2009}]{2009A&A...494..987P}
{Pomar{\`e}s} M.,  et~al., 2009, \mn@doi [\aap] {10.1051/0004-6361:200811050},
  \href {http://adsabs.harvard.edu/abs/2009A\%26A...494..987P} {494, 987}

\bibitem[\protect\citeauthoryear{{Preibisch}}{{Preibisch}}{1997}]{1997A&A...324..690P}
{Preibisch} T.,  1997, \aap, \href
  {http://adsabs.harvard.edu/abs/1997A\%26A...324..690P} {324, 690}

\bibitem[\protect\citeauthoryear{{Preibisch} \& {Feigelson}}{{Preibisch} \&
  {Feigelson}}{2005}]{2005ApJS..160..390P}
{Preibisch} T.,  {Feigelson} E.~D.,  2005, \mn@doi [\apjs] {10.1086/432094},
  \href {http://adsabs.harvard.edu/abs/2005ApJS..160..390P} {160, 390}

\bibitem[\protect\citeauthoryear{{Preibisch} \& {Zinnecker}}{{Preibisch} \&
  {Zinnecker}}{2002}]{pre+02}
{Preibisch} T.,  {Zinnecker} H.,  2002, \mn@doi [\aj] {10.1086/338851}, \href
  {http://adsabs.harvard.edu/abs/2002AJ....123.1613P} {123, 1613}

\bibitem[\protect\citeauthoryear{{Preibisch} et~al.,}{{Preibisch}
  et~al.}{2005a}]{pre+05}
{Preibisch} T.,  et~al., 2005a, \mn@doi [\apjs] {10.1086/432891}, \href
  {http://adsabs.harvard.edu/abs/2005ApJS..160..401P} {160, 401}

\bibitem[\protect\citeauthoryear{{Preibisch} et~al.,}{{Preibisch}
  et~al.}{2005b}]{2005ApJS..160..401P}
{Preibisch} T.,  et~al., 2005b, \mn@doi [\apjs] {10.1086/432891}, \href
  {http://adsabs.harvard.edu/abs/2005ApJS..160..401P} {160, 401}

\bibitem[\protect\citeauthoryear{{Preibisch} et~al.,}{{Preibisch}
  et~al.}{2011}]{2011AA...530A..34P}
{Preibisch} T.,  et~al., 2011, \mn@doi [\aap] {10.1051/0004-6361/201116781},
  \href {http://adsabs.harvard.edu/abs/2011A\%26A...530A..34P} {530, A34}

\bibitem[\protect\citeauthoryear{{Puga} et~al.,}{{Puga}
  et~al.}{2010}]{2010A&A...517A...2P}
{Puga} E.,  et~al., 2010, \mn@doi [\aap] {10.1051/0004-6361/200913294}, \href
  {http://adsabs.harvard.edu/abs/2010A\%26A...517A...2P} {517, A2}

\bibitem[\protect\citeauthoryear{{Rana}}{{Rana}}{1991}]{1991ARAA..29..129R}
{Rana} N.~C.,  1991, \mn@doi [\araa] {10.1146/annurev.aa.29.090191.001021},
  \href {http://adsabs.harvard.edu/abs/1991ARA\%26A..29..129R} {29, 129}

\bibitem[\protect\citeauthoryear{{Rauw}, {Manfroid}  \& {De Becker}}{{Rauw}
  et~al.}{2010}]{2010AA...511A..25R}
{Rauw} G.,  {Manfroid} J.,   {De Becker} M.,  2010, \mn@doi [\aap]
  {10.1051/0004-6361/200912780}, \href
  {http://adsabs.harvard.edu/abs/2010A\%26A...511A..25R} {511, A25}

\bibitem[\protect\citeauthoryear{{Rivera-Ingraham}, {Martin}, {Polychroni}  \&
  {Moore}}{{Rivera-Ingraham} et~al.}{2011}]{2011ApJ...743...39R}
{Rivera-Ingraham} A.,  {Martin} P.~G.,  {Polychroni} D.,   {Moore} T.~J.~T.,
  2011, \mn@doi [\apj] {10.1088/0004-637X/743/1/39}, \href
  {http://adsabs.harvard.edu/abs/2011ApJ...743...39R} {743, 39}

\bibitem[\protect\citeauthoryear{{Robitaille}, {Whitney}, {Indebetouw}, {Wood}
  \& {Denzmore}}{{Robitaille} et~al.}{2006}]{2006ApJS..167..256R}
{Robitaille} T.~P.,  {Whitney} B.~A.,  {Indebetouw} R.,  {Wood} K.,
  {Denzmore} P.,  2006, \mn@doi [\apjs] {10.1086/508424}, \href
  {http://adsabs.harvard.edu/abs/2006ApJS..167..256R} {167, 256}

\bibitem[\protect\citeauthoryear{{Robitaille}, {Whitney}, {Indebetouw}  \&
  {Wood}}{{Robitaille} et~al.}{2007}]{2007ApJS..169..328R}
{Robitaille} T.~P.,  {Whitney} B.~A.,  {Indebetouw} R.,   {Wood} K.,  2007,
  \mn@doi [\apjs] {10.1086/512039}, \href
  {http://adsabs.harvard.edu/abs/2007ApJS..169..328R} {169, 328}

\bibitem[\protect\citeauthoryear{{Robitaille} et~al.,}{{Robitaille}
  et~al.}{2008}]{2008AJ....136.2413R}
{Robitaille} T.~P.,  et~al., 2008, \mn@doi [\aj]
  {10.1088/0004-6256/136/6/2413}, \href
  {http://adsabs.harvard.edu/abs/2008AJ....136.2413R} {136, 2413}

\bibitem[\protect\citeauthoryear{{Rowan-Robinson}, {Gonzalez-Solares},
  {Vaccari}  \& {Marchetti}}{{Rowan-Robinson}
  et~al.}{2013}]{2013MNRAS.428.1958R}
{Rowan-Robinson} M.,  {Gonzalez-Solares} E.,  {Vaccari} M.,   {Marchetti} L.,
  2013, \mn@doi [\mnras] {10.1093/mnras/sts163}, \href
  {http://adsabs.harvard.edu/abs/2013MNRAS.428.1958R} {428, 1958}

\bibitem[\protect\citeauthoryear{{Sagar} \& {Richtler}}{{Sagar} \&
  {Richtler}}{1991}]{1991A&A...250..324S}
{Sagar} R.,  {Richtler} T.,  1991, \aap, \href
  {http://adsabs.harvard.edu/abs/1991A%26A...250..324S} {250, 324}

\bibitem[\protect\citeauthoryear{{Salpeter}}{{Salpeter}}{1955}]{1955ApJ...121..161S}
{Salpeter} E.~E.,  1955, \mn@doi [\apj] {10.1086/145971}, \href
  {http://adsabs.harvard.edu/abs/1955ApJ...121..161S} {121, 161}

\bibitem[\protect\citeauthoryear{{Samal}, {Pandey}, {Ojha}, {Chauhan}, {Jose}
  \& {Pandey}}{{Samal} et~al.}{2012}]{2012ApJ...755...20S}
{Samal} M.~R.,  {Pandey} A.~K.,  {Ojha} D.~K.,  {Chauhan} N.,  {Jose} J.,
  {Pandey} B.,  2012, \mn@doi [\apj] {10.1088/0004-637X/755/1/20}, \href
  {http://adsabs.harvard.edu/abs/2012ApJ...755...20S} {755, 20}

\bibitem[\protect\citeauthoryear{{Sandell} \& {Sievers}}{{Sandell} \&
  {Sievers}}{2004}]{2004ApJ...600..269S}
{Sandell} G.,  {Sievers} A.,  2004, \mn@doi [\apj] {10.1086/379646}, \href
  {http://adsabs.harvard.edu/abs/2004ApJ...600..269S} {600, 269}

\bibitem[\protect\citeauthoryear{{Sandell} \& {Wright}}{{Sandell} \&
  {Wright}}{2010}]{2010ApJ...715..919S}
{Sandell} G.,  {Wright} M.,  2010, \mn@doi [\apj]
  {10.1088/0004-637X/715/2/919}, \href
  {http://adsabs.harvard.edu/abs/2010ApJ...715..919S} {715, 919}

\bibitem[\protect\citeauthoryear{{Sandell}, {Goss}, {Wright}  \&
  {Corder}}{{Sandell} et~al.}{2009}]{2009ApJ...699L..31S}
{Sandell} G.,  {Goss} W.~M.,  {Wright} M.,   {Corder} S.,  2009, \mn@doi
  [\apjl] {10.1088/0004-637X/699/1/L31}, \href
  {http://adsabs.harvard.edu/abs/2009ApJ...699L..31S} {699, L31}

\bibitem[\protect\citeauthoryear{{Sandford}, {Whitaker}  \& {Klein}}{{Sandford}
  et~al.}{1982}]{1982ApJ...260..183S}
{Sandford} II M.~T.,  {Whitaker} R.~W.,   {Klein} R.~I.,  1982, \mn@doi [\apj]
  {10.1086/160245}, \href {http://adsabs.harvard.edu/abs/1982ApJ...260..183S}
  {260, 183}

\bibitem[\protect\citeauthoryear{{Scalo}}{{Scalo}}{1986}]{1986FCPh...11....1S}
{Scalo} J.~M.,  1986, \fcp, \href
  {http://adsabs.harvard.edu/abs/1986FCPh...11....1S} {11, 1}

\bibitem[\protect\citeauthoryear{{Scalo}}{{Scalo}}{1998}]{1998ASPC..142..201S}
{Scalo} J.,  1998, in {Gilmore} G.,  {Howell} D.,  eds,  Astronomical Society
  of the Pacific Conference Series Vol. 142, The Stellar Initial Mass Function
  (38th Herstmonceux Conference). p.~201 (\mn@eprint {}
  {arXiv:astro-ph/9712317})

\bibitem[\protect\citeauthoryear{{Schmidt-Kaler}}{{Schmidt-Kaler}}{1982}]{Schmidt-Kaler1982}
{Schmidt-Kaler} T.,  1982, {in Landolt-B{\"o}rnstein: Numerical Data and
  Functional Relationship in Science and Technology, Vol. 2b. eds. Schaifers
  K., Voigt H. H., Landolt H. (Springer-Verlag), Berlin, p. 19}

\bibitem[\protect\citeauthoryear{{Sharma}, {Pandey}, {Ojha}, {Chen}, {Ghosh},
  {Bhatt}, {Maheswar}  \& {Sagar}}{{Sharma} et~al.}{2007}]{2007MNRAS.380.1141S}
{Sharma} S.,  {Pandey} A.~K.,  {Ojha} D.~K.,  {Chen} W.~P.,  {Ghosh} S.~K.,
  {Bhatt} B.~C.,  {Maheswar} G.,   {Sagar} R.,  2007, \mn@doi [\mnras]
  {10.1111/j.1365-2966.2007.12156.x}, \href
  {http://adsabs.harvard.edu/abs/2007MNRAS.380.1141S} {380, 1141}

\bibitem[\protect\citeauthoryear{{Sharma}, {Pandey}, {Ogura}, {Aoki}, {Pandey},
  {Sandhu}  \& {Sagar}}{{Sharma} et~al.}{2008}]{2008AJ....135.1934S}
{Sharma} S.,  {Pandey} A.~K.,  {Ogura} K.,  {Aoki} T.,  {Pandey} K.,  {Sandhu}
  T.~S.,   {Sagar} R.,  2008, \mn@doi [\aj] {10.1088/0004-6256/135/5/1934},
  \href {http://adsabs.harvard.edu/abs/2008AJ....135.1934S} {135, 1934}

\bibitem[\protect\citeauthoryear{{Sharma} et~al.,}{{Sharma}
  et~al.}{2012}]{2012PASJ...64..107S}
{Sharma} S.,  et~al., 2012, \pasj, \href
  {http://adsabs.harvard.edu/abs/2012PASJ...64..107S} {64, 107}

\bibitem[\protect\citeauthoryear{{Sharma} et~al.,}{{Sharma}
  et~al.}{2016}]{2016AJ....151..126S}
{Sharma} S.,  et~al., 2016, \mn@doi [\aj] {10.3847/0004-6256/151/5/126}, \href
  {http://adsabs.harvard.edu/abs/2016AJ....151..126S} {151, 126}

\bibitem[\protect\citeauthoryear{{Siess}, {Dufour}  \& {Forestini}}{{Siess}
  et~al.}{2000}]{2000AA...358..593S}
{Siess} L.,  {Dufour} E.,   {Forestini} M.,  2000, \aap, \href
  {http://adsabs.harvard.edu/abs/2000A\%26A...358..593S} {358, 593}

\bibitem[\protect\citeauthoryear{{Smith}, {Brickhouse}, {Liedahl}  \&
  {Raymond}}{{Smith} et~al.}{2001}]{smi+01}
{Smith} R.~K.,  {Brickhouse} N.~S.,  {Liedahl} D.~A.,   {Raymond} J.~C.,  2001,
  \mn@doi [\apjl] {10.1086/322992}, \href
  {http://adsabs.harvard.edu/abs/2001ApJ...556L..91S} {556, L91}

\bibitem[\protect\citeauthoryear{{Spitzer}}{{Spitzer}}{1978}]{1978ppim.book.....S}
{Spitzer} L.,  1978, {Physical processes in the interstellar medium}

\bibitem[\protect\citeauthoryear{{Stahler} \& {Palla}}{{Stahler} \&
  {Palla}}{2005}]{2005fost.book.....S}
{Stahler} S.~W.,  {Palla} F.,  2005, {The Formation of Stars}

\bibitem[\protect\citeauthoryear{{Stassun}, {Ardila}, {Barsony}, {Basri}  \&
  {Mathieu}}{{Stassun} et~al.}{2004}]{sta+04}
{Stassun} K.~G.,  {Ardila} D.~R.,  {Barsony} M.,  {Basri} G.,   {Mathieu}
  R.~D.,  2004, \mn@doi [\aj] {10.1086/420989}, \href
  {http://adsabs.harvard.edu/abs/2004AJ....127.3537S} {127, 3537}

\bibitem[\protect\citeauthoryear{{Stelzer} \& {Neuh{\"a}user}}{{Stelzer} \&
  {Neuh{\"a}user}}{2001a}]{ste+01}
{Stelzer} B.,  {Neuh{\"a}user} R.,  2001a, \mn@doi [\aap]
  {10.1051/0004-6361:20011093}, \href
  {http://adsabs.harvard.edu/abs/2001A\%26A...377..538S} {377, 538}

\bibitem[\protect\citeauthoryear{{Stelzer} \& {Neuh{\"a}user}}{{Stelzer} \&
  {Neuh{\"a}user}}{2001b}]{2001A&A...377..538S}
{Stelzer} B.,  {Neuh{\"a}user} R.,  2001b, \mn@doi [\aap]
  {10.1051/0004-6361:20011093}, \href
  {http://adsabs.harvard.edu/abs/2001A\%26A...377..538S} {377, 538}

\bibitem[\protect\citeauthoryear{{Stetson}}{{Stetson}}{1987}]{1987PASP...99..191S}
{Stetson} P.~B.,  1987, \mn@doi [\pasp] {10.1086/131977}, \href
  {http://adsabs.harvard.edu/abs/1987PASP...99..191S} {99, 191}

\bibitem[\protect\citeauthoryear{{Stetson}}{{Stetson}}{1992}]{1992ASPC...25..297S}
{Stetson} P.~B.,  1992, in {Worrall} D.~M.,  {Biemesderfer} C.,   {Barnes} J.,
  eds,  Astronomical Society of the Pacific Conference Series Vol. 25,
  Astronomical Data Analysis Software and Systems I. p.~297

\bibitem[\protect\citeauthoryear{{Sung}, {Bessell}  \& {Lee}}{{Sung}
  et~al.}{1997}]{1997AJ....114.2644S}
{Sung} H.,  {Bessell} M.~S.,   {Lee} S.-W.,  1997, \mn@doi [\aj]
  {10.1086/118674}, \href {http://adsabs.harvard.edu/abs/1997AJ....114.2644S}
  {114, 2644}

\bibitem[\protect\citeauthoryear{{Telleschi}, {G{\"u}del}, {Briggs}, {Audard}
  \& {Palla}}{{Telleschi} et~al.}{2007a}]{2007A&A...468..425T}
{Telleschi} A.,  {G{\"u}del} M.,  {Briggs} K.~R.,  {Audard} M.,   {Palla} F.,
  2007a, \mn@doi [\aap] {10.1051/0004-6361:20066565}, \href
  {http://adsabs.harvard.edu/abs/2007A\%26A...468..425T} {468, 425}

\bibitem[\protect\citeauthoryear{{Telleschi}, {G{\"u}del}, {Briggs}, {Audard}
  \& {Scelsi}}{{Telleschi} et~al.}{2007b}]{tel+07}
{Telleschi} A.,  {G{\"u}del} M.,  {Briggs} K.~R.,  {Audard} M.,   {Scelsi} L.,
  2007b, \mn@doi [\aap] {10.1051/0004-6361:20066193}, \href
  {http://adsabs.harvard.edu/abs/2007A\%26A...468..443T} {468, 443}

\bibitem[\protect\citeauthoryear{{Thompson}, {White}, {Morgan}, {Miao},
  {Fridlund}  \& {Huldtgren-White}}{{Thompson}
  et~al.}{2004}]{2004A&A...414.1017T}
{Thompson} M.~A.,  {White} G.~J.,  {Morgan} L.~K.,  {Miao} J.,  {Fridlund}
  C.~V.~M.,   {Huldtgren-White} M.,  2004, \mn@doi [\aap]
  {10.1051/0004-6361:20031680}, \href
  {http://adsabs.harvard.edu/abs/2004A\%26A...414.1017T} {414, 1017}

\bibitem[\protect\citeauthoryear{{Thompson}, {Urquhart}, {Moore}  \&
  {Morgan}}{{Thompson} et~al.}{2012}]{2012MNRAS.421..408T}
{Thompson} M.~A.,  {Urquhart} J.~S.,  {Moore} T.~J.~T.,   {Morgan} L.~K.,
  2012, \mn@doi [\mnras] {10.1111/j.1365-2966.2011.20315.x}, \href
  {http://adsabs.harvard.edu/abs/2012MNRAS.421..408T} {421, 408}

\bibitem[\protect\citeauthoryear{{Tsujimoto}, {Townsley}, {Feigelson}, {Broos},
  {Getman}  \& {Garmire}}{{Tsujimoto} et~al.}{2005}]{2005prpl.conf.8307T}
{Tsujimoto} M.,  {Townsley} L.,  {Feigelson} E.~D.,  {Broos} P.,  {Getman}
  K.~V.,   {Garmire} G.,  2005, in Protostars and Planets V. p.~8307

\bibitem[\protect\citeauthoryear{{Vacca}, {Garmany}  \& {Shull}}{{Vacca}
  et~al.}{1996}]{1996ApJ...460..914V}
{Vacca} W.~D.,  {Garmany} C.~D.,   {Shull} J.~M.,  1996, \mn@doi [\apj]
  {10.1086/177020}, \href {http://adsabs.harvard.edu/abs/1996ApJ...460..914V}
  {460, 914}

\bibitem[\protect\citeauthoryear{{Wang}, {Townsley}, {Feigelson}, {Getman},
  {Broos}, {Garmire}  \& {Tsujimoto}}{{Wang}
  et~al.}{2007}]{2007ApJS..168..100W}
{Wang} J.,  {Townsley} L.~K.,  {Feigelson} E.~D.,  {Getman} K.~V.,  {Broos}
  P.~S.,  {Garmire} G.~P.,   {Tsujimoto} M.,  2007, \mn@doi [\apjs]
  {10.1086/509147}, \href {http://adsabs.harvard.edu/abs/2007ApJS..168..100W}
  {168, 100}

\bibitem[\protect\citeauthoryear{{Ward-Thompson} \&
  {Whitworth}}{{Ward-Thompson} \& {Whitworth}}{2011}]{2011isf..book.....W}
{Ward-Thompson} D.,  {Whitworth} A.~P.,  2011, {An Introduction to Star
  Formation}

\bibitem[\protect\citeauthoryear{{Werner}, {Becklin}, {Gatley}, {Matthews},
  {Neugebauer}  \& {Wynn-Williams}}{{Werner}
  et~al.}{1979}]{1979MNRAS.188..463W}
{Werner} M.~W.,  {Becklin} E.~E.,  {Gatley} I.,  {Matthews} K.,  {Neugebauer}
  G.,   {Wynn-Williams} C.~G.,  1979, \mnras, \href
  {http://adsabs.harvard.edu/abs/1979MNRAS.188..463W} {188, 463}

\bibitem[\protect\citeauthoryear{{White} et~al.,}{{White}
  et~al.}{1999}]{1999A&A...342..233W}
{White} G.~J.,  et~al., 1999, \aap, \href
  {http://adsabs.harvard.edu/abs/1999A\%26A...342..233W} {342, 233}

\bibitem[\protect\citeauthoryear{{White}, {Greene}, {Doppmann}, {Covey}  \&
  {Hillenbrand}}{{White} et~al.}{2007}]{2007prpl.conf..117W}
{White} R.~J.,  {Greene} T.~P.,  {Doppmann} G.~W.,  {Covey} K.~R.,
  {Hillenbrand} L.~A.,  2007, Protostars and Planets V, \href
  {http://adsabs.harvard.edu/abs/2007prpl.conf..117W} {pp 117--132}

\bibitem[\protect\citeauthoryear{{Whitney}, {Wood}, {Bjorkman}  \&
  {Wolff}}{{Whitney} et~al.}{2003a}]{2003ApJ...591.1049W}
{Whitney} B.~A.,  {Wood} K.,  {Bjorkman} J.~E.,   {Wolff} M.~J.,  2003a,
  \mn@doi [\apj] {10.1086/375415}, \href
  {http://adsabs.harvard.edu/abs/2003ApJ...591.1049W} {591, 1049}

\bibitem[\protect\citeauthoryear{{Whitney}, {Wood}, {Bjorkman}  \&
  {Cohen}}{{Whitney} et~al.}{2003b}]{2003ApJ...598.1079W}
{Whitney} B.~A.,  {Wood} K.,  {Bjorkman} J.~E.,   {Cohen} M.,  2003b, \mn@doi
  [\apj] {10.1086/379068}, \href
  {http://adsabs.harvard.edu/abs/2003ApJ...598.1079W} {598, 1079}

\bibitem[\protect\citeauthoryear{{Whitney}, {Indebetouw}, {Bjorkman}  \&
  {Wood}}{{Whitney} et~al.}{2004}]{2004ApJ...617.1177W}
{Whitney} B.~A.,  {Indebetouw} R.,  {Bjorkman} J.~E.,   {Wood} K.,  2004,
  \mn@doi [\apj] {10.1086/425608}, \href
  {http://adsabs.harvard.edu/abs/2004ApJ...617.1177W} {617, 1177}

\bibitem[\protect\citeauthoryear{{Whittet}}{{Whittet}}{2003}]{2003dge..conf.....W}
{Whittet} D.~C.~B.,  ed. 2003, {Dust in the galactic environment}

\bibitem[\protect\citeauthoryear{{Whitworth}, {Bhattal}, {Chapman}, {Disney}
  \& {Turner}}{{Whitworth} et~al.}{1994}]{1994MNRAS.268..291W}
{Whitworth} A.~P.,  {Bhattal} A.~S.,  {Chapman} S.~J.,  {Disney} M.~J.,
  {Turner} J.~A.,  1994, \mn@doi [\mnras] {10.1093/mnras/268.1.291}, \href
  {http://adsabs.harvard.edu/abs/1994MNRAS.268..291W} {268, 291}

\bibitem[\protect\citeauthoryear{{Williams} \& {Cieza}}{{Williams} \&
  {Cieza}}{2011}]{2011ARA&A..49...67W}
{Williams} J.~P.,  {Cieza} L.~A.,  2011, \mn@doi [\araa]
  {10.1146/annurev-astro-081710-102548}, \href
  {http://adsabs.harvard.edu/abs/2011ARA%26A..49...67W} {49, 67}

\bibitem[\protect\citeauthoryear{{Willis}, {Marengo}, {Allen}, {Fazio}, {Smith}
   \& {Carey}}{{Willis} et~al.}{2013}]{2013ApJ...778...96W}
{Willis} S.,  {Marengo} M.,  {Allen} L.,  {Fazio} G.~G.,  {Smith} H.~A.,
  {Carey} S.,  2013, \mn@doi [\apj] {10.1088/0004-637X/778/2/96}, \href
  {http://adsabs.harvard.edu/abs/2013ApJ...778...96W} {778, 96}

\bibitem[\protect\citeauthoryear{{Wynn-Williams}, {Becklin}  \&
  {Neugebauer}}{{Wynn-Williams} et~al.}{1974}]{1974ApJ...187..473W}
{Wynn-Williams} C.~G.,  {Becklin} E.~E.,   {Neugebauer} G.,  1974, \mn@doi
  [\apj] {10.1086/152656}, \href
  {http://adsabs.harvard.edu/abs/1974ApJ...187..473W} {187, 473}

\bibitem[\protect\citeauthoryear{{Yadav} et~al.,}{{Yadav}
  et~al.}{2016}]{2016MNRAS.461.2502Y}
{Yadav} R.~K.,  et~al., 2016, \mn@doi [\mnras] {10.1093/mnras/stw1356}, \href
  {http://adsabs.harvard.edu/abs/2016MNRAS.461.2502Y} {461, 2502}

\bibitem[\protect\citeauthoryear{{Zavagno}, {Deharveng}, {Comer{\'o}n},
  {Brand}, {Massi}, {Caplan}  \& {Russeil}}{{Zavagno}
  et~al.}{2006}]{2006A&A...446..171Z}
{Zavagno} A.,  {Deharveng} L.,  {Comer{\'o}n} F.,  {Brand} J.,  {Massi} F.,
  {Caplan} J.,   {Russeil} D.,  2006, \mn@doi [\aap]
  {10.1051/0004-6361:20053952}, \href
  {http://adsabs.harvard.edu/abs/2006A\%26A...446..171Z} {446, 171}

\makeatother
\end{thebibliography}
\bibliographystyle{mnras}

\appendix

\section{Distance and reddening of NGC 7538}

NGC 7538 is located in the second quadrant of the Galaxy in the Perseus arm.
Its distance estimates in literature ranges from 2.2 kpc \citep{1986A&A...161..130M} to 2.8 kpc \citep{1978A&A....66....1C}.
The accuracy of these photometrically determined distances are typically 10\%-20\%. 
\citet{2010A&A...517A...2P} reported a spectro-photometric distance of 2.7 $\pm$ 0.5 kpc to this region.
\citet{2009ApJ...693..406M}, using the trigonometric parallaxes of methanol masers
which are usually associated with high-mass SFRs,
derived a most accurate distance  of this region as 2.65$^{+0.12}_{-0.11}$.
Therefore, we have adopted 2.65 kpc as the distance for NGC 7538 in our analyses.

\begin{figure}
\centering
\includegraphics[height=7cm,width=7cm]{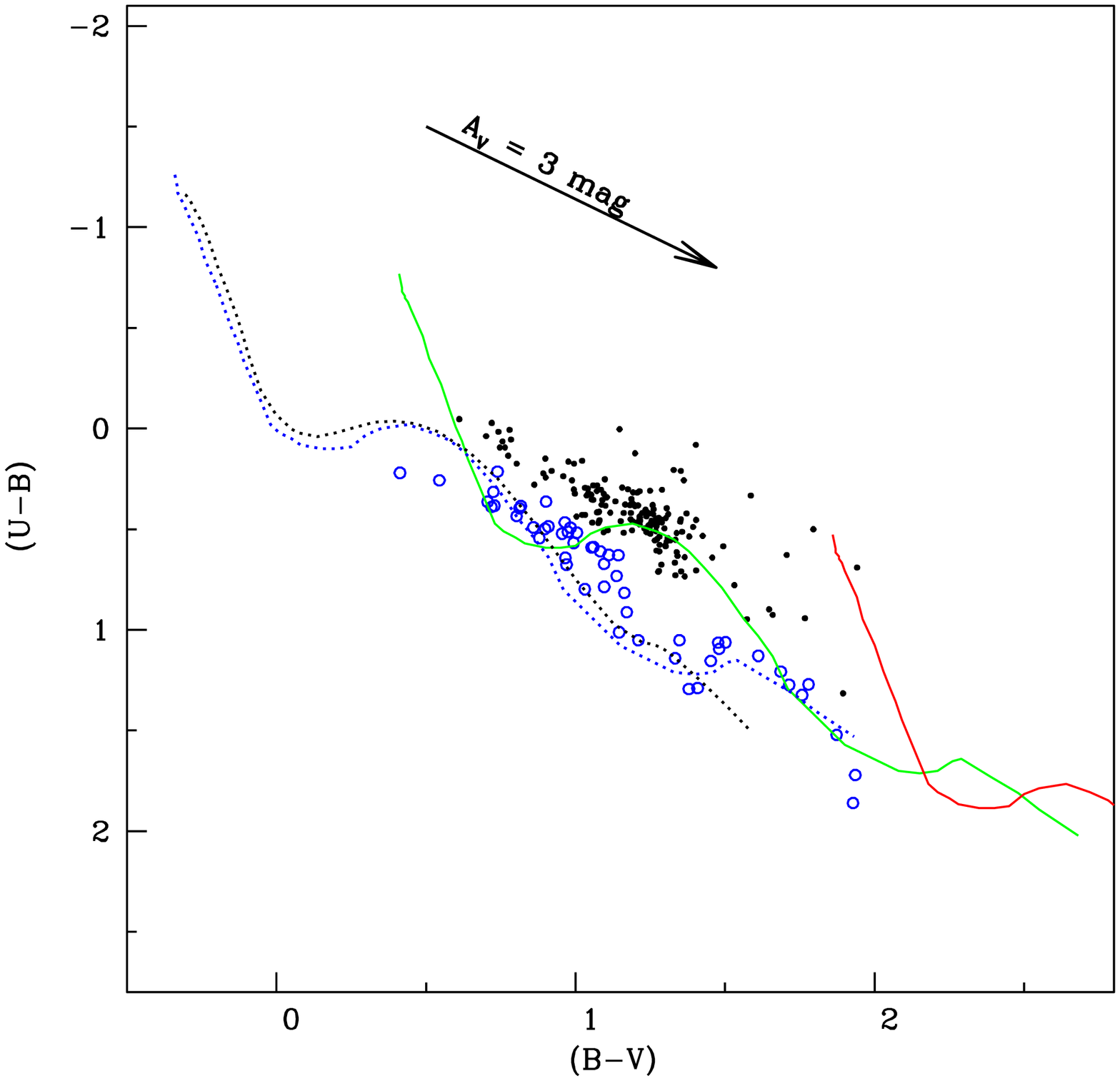}
\caption{\label{ccd} $(U-B)/(B-V)$ TCD for the sources in the NGC 7538 region. The dotted blue curve represents the intrinsic
ZAMS for $Z=0.02$ by \citet{Schmidt-Kaler1982} along with the selected foreground stars (blue open circles). 
1 Myr isochrone (equivalent to ZAMS) from \citet{2008AA...482..883M} is also shown as black dotted curve.
The continuous curves represent the \citet{Schmidt-Kaler1982} ZAMS shifted along 
the reddening vector (see text for details) by $E(B-V)_{min}$ = 0.75 mag  (green curve) and  
$E(B-V)_{max}$ = 2.2 mag (red curve) for references to the stars embedded in the nebulosity of NGC 7538 (black dots). 
}
\end{figure}

A $(U - B)/(B - V)$ TCD has been used to estimate the amount of reddening
towards the NGC 7538 region. In Fig.~\ref{ccd} we show the TCD with the intrinsic zero-age-main-sequence
(ZAMS, blue dotted curve) taken from \citet{Schmidt-Kaler1982} along with the identified stars (black dots).
For comparison, we have also overplotted in Fig.~\ref{ccd}, the 1 Myr isochrone (equivalent to ZAMS) from \citet{2008AA...482..883M}
which agrees well with that of \citet{Schmidt-Kaler1982}.
The distribution of the stars shows a large spread along the reddening line, indicating 
heavy differential reddening in this region.
It reveals two different populations, one (blue open circles) distributing along the ZAMS and another (black dots) 
showing a large spread in their reddening value.       
The former having negligible reddening must be the foreground population and the latter could be member stars.
The both populations are selected visually on the basis of their locations with respect to the ZAMS 
\citep[for detail cf.,][]{1974ASSL...41.....G,1994ApJS...90...31P}. 
If we look at the MIR image of the NGC 7538 region (Fig.~\ref{color}), 
we see  several dust lanes along with enhancements
of nebular emission at many places; the both are likely responsible for the large spread of reddening in the latter population.

The ZAMS from \citet{Schmidt-Kaler1982} is  shifted along the reddening vector with a slope of
$E(U - B)/E(B - V)$ = $0.72\times0.91$ (corresponding to $R_V$ = 2.82) to match the distribution of the stars showing the minimum 
reddening among the member population (green curve); this gives $E(B-V)_{min}$ =0.75 and  A$_V\simeq$ 0.75$\times$2.82 = 2.1 mag. 
The others may be embedded in the nebulosity of the H\,{\sevensize II}.
The ZAMS is further shifted along the reddening vector with a slope of
$E(U - B)/E(B - V)$ = $0.72\times1.24$ (corresponding to $R_V$ = 3.85) to match the distribution of these embedded stars showing the maximum 
reddening value, $E(B-V)_{max}$=2.2 mag (red curve,  A$_V\simeq$ (0.75$\times$2.82 + 1.45$\times$3.85) = 7.7 mag). 
The approximate error in the reddening measurement `$E(B-V)$' is 0.2 mag, as has been determined by the procedure
 outlined in \citet{1994ApJS...90...31P}.

\section{Reddening Law}

We have used the technique as described by \citet{2003AA...397..191P} to study the nature of the
diffuse interstellar medium (ISM) associated with the NGC 7538 region.
This can be represented by the ratio of total-to-selective extinction $R_V$ = $A_V$/$E(B-V)$.
The normal reddening law for the solar neighborhood gives the 
value $R_V$ = 3.1$\pm$ 0.2 \citep{2003dge..conf.....W, 1989AJ.....98..611G, 2011JKAS...44...39L},
but in the case of several SFRs, it is found to be
anomalously high \citep[see e.g.,][]{ 2000PASJ...52..847P, 2008MNRAS.383.1241P, 2012AJ....143...41H, 2013ApJ...764..172P, 2014A&A...567A.109K}. 

\begin{figure}
\centering\includegraphics[height=6cm,width=7cm,angle=0]{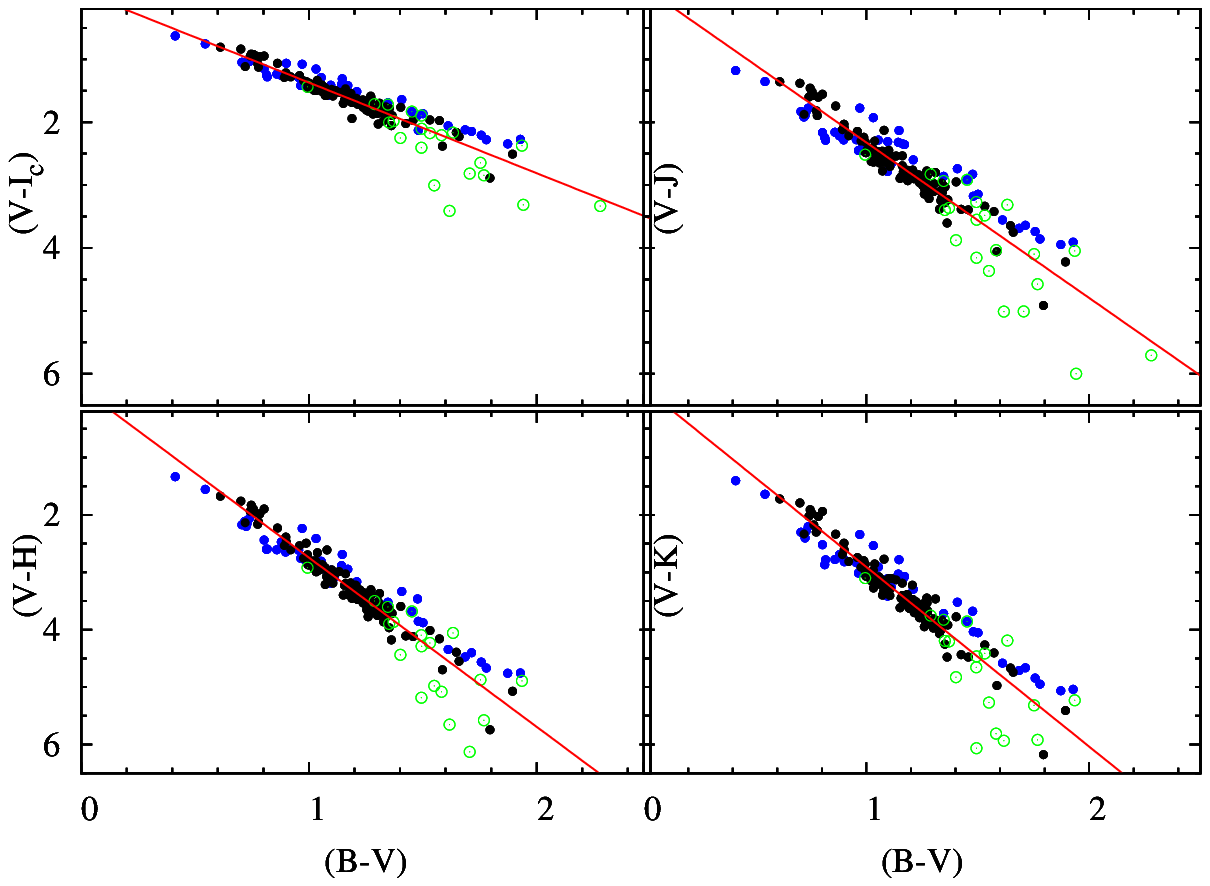}
\caption{\label{2color} $(V-I_c),(V - J), (V -H), (V - K)$ versus $(B - V)$ TCDs
for the stars associated with the NGC 7538 region (black dots) and for the foreground populations (blue dots) (cf. Appendix A). 
Open green circles are identified YSOs (cf. Section 3.1.4) and are not used in the analysis. 
Straight lines show the least-square fit to the stars in the NGC 7538 region.}
\end{figure}

The TCDs of the form of $(V - \lambda)$ versus $(B - V)$,
where $\lambda$ indicates one of the wavelengths of the broad-band filters ($R, I, J, H, K, L$), 
provide an effective method for separating the influence of the normal extinction produced by
the general ISM from that of the abnormal extinction arising within regions
having a peculiar distribution of dust sizes \citep[cf.][]{1990A&A...227L...5C, 2000PASJ...52..847P}.
We have selected all the stars having optical and NIR detections and plotted
their  $(V - \lambda)$ versus $(B - V)$ TCDs in Fig.~\ref{2color}.
It reveals two distributions having different slopes. Presumably, blue 
and black dots are the foreground population and the stars associated with NGC 7538, respectively, selected on the basis of their reddening values (cf. Appendix A).
Since YSOs (open circles) show excess IR emission, their position can deviate from those of the MS stars in the above TCDs, therefore
they  have not been used in the calculation of the reddening law.
The slopes of the least square fit to the distribution of the MS member 
stars (black dots) in the $(V-I_c),(V-J),(V-H)$ and $(V-K)$ versus $(B-V)$ TCDs are found to be
$1.44\pm0.04, 2.47\pm0.07, 2.95\pm0.07$ and $3.03\pm0.06$, respectively, which are higher
than those found for the general ISM \citep[1.10, 1.96, 2.42 and 2.60; cf.,][]{2003AA...397..191P}. On the other hand, 
the field population (blue dots) gives lower values for them (i.e., $1.12\pm0.04, 1.75\pm0.08, 2.26\pm0.07$ and $2.36\pm0.06$).

The slopes for the MS stars associated with the NGC 7538 region estimated as above yield a higher 
value for $R_V$ ($\sim3.85\pm0.15$) \citep[for a description on reddening law estimation, see][]{2003AA...397..191P}, 
indicating larger grain sizes of the material in this region as compared to the general ISM.
In many SFRs, $R_V$s tend to deviate from the normal value, preferably towards the higher ones,
for example: $R_V$ = 3.7 \citep[][the Carina region]{2014A&A...567A.109K}, $R_V$= 3.3 \citep[][NGC 1931]{2013ApJ...764..172P},
$R_V$ = 3.5 \citep[][NGC  281]{2012PASJ...64..107S} and $R_V$ = 3.7 \citep[][Be 59]{2008MNRAS.383.1241P}.
Within dense dark clouds, the accretion of ice
mantles on grains and the coagulation due to grain collision can change the size distribution
leading to higher $R_V$ values \citep{1989ApJ...345..245C}.
The value of $R_V$ for the foreground population (blue dots in Fig. \ref{2color}) towards the direction of NGC 7538 
comes out to be $\sim2.82\pm0.10$,
indicating a slightly smaller grain size in the foreground medium of NGC 7538 as compared to the general ISM.
It is interesting to point out that 
\citet{2012MNRAS.419.2587E} have reported the mean value of $R_V$ as $2.79\pm0.18$ for the general ISM towards the Be 59 SFR ($l=118.22^\circ$, $b=5.00^\circ$), which is in a similar direction to NGC 7538, on the basis of polarimetric observations.

\bsp    
\label{lastpage}
\end{document}